\documentclass[twoside,11pt]{article}
\pdfoutput=1 %

\usepackage{jmlr2e}
\hypersetup{hidelinks} %

\usepackage{xr}
\usepackage{comment}
\usepackage{subcaption}
\usepackage{caption}
\usepackage{graphicx}
\usepackage{amsmath}
\usepackage{amssymb}
\usepackage{thmtools,thm-restate}
\usepackage{mathtools}
\usepackage{bbm}
\usepackage{wrapfig,lipsum,booktabs}
\usepackage{nameref}
\usepackage{tabularx}
\usepackage{multirow}
\usepackage{mathrsfs}
\usepackage{xcolor}
\usepackage[ruled,vlined]{algorithm2e}
\usepackage{multicol}
\newcommand{\R}{\mathbb{R}}

\newcommand{\abs}[1]{|#1|}

\newcommand{\E}{\mathbb{E}}
\newcommand{\Prob}{\mathbb{P}}

\newcommand{\eqas}{\overset{\mathrm{a.s.}}{=}}
\newcommand{\ind}{\mathbbm{1}}
\newcommand{\Law}{\mathrm{Law}}

\DeclarePairedDelimiterX{\infdivx}[2]{(}{)}{#1\;\delimsize\|\;#2}

\newcommand\ci{\perp\!\!\!\perp}

\newcommand{\dee}{\mathrm{d}}

\newcommand{\Xspacedim}{d}
\newcommand{\Xspace}{\mathcal{X}}
\newcommand{\Aspace}{\mathcal{A}}
\newcommand{\A}{A}
\newcommand{\X}{X}
\newcommand{\xx}{x}
\newcommand{\Xt}{\widehat{\X}}

\newcommand{\ax}{a}

\newcommand{\sx}{s}

\newcommand{\nx}{n}

\newcommand{\gx}{g}

\newcommand{\tx}{t}
\newcommand{\B}{B}
\newcommand{\N}{N}
\newcommand{\T}{T}

\newcommand{\fx}{f}

\newcommand{\lo}[1]{#1_{\mathrm{lo}}}
\newcommand{\up}[1]{#1_{\mathrm{up}}}

\newcommand{\Y}{Y}
\newcommand{\Yt}{\widehat{Y}}
\newcommand{\ylo}{\lo{y}}
\newcommand{\yup}{\up{y}}

\newcommand{\Ylo}{\lo{\Y}}
\newcommand{\Yup}{\up{\Y}}

\newcommand{\Q}{Q}
\newcommand{\Qt}{\widehat{\Q}}
\newcommand{\Qlo}{\lo{\Q}}
\newcommand{\Qup}{\up{\Q}}

\newcommand{\qlo}[1]{\lo{R}^{#1}}
\newcommand{\qup}[1]{\up{R}^{#1}}
\newcommand{\qt}[1]{\widehat{R}^{#1}}

\newcommand{\YClmean}{{\lo{\mu}}}

\newcommand{\Ytmean}{\widehat{\mu}}
\newcommand{\CIlen}{\Delta}

\newcommand{\Hyp}{\mathcal{H}}
\newcommand{\Hlo}{\lo{\Hyp}}
\newcommand{\plo}{p_{\textup{lo}}}
\newcommand{\Hup}{\up{\Hyp}}
\newcommand{\pup}{p_{\textup{up}}}

\newcommand{\D}{\mathcal{D}}
\newcommand{\Dt}{\widehat{\D}}

\newcommand{\twinfunction}{h}
\newcommand{\twinnoise}{U}
\newcommand{\ux}{u}

\newcommand{\AppendixName}{Appendix\xspace}

\makeatletter
\newcommand{\runinheadcore}{\@startsection{paragraph}{4}{\z@}{3.25ex \@plus1ex \@minus.2ex}{-1em}{\normalfont\normalsize\itshape}}
\makeatother
\newcommand{\runinhead}[1]{\runinheadcore{#1.}}
 
\usepackage{lastpage}
\jmlrheading{27}{2026}{1-\pageref{LastPage}}{10/24; Revised 4/26}{8/26}{24-1752}{Rob Cornish, Muhammad Faaiz Taufiq, Arnaud Doucet, and Chris Holmes}

\ShortHeadings{Causal Falsification of Digital Twins}{Cornish, Taufiq, Doucet, and Holmes}
\firstpageno{1}

\begin{document}

\title{Causal Falsification of Digital Twins}

\author{\name Rob Cornish$^\ast$ \email rob.cornish@ntu.edu.sg \\
       \addr College of Computing and Data Science\\
       Nanyang Technological University\\
       Singapore
       \AND
       \name Muhammad Faaiz Taufiq$^\ast$ \email muhammad.taufiq@stats.ox.ac.uk \\
       \addr Department of Statistics\\
       University of Oxford\\
       Oxford, United Kingdom
	   \AND
       \name Arnaud Doucet \email doucet@stats.ox.ac.uk \\
       \addr Department of Statistics\\
	   University of Oxford\\
	   Oxford, United Kingdom
	   \AND
       \name Chris Holmes \email cholmes@stats.ox.ac.uk \\
       \addr Department of Statistics\\
	   University of Oxford\\
	   Oxford, United Kingdom
	   }

\editor{Ilya Shpitser}

\maketitle

\def\thefootnote{*}\footnotetext{Both authors contributed equally to this work.}\def\thefootnote{\arabic{footnote}}

\begin{abstract}%
\emph{Digital twins} are simulation-based models designed to predict how a real-world process will evolve in response to interventions.
This modelling paradigm holds substantial promise in many applications, but rigorous procedures for assessing their accuracy are essential for safety-critical settings.
We consider how to assess the accuracy of a digital twin using real-world data.
We formulate this as a causal inference problem, which leads to a precise definition of what it means for a twin to be ``correct''. %
Unfortunately, fundamental results from causal inference mean observational data cannot be used to certify a twin in this sense unless potentially tenuous assumptions are made, such as that the data are unconfounded.
To avoid these assumptions, we propose instead to find situations in which the twin \emph{is not} correct, and present a general-purpose statistical procedure for doing so.
Our approach yields reliable and actionable information about the twin under only the assumption of an i.i.d.\ dataset of observational trajectories, and remains sound even if the data are confounded.
We apply our methodology to a large-scale, real-world case study involving sepsis modelling within the Pulse Physiology Engine, which we assess using the MIMIC-III dataset of ICU patients.
\end{abstract} 

\section{Introduction}

There is increasing interest in the use of simulation-based models for obtaining \emph{causal} insights.
Such models aim to describe what \emph{would} occur when different actions or interventions are applied to some real-world process of interest, thereby allowing planning and decision-making to be done with a fuller understanding of the different outcomes that may result.
In many applications, models of this kind are referred to as \emph{digital twins} \citep{barricelli2019survey,jones2020characterising,niederer2021scaling}.
These have been considered for a wide range of use-cases including aviation \citep{tuegel2011reengineering}, manufacturing \citep{lu2020digital}, healthcare \citep{corral2020digital,coorey2022health}, civil engineering \citep{sacks2020construction}, and agriculture \citep{jans2020digital}.

Many applications of digital twins are considered safety-critical, which means the cost of deploying an inaccurate twin to production is potentially very high.
As such, methodology for assessing the performance of a twin before its deployment is essential for the safe, widespread adoption of digital twins in practice \citep{niederer2021scaling}.
In this work, we consider the problem of assessing twin accuracy and propose a concrete, theoretically grounded, and general-purpose methodology to this end.
We focus specifically on the use of statistical methods that leverage data obtained from the real-world process that the twin is designed to model.
Such strategies are increasingly viable for many applications as datasets grow larger, and offer the promise of lower overheads compared with alternative strategies that rely for instance on domain expertise.

We formulate twin assessment as a problem of \emph{causal inference} \citep{rubin1974estimating,rubin2005causal,pearl2009causality,hernan2020causal}.
In particular, we consider a twin to be accurate if it correctly captures the behaviour of a real-world process of interest in response to certain interventions, rather than the behaviour of the process as it evolves on its own.
This seems in keeping with the overall (if often implicit) design objectives that underlie many applications, including those cited above.
Our approach also highlights certain pitfalls associated with conventional methods that do not account for causal factors, and that can give rise to misleading inferences about the twin as a result.

In most applications, it is desirable for an assessment procedure to be reliable and robust, and for its conclusions about the twin to be highly trustworthy.
As such, our goal in this paper is to obtain a methodology that is always \emph{sound}, even possibly at the expense of being conservative: we prefer not to draw any conclusion about the accuracy of the twin at all than to draw some conclusion that is potentially misleading.
To this end, we rely on minimal assumptions about the twin and the real-world process of interest.
In addition to improving robustness, this also means our resulting methodology is very general, and may be applied to a wide variety of twins across application domains.

\subsection{Contribution}

We begin by providing a causal model for a general-purpose twin and the data we have available for assessment.
We use this to show precisely that it is not possible to use observational data to \emph{certify} that the twin is causally accurate unless strong and often tenuous assumptions are made about the data-generating process, such as that the data are free of unmeasured confounding.
To avoid these assumptions, we propose an assessment paradigm instead based on \emph{falsification}: we search for specific cases when the twin fails to be correct, rather than trying to quantify its overall accuracy in a more holistic sense. %

To obtain a practical methodology suitable for real twins, we provide a novel set of longitudinal causal bounds that hold without additional assumptions.
These bounds generalize the classical result of \citet{manski}, and can be considerably more informative in comparison.
We use this result as the basis for a general-purpose statistical testing procedure for falsifying a twin.
Overall, our method relies on only the assumption of an i.i.d.\ dataset of observational trajectories: it does not require modelling the dynamics of the real-world process or any internal implementation details of the twin, and remains sound in the presence of arbitrary unmeasured confounding. %
We demonstrate the effectiveness of our procedure through a large-scale, real-world case study in which we use the MIMIC-III ICU dataset \citep{mimic} to assess the Pulse Physiology Engine \citep{pulse}, an open-source model for human physiology simulation.

\subsection{Related Work}

Various high-level guidelines and workflows have been proposed for the assessment of digital twins in the literature to date \citep{roy2011comprehensive,grieves2017digital,khan2018digital,corral2020digital,kochunas2021digital,niederer2021scaling,dahmen2022verification}. %
In some cases, these guidelines have been codified as standards: for example, the ASME V\&V40 Standard \citep{amse2018assessing} provides a risk-based framework for assessing the credibility of a model from a variety of factors that include source code quality and the mathematical form of the model \citep{galappaththige2022credibility}.
However, a significant gap still exists between these guidelines and a practical implementation that could be deployed for real twins, and the need for a rigorous lower-level framework to enable the systematic assessment of twins has been noted in this literature \citep{corral2020digital,niederer2021scaling,kapteyn2021probabilistic,masison2021modular}.
We contribute towards this effort by describing a precise statistical methodology for twin assessment that can be readily implemented in practice, and which is accompanied by theoretical guarantees of robustness that hold under minimal assumptions.

In addition, a variety of concrete assessment procedures have been applied to certain specific digital twin models in the literature.
For example, the Pulse Physiology Engine \citep{pulse}, which we consider in our empirical case study, as well as the related BioGears \citep{sepsis-modelling,biogears} were both assessed by comparing their outputs with ad hoc values based on available medical literature and the opinions of subject matter experts.
Other twins have been assessed by comparing their outputs with real-world data through a variety of bespoke numerical schemes \citep{larrabide2012fast,hemmler2019patient,DT-patient,jans2020digital,galappaththige2022credibility}.
In contrast, our paper proposes a general-purpose statistical procedure for assessing twins that may be applied generically across many applications and architectures. %
To the best of our knowledge, we are also the first to identify the need for a causal approach to twin assessment and the pitfalls that arise when causal considerations are not accounted for.

Overall, our assessment approach follows the guidelines for empirical analysis proposed by econometrician Charles Manski \citep{manski1995identification}, who advocates for researchers to establish a ``domain of consensus'' \citep{manski2009identification} by analysing their data initially under minimal assumptions, and only afterwards to consider what follows as additional assumptions are introduced.
To this end, Manski and others developed a literature on \emph{partial identification} \citep{manski1989anatomy,manski,balke1997bounds,manski1997monotone,manski2000monotone,manski2003partial} to provide statistical tools with which this high-level strategy may be implemented.
This literature provides a suite of tools for reasoning about models whose parameters cannot be point-identified using observational data.
Similar ideas and techniques have also appeared in more recent work \citep{bareinboim,namkoong2020offpolicy,hussain2022falsification,zhang2022partial}.
However, to the best of our knowledge, existing results in this vast literature are largely not tailored towards assessing twin models, due in particular to their online and longitudinal structure, which complicates analysis.
In particular, a direct application of Manski's classical bounds \citep{manski} to our conditional longitudinal setting would not be identifiable in general.
Accordingly, in Section \ref{sec:causal-bounds} we provide a novel set of identifiable bounds on Manski's bounds, which we use as the basis for our practical methodology in Section \ref{sec:hypotheses-from-causal-bounds}. 
Our bounds require minimal assumptions, and we provide various analyses showing they cannot be improved upon unless additional assumptions are added.

The partial identification literature has developed a suite of inference techniques for parameters that are not point-identified.
A major line of work provides confidence sets for individual parameters in these settings \citep{imbens2004confidence,rosen2008confidence,romano2014practical}, while another constructs confidence sets for the entire identification region \citep{chernozhukov2007estimation,romano2010inference}.
These methods often seek a high level of generality, which has led to various inferential machinery based on subsampling, moment selection, and related asymptotic techniques \citep{romano2008inference,andrews2010inference} (see \citealt{molinari2020chapter} and \citealt{canay2023users} for reviews).
Our testing procedure described in Section~\ref{sec:hypotheses-from-causal-bounds} follows the same overall logic as this second line of work, but with our end goal being testing rather than obtaining confidence sets.
This reduces the problem to one-sided tests of conditional expectations, for which Hoeffding's inequality yields valid finite-sample type~I error control without asymptotic approximations.
A separate line of work considers specification tests for whether a partially identified region is non-empty \citep{bugni2015specification,kitagawa2015test,mourifie2017testing}.
This is distinct from our problem, since our bounds are always guaranteed to be nonempty by construction.

\section{Causal Formulation} \label{sec:causal-formulation}

\subsection{The Real-World Process} \label{sec:real-world-process}

We begin by providing a causal model of the real-world process that the twin is designed to simulate.
We do so in the language of \emph{potential outcomes} \citep{rubin1974estimating,rubin2005causal}, although we could have used the alternative framework of directed acyclic graphs and structural causal models \citep{pearl2009causality}. %
We assume the real-world process operates over a fixed time horizon $\T \in \{1, 2, \ldots\}$.
This simplifies our presentation in what follows, and it is straightforward to generalize our methodology to variable length time horizons if needed.
For each $\tx \in \{0, \ldots, \T\}$, we assume the process gives rise to an \emph{observation} at time $\tx$, which takes values in some real-valued space $\Xspace_\tx \coloneqq \R^{\Xspacedim_\tx}$.
We also assume that the process can be influenced by some \emph{action} taken at each time $\tx \in \{1, \ldots, \T\}$.
We denote the space of actions available at time $\tx$ by $\Aspace_\tx$, which in this work we assume is always finite.
For example, in a robotics context, the observations may consist of all the readings of all the sensors of the robot, and the actions may consist of commands that can be input by an external user.
In a medical context, the observations may consist of the vital signs of a patient, and the actions may consist of possible treatments or interventions.
To streamline notation, we will index these spaces using vector notation, so that e.g.\ $\Aspace_{1:\tx}$ denotes the cartesian product $\Aspace_\sx \times \cdots \times \Aspace_\tx$, and $\ax_{\sx:\tx} \in \Aspace_{\sx:\tx}$ is a choice of $\ax_\sx \in \Aspace_\sx, \ldots, \ax_\tx \in \Aspace_\tx$.
We also allow $\sx > \tx$ in this notation, so that e.g.\ $\ax_{\sx:\tx}$ denotes the empty sequence $\emptyset$, and $\Aspace_{\sx:\tx}$ denotes the set $\{\emptyset\}$ containing just the empty sequence and nothing else (so we still have $\ax_{\sx:\tx} \in \Aspace_{\sx:\tx}$).

We model the dynamics of the real-world process via the longitudinal potential outcomes framework proposed by \citet{robins1986new}, which imposes only a weak temporal structure on the underlying phenomena of interest and so may be applied across a wide range of applications in practice.
In particular, for each $\ax_{1:\T} \in \Aspace_{1:\T}$, we posit the existence of random variables or \emph{potential outcomes} $\X_0, \X_1(\ax_1), \ldots, \X_\T(\ax_{1:\T})$, where $\X_\tx(\ax_{1:\tx})$ takes values in $\Xspace_\tx$.
We will denote this sequence more concisely as $\X_{0:\T}(\ax_{1:\T})$.
Intuitively, $\X_0$ represents data available before the first action, while $\X_{1:\T}(\ax_{1:\T})$ represents the sequence of real-world outcomes that would occur if actions $\ax_{1:\T}$ were taken successively.
These quantities are therefore of fundamental interest for planning a course of actions to achieve some desired result.

As random variables, each $\X_{\tx}(\ax_{1:\tx})$ may depend on additional randomness that is not explicitly modelled, and so in particular may be influenced by all the previous potential outcomes $\X_{0:\tx-1}(\ax_{1:\tx-1})$, and possibly other random quantities.
Notice however that each $\X_{\sx}(\ax_{1:\sx})$ does \emph{not} depend on the subsequent actions $\ax_{\sx+1:\T}$, which amounts to saying that future actions do not affect past outcomes.
This models a process whose initial state is determined by external factors, such as when a patient from some population first presents at a hospital, and where the process then evolves according to specific actions chosen from $\Aspace_{1:\T}$ as well as exogenous factors.
It is clear that this structure applies to a wide range of phenomena occurring in practice.

\subsection{The Digital Twin}

We think of the twin as a computational device that, when executed, outputs a sequence of values intended to simulate a possible future trajectory of the real-world process when certain actions in $\Aspace_{1:\T}$ are chosen, conditional on some initial data in $\Xspace_0$.
We allow the twin to make use of an internal random number generator to produce outputs that vary stochastically even under fixed inputs (although our framework encompasses twins that evolve deterministically also).
By executing the twin repeatedly, a user may therefore estimate the range of behaviours that the real-world process may exhibit under different action sequences, which can then inform planning and decision-making downstream.

Precisely, we model the output that the twin would produce at timestep $\tx \in \{1, \ldots, \T\}$ after receiving initialisation $\xx_0 \in \Xspace_0$ and successive inputs $\ax_{1:\tx} \in \Aspace_{1:\tx}$ as the quantity $\twinfunction_\tx(\xx_0, \ax_{1:\tx}, \twinnoise_{1:\tx})$, where $\twinfunction_\tx$ is a measurable function taking values in $\Xspace_{\tx}$, and each $\twinnoise_\sx$ is some random variable.
We will denote $\Xt_{\tx}(\xx_0, \ax_{1:\tx}) \coloneqq \twinfunction_\tx(\xx_0, \ax_{1:\tx}, \twinnoise_{1:\tx})$, which we also refer to as a potential outcome.
A full twin trajectory therefore consists of $\Xt_1(\xx_0, \ax_1), \ldots, \Xt_\T(\xx_0, \ax_{1:\T})$, which we write more compactly by $\Xt_{1:\T}(\xx_0, \ax_{1:\T})$.
Conceptually, $\twinfunction_1, \ldots, \twinfunction_\T$ constitute the program that executes inside the twin, and $\twinnoise_{1:\T}$ may be thought of as the collection of all outputs of the internal random number generator that the twin uses.
We assume these internal random numbers $\twinnoise_{1:\T}$ and the initial real-world observations $\X_0$ are independent, which is mild in practice.
We also assume that repeated executions of the twin give rise to i.i.d.\ copies of $\twinnoise_{1:\T}$.
This means that, given fixed inputs $\xx_0$ and $\ax_{1:\T}$, repeated executions of the twin produce i.i.d.\ copies of $\Xt_{1:\T}(\xx_0, \ax_{1:\T})$.
Otherwise, we make no assumptions about the precise form of either the $\twinfunction_\tx$ or the $\twinnoise_\tx$, which allows our model to encompass a wide variety of possible twin implementations.

\subsection{Correctness}

Before we can consider how to assess the twin, we must first define how we want the twin ideally to behave.
The following condition seems appropriate for many applications.

\begin{definition} \label{eq:interventional-correctness}
    The twin is \emph{interventionally correct} if, for $\Law[\X_0]$-almost all $\xx_0 \in \Xspace_0$ and $\ax_{1:\T} \in \Aspace_{1:\T}$, the distribution of $\Xt_{1:\T}(\xx_0, {\ax}_{1:\T})$ is equal to the conditional distribution of $\X_{1:\T}(\ax_{1:\T})$ given $\X_0 = \xx_0$.
\end{definition}

Operationally, if a twin is interventionally correct, then by repeatedly executing the twin and applying Monte Carlo techniques, it is possible to approximate arbitrarily well the conditional distribution of the future of the real-world process under each possible choice of action sequence.
The same can also be shown to hold when each action at each time $\tx$ is chosen dynamically on the basis of previous observations in $\Xspace_{0:\tx}$. %
As a result, an interventionally correct twin may be used for \emph{planning}, or in other words may be used to select a policy for choosing actions that will yield a desirable distribution over observations at each step.
We emphasise that interventional correctness does not mean the twin will accurately predict the behaviour of any \emph{specific} trajectory of the real-world process in an almost sure sense (unless the real-world process is deterministic), but only the distribution of outcomes that will be observed over repeated independent trajectories.
However, this is sufficient for many applications, and appears to be the strongest guarantee possible when dealing with real-world phenomena whose underlying behaviour is stochastic.

Definition \ref{eq:interventional-correctness} introduces some technical difficulties that arise in the general case when conditioning on events with probability zero (e.g.\ $\{\X_0 = \xx_0\}$ if $\X_0$ is continuous).
In what follows, it is more convenient to consider an unconditional formulation of interventional correctness.
This is supplied by the following straightforward result, which considers the behaviour of the twin when it is initialised with the (random) value of $\X_0$ taken from the real-world process, rather than with a fixed choice of $\xx_0$.
See Section \ref{sec:unconditional-interventional-correctness-proof} of the \AppendixName for a proof.

\begin{proposition} \label{prop:interventional-correctness-alternative-characterisation}
    The twin is interventionally correct if and only if, for all choices of $\ax_{1:\T} \in \Aspace_{1:\T}$, the distribution of $(\X_0, \Xt_{1:\T}(\X_0, \ax_{1:\T}))$ is equal to the distribution of $\X_{0:\T}(\ax_{1:\T})$.
\end{proposition}

\subsection{Online Prediction}

Our model here represents a twin at time $\tx = 0$ making predictions about all future timesteps $\tx \in \{1, \ldots, \T\}$ under different choices of inputs $\ax_{1:\T}$.
In practice, many twins are designed to receive new information at each timestep in an online fashion and update their predictions for subsequent timesteps accordingly \citep{grieves2017digital,niederer2021scaling}.
Various notions of correctness can be devised for this online setting.
We describe two possibilities in Section \ref{sec:online-prediction} of the \AppendixName, and show that these notions of correctness essentially reduce to Definition \ref{eq:interventional-correctness}, which motivates our focus on that notion in what follows.

\section{Data-Driven Twin Assessment} \label{sec:data-driven-twin-assessment}

There are many conceivable methods for assessing the accuracy of a twin, including static analysis of the twin's source code and the solicitation of domain expertise, and in practice it seems most robust to use a combination of different techniques rather than relying on any single one \citep{amse2018assessing,niederer2021scaling}.
However, in this paper, we focus on what may be called \emph{data-driven assessment}, which we see as an important component of a larger assessment pipeline. %
That is, we consider the use of statistical methods that rely solely on a dataset of trajectories obtained from the real-world process and the twin.
Unfortunately, as we now explain, fundamental results from the causal inference literature mean that it is not possible to obtain a data-driven assessment procedure that can \emph{certify} that a twin is interventionally correct without further assumptions.
Accordingly, we propose a strategy based on \emph{falsifying} the twin, which we develop into a concrete statistical testing procedure in later sections.

We will assume access to a dataset of trajectories obtained by observing the interaction of some behavioural agents with the real-world process.
We model each trajectory as follows.
First, we represent the action chosen by the agent at time $\tx \in \{1, \ldots, \T\}$ as an $\Aspace_\tx$-valued random variable $\A_\tx$.
We then obtain a trajectory in our dataset by recording at each step the action $\A_\tx$ chosen and the observation $\X_\tx(\A_{1:\tx})$ corresponding to this choice of action.
As a result, each observed trajectory has the following form:
\begin{equation} \label{eq:observed-data-trajectory}
    \X_0, \A_1, \X_1(\A_1), \ldots, \A_\T, \X_\T(\A_{1:\T}).
\end{equation}
This is the standard \emph{consistency} assumption in causal inference \citep{hernan2020causal}, and intuitively means the potential outcome $\X_\tx(\ax_{1:\tx})$ is observed in the data when the agent actually chose $\A_{1:\tx} = \ax_{1:\tx}$.
We model our full dataset as a set of i.i.d.\ copies of \eqref{eq:observed-data-trajectory}. %

\subsection{Certification Is Unsound in General}

A natural high-level strategy for twin assessment has the following structure.
First, some hypothesis $\Hyp$ is chosen with the following property:
 \begin{equation} \label{eq:verification-hypothesis-property}
    \text{If $\Hyp$ is true, then the twin is interventionally correct.}%
\end{equation}
Data is then used to try to show $\Hyp$ is true, perhaps up to some level of confidence. %
If successful, it follows by construction that the twin is interventionally correct.
Assessment procedures designed to \emph{certify} the twin in this way are appealing because they promise a strong guarantee of accuracy for certified twins. %
Unfortunately, the following foundational result from the causal inference literature (often referred to as the \emph{fundamental problem of causal inference}; \citealp{holland1986statistics}) means that data-driven certification procedures of this kind are in general unsound, as we explain next.
This is a standard and well-known result within the field of causal inference, but appears to be less widely known within the digital twin literature.
For completeness, Section \ref{sec:non-identifiability-result-proof-supp} of the \AppendixName includes a self-contained proof of this result in our notation.

\begin{theorem} \label{prop:nonidentifiability}
    If $\Prob(\A_{1:\T} \neq \ax_{1:\T}) > 0$, then the distribution of $\X_{0:\T}(\ax_{1:\T})$ is not uniquely identified by the distribution of the data in \eqref{eq:observed-data-trajectory} without further assumptions.
\end{theorem}

Since the distribution of the data encodes the information that would be contained in an infinitely large dataset of trajectories, Theorem \ref{prop:nonidentifiability} imposes a fundamental limit on what can be learned about the distribution of $\X_{0:\T}(\ax_{1:\T})$ from the data we have assumed.
It follows that if $\Hyp$ is any hypothesis satisfying \eqref{eq:verification-hypothesis-property}, then $\Hyp$ cannot be determined to be true from even an infinitely large dataset.
This is because, if we could do so, then we could also determine the distribution of $\X_{0:\T}(\ax_{1:\T})$, since by Proposition \ref{prop:interventional-correctness-alternative-characterisation} this would be equal to the distribution of $(\X_0, \Xt_{\T}(\X_0, \ax_{1:\T}))$.
In other words, we cannot use the data alone to certify that the twin is interventionally correct.

\subsection{The Assumption of No Unmeasured Confounding} \label{sec:no-unmeasured-confounding-assumption}

Theorem \ref{prop:nonidentifiability} is true in the general case, when no additional assumptions about the data-generating process are made.
One way forward is therefore to introduce assumptions under which the distribution of $\X_{0:\T}(\ax_{1:\T})$ \emph{can} be identified.
This would mean it is possible to certify that the twin is interventionally correct, since, at least in principle, we could simply check whether this matches the distribution of $(\X_0, \Xt_{1:\T}(\X_{0}, \ax_{1:\T}))$ produced by the twin.

The most common such assumption in the causal inference literature is that the data are free of \emph{unmeasured confounding}.
Informally, this holds when each action $\A_\tx$ is chosen by the behavioural agent solely on the basis of the information available at time $\tx$ that is actually recorded in the dataset, namely $\X_{0}, \A_1, \X_1(\A_1), \ldots, \A_{\tx-1}, \X_{\tx-1}(\A_{1:\tx-1})$, as well as possibly some additional randomness that is independent of the real-world process, such as the outcome of a coin toss. (This can be made precise via the \emph{sequential randomisation assumption} introduced by \citet{robins1986new}.)
Unobserved confounding is present whenever this does not hold, i.e.\ whenever some unmeasured factor simultaneously influences both the agent's choice of action and the observation produced by the real-world process.

It is reasonable to assume that the data are unconfounded in certain contexts.
For example, it may be possible to gather data in a way that specifically guarantees there is no confounding.
Randomised controlled trials, which ensure that each $\A_\tx$ is chosen via a carefully designed randomisation procedure \citep{lavori2004dynamic,murphy2005experimental}, constitute a widespread example of this approach.
Likewise, it is possible to show that the data are unconfounded if each $\X_\tx(\ax_{1:\tx})$ is a deterministic function of $\X_{0:\tx-1}(\ax_{1:\tx-1})$ and $\ax_{\tx}$, which may be reasonable to assume for example in certain low-level physics or engineering contexts.
(See Section \ref{eq:deterministic-potential-outcomes-are-unconfounded} of the \AppendixName for a proof.)
However, for stochastic phenomena and for typical datasets, it is widely acknowledged that the assumption of no unmeasured confounding will rarely hold \citep{murphy2003optimal,tsiatis2019dynamic}, and so assessment procedures based on this assumption may yield unreliable results in practice.
Section \ref{sec:motivating-example} of the \AppendixName illustrates this concretely with a toy scenario.

\subsection{General-Purpose Assessment via Falsification}

Our goal is to obtain an assessment methodology that is general-purpose, and as such we would like to avoid introducing assumptions such as unconfoundedness that do not hold in general.
To achieve this, borrowing philosophically from \citet{popper2005logic}, we propose a strategy that replaces the goal of verifying the interventional correctness of the twin with that of \emph{falsifying} it.
Specifically, we consider hypotheses $\Hyp$ with the dual property to \eqref{eq:verification-hypothesis-property}:
\begin{equation} \label{eq:falsification-hypothesis-property}
    \text{If the twin is interventionally correct, then $\Hyp$ is true.}
\end{equation}
We will then try to show that each such $\Hyp$ is \emph{false}.
Whenever we are successful, we will thereby have gained some knowledge about a failure mode of the twin, since by construction the twin can only be correct if $\Hyp$ is true.
In effect, each $\Hyp$ we falsify will constitute a reason that the twin is not correct, and may suggest concrete improvements to its design, or may identify cases where its output should not be trusted.

Importantly, unlike for \eqref{eq:verification-hypothesis-property}, Theorem \ref{prop:nonidentifiability} does not preclude the possibility of data-driven assessment procedures based on \eqref{eq:falsification-hypothesis-property}.
As we show below, there do exist hypotheses $\Hyp$ satisfying \eqref{eq:falsification-hypothesis-property} that can in principle be determined to be false from the data alone without additional assumptions.
In this sense, falsification provides a means for \emph{sound} data-driven twin assessment, %
whose results can be relied upon across a wide range of circumstances.
On the other hand, falsification approaches cannot provide a \emph{complete} guarantee about the accuracy of a twin: even if we fail to falsify many $\Hyp$ satisfying \eqref{eq:falsification-hypothesis-property}, we cannot then infer that the twin is correct.
As such, in situations where (for example) it is reasonable to believe that the data are in fact unconfounded, it may be desirable to use this assumption to obtain additional information about the twin than is possible from falsification alone.

\section{Longitudinal Causal Bounds} \label{sec:causal-bounds}

\subsection{Manski's Bounds}

A classical result by \citet{manski} gives a worst-case bound on the average interventional behaviour of a causal process.
In our context, and using the notation established above, this result gives the following:

\begin{theorem}[\citealp{manski}] \label{thm:manski-bounds}
    Suppose $(\Y(\ax_{1:\tx}') : \ax_{1:\tx}' \in \Aspace_{1:\tx})$ are real-valued potential outcomes defined jointly with $(\X_{0:\T}(\ax_{1:\T}') : \ax_{1:\T}' \in \Aspace_{1:\T})$ and $\A_{1:\T}$.
    Suppose that for some $\tx \in \{1, \ldots, \T\}$, $\ax_{1:\tx} \in \Aspace_{1:\tx}$, measurable $\B_{0:\tx} \subseteq \Xspace_{0:\tx}$, and $\ylo, \yup \in \R$ we have
    \begin{gather}
        \Prob(\X_{0:\tx}(\ax_{1:\tx}) \in \B_{0:\tx}) > 0 \label{eq:B-positivity-assumption} \\
        \Prob(\ylo \leq \Y(\ax_{1:\tx}) \leq \yup \mid \X_{0:\tx}(\ax_{1:\tx}) \in \B_{0:\tx}) = 1.  \label{eq:Y-boundedness-assumption}
    \end{gather}
    It then holds that
    \begin{equation} \label{eq:manski-bounds-fully-written}
        \E[\Ylo \mid \X_{0:\tx}(\ax_{1:\tx}) \in \B_{0:\tx}]
        \leq \E[\Y(\ax_{1:\tx}) \mid \X_{0:\tx}(\ax_{1:\tx}) \in \B_{0:\tx}]
        \leq  \E[\Yup \mid \X_{0:\tx}(\ax_{1:\tx}) \in \B_{0:\tx}],
    \end{equation}
    where we denote
    \begin{align*}
        \Ylo \coloneqq \ind(\A_{1:\tx} &= \ax_{1:\tx}) \, \Y(\A_{1:\tx}) + \ind(\A_{1:\tx} \neq \ax_{1:\tx}) \, \ylo \\
        \Yup \coloneqq \ind(\A_{1:\tx} &= \ax_{1:\tx}) \, \Y(\A_{1:\tx}) + \ind(\A_{1:\tx} \neq \ax_{1:\tx}) \, \yup.
    \end{align*}
\end{theorem}

Specifically, this follows by instantiating the main result of \citet{manski} with everything conditional on the event $\{\X_{0:\tx}(\ax_{1:\tx}) \in \B_{0:\tx}\}$, which has nonzero probability by \eqref{eq:B-positivity-assumption}.

The potential outcome $\Y(\ax_{1:\tx})$ may be thought of as some quantitative outcome of interest.
For example, in a medical context, $\Y(\ax_{1:\tx})$ might represent the heart rate of a patient at time $\tx$ after receiving some treatments $\ax_{1:\tx}$.
When defining our hypotheses below, we consider the specific form $\Y(\ax_{1:\tx}) \coloneqq \fx(\X_{0:\tx}(\ax_{1:\tx}))$, where $\fx : \Xspace_{0:\tx} \to \R$ is some scalar function.
The bounded quantity $\E[\Y(\ax_{1:\tx}) \mid \X_{0:\tx}(\ax_{1:\tx}) \in \B_{0:\tx}]$ in \eqref{eq:manski-bounds-fully-written} then just gives the average behaviour of this outcome.

From a methodological perspective, Theorem \ref{thm:manski-bounds} has the following well-known consequence.
By taking $\B_{1:\tx} = \Xspace_{1:\tx}$ to be the whole space for all but the initial timestep, the bounds in Theorem \ref{thm:manski-bounds} specialise to the following:
\begin{equation} \label{eq:manski-bounds-fully-written-special-case}
    \E[\Ylo \mid \X_{0} \in \B_{0}] \leq \E[\Y(\ax_{1:\tx}) \mid \X_{0} \in \B_{0}] \leq  \E[\Yup \mid \X_{0} \in \B_{0}].
\end{equation}
By Theorem \ref{prop:nonidentifiability}, the middle term is in general not identified by the data since it depends on $\Y(\ax_{1:\tx})$, which is not observed unless $\A_{1:\tx} = \ax_{1:\tx}$.
On the other hand, both the lower and upper bounds \emph{are} identified, since the relevant random variables $\Ylo$, $\Yup$, and $\X_0$ can all be expressed as functions of the observed data $\X_{0:\tx}(\A_{1:\tx})$ and $\A_{1:\tx}$.
In this way, Theorem \ref{thm:manski-bounds} bounds the behaviour of a non-identifiable quantity in terms of identifiable ones.
At a high level, this is achieved by replacing $\Y(\ax_{1:\tx})$, whose value is only observed on the event $\{\A_{1:\tx} = \ax_{1:\tx}\}$, with $\Ylo$ and $\Yup$, which are equal to $\Y(\ax_{1:\tx})$ when its value is observed (i.e.\ when $\A_{1:\tx} = \ax_{1:\tx}$), and which fall back to the worst-case values of $\ylo$ and $\yup$ otherwise (see e.g.\ \citealt[Section 6.1]{imbens2009recent}; \citealt[Sections 2.1--2.2]{molinari2020chapter}).
This does not require any additional causal assumptions beyond \eqref{eq:B-positivity-assumption} and \eqref{eq:Y-boundedness-assumption}, and in particular remains true under arbitrary unmeasured confounding.

However, given general conditioning sets $\B_{1:\tx} \subsetneq \Xspace_{1:\tx}$, the bounds in Theorem \ref{thm:manski-bounds} are usually \emph{not} identifiable.
This is because they depend on the quantity $\X_{0:\tx}(\ax_{1:\tx})$, which is not observed in the data when $\A_{1:\tx} \neq \ax_{1:\tx}$.
In other words, Manski's bounds do not allow for conditioning on the intermediate interventional behaviour of the real-world process: they can only be applied to bound the average interventional behaviour of the real-world process with all longitudinal information ignored except for the initial observation $\X_0$.
For the falsification methodology we develop in Section \ref{sec:hypotheses-from-causal-bounds}, this is a significant limitation, since conditioning on intermediate observations yields a richer family of hypotheses and more granular information about the twin's accuracy (see Section \ref{sec:informativeness-of-our-bounds}).

\subsection{Identifiable Conditional Bounds} \label{sec:causal-bounds-statement}

As a way forward, we provide \emph{bounds on the bounds} \eqref{eq:manski-bounds-fully-written} in terms of observed quantities, thereby permitting their use in our falsification procedure below.
In particular, we have the following key result:

\begin{lemma} \label{lem:causal-bounds}
Consider the same setup as Theorem \ref{thm:manski-bounds}, and define
\[
    \N \coloneqq \max \{0 \leq \sx \leq \tx \mid \A_{1:\sx} = \ax_{1:\sx}\},
\]
where this maximum is well-defined since $\A_{1:0} = \emptyset = \ax_{1:0}$ by our convention from Section \ref{sec:real-world-process}.
The upper and lower bounds in \eqref{eq:manski-bounds-fully-written} are then themselves bounded as follows:
\begin{align}
    \E[\Ylo \mid \X_{0:\tx}(\ax_{1:\tx}) \in \B_{0:\tx}] &\geq \E[\Ylo \mid \X_{0:\N}(\A_{1:\N}) \in \B_{0:\N}] \label{eq:causal-bounds-lower} \\
    \E[\Yup \mid \X_{0:\tx}(\ax_{1:\tx}) \in \B_{0:\tx}]
        &\leq \E[\Yup \mid \X_{0:\N}(\A_{1:\N}) \in \B_{0:\N}]. \notag
\end{align}
\end{lemma}

See Appendix \ref{sec:causal-bounds-proofs} for a proof.
Intuitively, the random variable $\N$ encodes the largest value of $\sx \in \{0, \ldots, \tx\}$ for which the prefix $\ax_{1:\sx}$ of the full action sequence $\ax_{1:\tx}$ agrees with the actions $\A_{1:\sx}$ actually taken by the agent up to that timestep.
Another way to write this is as follows:
\[
    \N = \sum_{\sx=1}^{\tx} \ind(\A_{1:\sx} = \ax_{1:\sx}).
\]
It follows that $\N$ and hence $\X_{0:\N}(\A_{1:\N})$ are always observed by construction, making the right-hand sides of \eqref{eq:causal-bounds-lower} identifiable.

By combining Lemma \ref{lem:causal-bounds} with \eqref{eq:manski-bounds-fully-written}, we immediately obtain the following identifiable counterpart of Theorem \ref{thm:manski-bounds}.
This result will form the basis for the falsification procedure we develop in Section \ref{sec:hypotheses-from-causal-bounds}.

\begin{theorem} \label{thm:causal-bounds}
    Under the same setup as Theorem \ref{thm:manski-bounds}, and with $\N$ as defined in Lemma \ref{lem:causal-bounds}, it holds that
   \begin{equation} \label{eq:causal-bounds-fully-written}
        \Qlo \leq \Q \leq \Qup,
    \end{equation}
    where we denote
    \begin{align*}
        \Q &\coloneqq \E[\Y(\ax_{1:\tx}) \mid \X_{0:\tx}(\ax_{1:\tx}) \in \B_{0:\tx}] \\
        \Qlo &\coloneqq \E[\Ylo \mid \X_{0:\N}(\A_{1:\N}) \in \B_{0:\N}] \\
        \Qup &\coloneqq \E[\Yup \mid \X_{0:\N}(\A_{1:\N}) \in \B_{0:\N}].
    \end{align*}
\end{theorem}

Another interpretation of this result is as follows.
Analogously to Theorem \ref{thm:manski-bounds}, by applying the classical result of \citet{manski} with everything conditioned on the event $\{\X_{0:\N}(\A_{1:\N}) \in \B_{0:\N}\}$, we obtain directly
\[
    \Qlo \leq \E[\Y(\ax_{1:\tx}) \mid \X_{0:\N}(\ax_{1:\N}) \in \B_{0:\N}] \leq \Qup.
\]
Theorem \ref{thm:causal-bounds} shows that these bounds are also valid when the middle term here is replaced by $\Q$.
This is a more useful quantity for practical purposes than $\E[Y(a_{1:t}) \mid X_{0:N}(a_{1:N}) \in B_{0:N}]$, since the interaction between the random quantity $N$ and the sequence of observations $X_{0:N}(a_{1:N})$ makes the latter harder to interpret and use in practice.

In the structural causal modelling framework \citep{pearl2009causality}, a related result to Theorem \ref{thm:causal-bounds} was given as Corollary 1 by \citet{bareinboim}.
However, their result involves a complicated ratio of unknown quantities that makes estimation of their bounds difficult, since it is not obvious how to obtain an unbiased estimator for their ratio term.
In contrast, our proposed bounds are considerably simpler, since both $\Qlo$ and $\Qup$ here are expressed as (conditional) expectations. 
This makes their unbiased estimation straightforward, which we use to obtain exact confidence intervals for both terms in Section \ref{sec:statistical-methodology}.

\subsection{Informativeness} \label{sec:informativeness-of-our-bounds}

For Theorem \ref{thm:causal-bounds} to be useful in practice, we would like the bounds $[\Qlo, \Qup]$ to be relatively narrow compared with the worst-case bounds $[\ylo, \yup]$ that are trivially implied by \eqref{eq:Y-boundedness-assumption}.
We can quantify the extent to which this occurs by the ratio
\begin{equation} \label{eq:tightness-of-bounds-measure}
    \frac{\Qup - \Qlo}{\yup - \ylo} = 1 - \Prob(\A_{1:\tx} = \ax_{1:\tx} \mid \X_{0:\N}(\A_{1:\N}) \in \B_{0:\N}),
\end{equation}
where the equality here follows from the definitions of $\Qlo$ and $\Qup$ together with some straightforward manipulations.
In other words, the (relative) tightness of our bounds is determined by the value of $\Prob(\A_{1:\tx} = \ax_{1:\tx} \mid \X_{0:\N}(\A_{1:\N}) \in \B_{0:\N})$, which is itself closely related to the classical \emph{propensity score} in the causal literature \citep{rosenbaum1983central}.
Intuitively, as this probability grows, so does $\Prob(\Y(\ax_{1:\tx}) = \Y(\A_{1:\tx}) \mid \X_{0:\N}(\A_{1:\N}) \in \B_{0:\N})$, which means the effect of unmeasured confounding on the value of $\Q$ is reduced, leading to tighter bounds.

The bounds \eqref{eq:manski-bounds-fully-written-special-case} that follow directly from the original result of \citet{manski} are often regarded as quite uninformative.
These correspond to the special case $\B_{1:\tx} = \Xspace_{1:\tx}$, and inspection of \eqref{eq:tightness-of-bounds-measure} shows their width is relatively large whenever $\Prob(\A_{1:\tx} = \ax_{1:\tx} \mid \X_0 \in \B_0)$ is small.
This occurs in many applications, particularly for longer action sequences.
On the other hand, it often seems reasonable to expect that certain longer action sequences will be fairly likely to occur when conditioned on some intermediate observations.
In other words, $\Prob(\A_{1:\tx} = \ax_{1:\tx} \mid \X_{0:\N}(\A_{1:\N}) \in \B_{0:\N})$ may be large (which means $[\Qlo, \Qup]$ is narrow), even if $\Prob(\A_{1:\tx} = \ax_{1:\tx} \mid \X_0 \in \B_0)$ is small.
By choosing $\B_{0:\tx}$ carefully, we therefore have an opportunity to obtain narrower bounds than those provided by Manski's original result.
The following result provides a straightforward sufficient condition for this to hold.

\begin{proposition} \label{prop:our-bounds-vs-manskis}
    Consider the same setup as Theorem \ref{thm:causal-bounds}, and suppose it also holds that $\Prob(\ylo \leq \Y(\ax_{1:\tx}) \leq \yup \mid \X_0 \in \B_0) = 1$.
    Then the width of Manski's bounds \eqref{eq:manski-bounds-fully-written-special-case} exceeds that of Theorem \ref{thm:causal-bounds}, i.e.\
    \[
        \E[\Yup \mid \X_0 \in \B_0] - \E[\Ylo \mid \X_0 \in \B_0] > \Qup - \Qlo,
    \]
    whenever we have $\Prob(\A_{1:\tx} = \ax_{1:\tx} \mid \X_{0:\N}(\A_{1:\N}) \in \B_{0:\N}) > \Prob(\A_{1:\tx} = \ax_{1:\tx} \mid \X_0 \in \B_0)$.
\end{proposition}

We note that Proposition \ref{prop:our-bounds-vs-manskis} compares bounds on two different quantities, namely $\Q$ (for our bounds) and $\E[\Y(\ax_{1:\tx}) \mid \X_{0} \in \B_{0}]$ (for Manski's bounds).
However, it still appears reasonable to compare the widths of both bounds, since these two quantities share the same worst-case behaviour of $[\ylo, \yup]$ in both cases.

Beyond the appeal of tightness, the conditional nature of Theorem \ref{thm:causal-bounds} also appears of interest simply for its own sake.
In particular, Theorem \ref{thm:causal-bounds} describes the interventional behaviour of $\Y(\ax_{1:\tx})$ conditional on the behaviour of the trajectory $\X_{0:\tx}(\ax_{1:\tx})$, thereby providing more granular information than can be obtained from Manski's bounds \eqref{eq:manski-bounds-fully-written-special-case} alone.
In Section \ref{sec:case-study} below, we also demonstrate empirically that conditioning on intermediate observations leads to substantially more informative bounds in our case study.

\subsection{Sharpness}

An immediate question is whether our bounds in Theorem \ref{thm:causal-bounds} are optimal.
For example, an alternative approach to obtaining bounds is to consider actions taken by the agent after time $\tx$.
This yields the following lower bound:
\begin{equation} \label{eq:alternative-bounds}
    \sup_{\ax_{\tx+1:\T}' \in \Aspace_{\tx+1:\T}}
    \E[\Y(\A_{1:\tx}) \ind\{\A_{1:\T} = (\ax_{1:\tx}, \ax_{\tx+1:\T}')\} + \ylo \ind\{\A_{1:\T} \neq (\ax_{1:\tx}, \ax_{\tx+1:\T}')\}]
    \leq \E[\Y(\ax_{1:\tx})],
\end{equation}
since the bound holds for any fixed $\ax_{\tx+1:\T}'$ when the supremum is removed from the left-hand side.
The cases of the upper bound and of nontrivial conditioning event are then also similar.

The following result shows that Theorem \ref{thm:causal-bounds} cannot be improved without further assumptions.
Intuitively speaking, the idea is that there always exists \emph{some} family of potential outcomes that produces the same observational data as our model, but that attains the worst-case bounds $\Qlo$ or $\Qup$.
Therefore, we cannot rule out the possibility that the true potential outcomes achieve $\Qlo$ or $\Qup$ from the observational data alone.
In particular this means that the alternative lower bound \eqref{eq:alternative-bounds} (which only depends on the observational distribution) is in fact not tighter than $\Qlo$.\footnote{This can also be seen directly as a consequence of the fact that $\Prob(\A_{1:\T} = (\ax_{1:\tx}, \ax_{\tx+1:\T}')) \leq \Prob(\A_{1:\tx} = \ax_{1:\tx})$ together with the assumption that $\Y(\ax_{1:\tx}) \geq \ylo$ almost surely.}

\begin{proposition} \label{prop:sharpness-of-bounds}
    Under the same setup as in Theorem \ref{thm:causal-bounds}, there always exists potential outcomes $(\tilde{\X}_{0:\T}(\ax_{1:\T}'), \tilde{\Y}(\ax_{1:\tx}') : \ax_{1:\T}' \in \Aspace_{1:\T})$ also satisfying \eqref{eq:B-positivity-assumption} and \eqref{eq:Y-boundedness-assumption} (mutatis mutandis) with
    \[
        (\tilde{\X}_{0:\T}(\A_{1:\T}), \tilde{\Y}(\A_{1:\tx}), \A_{1:\T}) \eqas (\X_{0:\T}(\A_{1:\T}), \Y(\A_{1:\tx}), \A_{1:\T})
    \]
    but for which $\E[\tilde{\Y}(\ax_{1:\tx}) \mid \tilde{\X}_{0:\tx}(\ax_{1:\tx}) \in \B_{0:\tx}] = \Qlo$.
    The corresponding statement is also true for $\Qup$.
\end{proposition}

\subsection{Continuous Bounds Must Be Trivial}

Instead of bounding $\Q$, in some cases we may wish to consider bounding the alternative quantity $\E[\Y(\ax_{1:\tx}) \mid \X_{0:\tx}(\ax_{1:\tx})]$ that conditions on the \emph{value} of $\X_{0:\tx}(\ax_{1:\tx})$ rather than on the event $\{\X_{0:\tx}(\ax_{1:\tx}) \in \B_{0:\tx}\}$.
To achieve this, it would be natural to generalize our assumption \eqref{eq:Y-boundedness-assumption} by supposing we now have measurable functions $\ylo, \yup : \Xspace_{0:\tx} \to \R$ such that
\begin{equation} \label{eq:worst-case-behaviour-functional-assumption}
    \ylo(\X_{0:\tx}(\ax_{1:\tx})) \leq \Y(\ax_{1:\tx}) \leq \yup(\X_{0:\tx}(\ax_{1:\tx})) \qquad \qquad \text{almost surely,}
\end{equation}
and taking as our goal to obtain measurable functions $\lo{\gx}, \up{\gx} : \Xspace_{0:\tx} \to \R$ such that
\begin{equation} \label{eq:almost-sure-bound-desired-result}
    \lo{\gx}(\X_{0:\tx}(\ax_{1:\tx})) \leq \E[\Y(\ax_{1:\tx}) \mid \X_{0:\tx}(\ax_{1:\tx})] \leq \up{\gx}(\X_{0:\tx}(\ax_{1:\tx})) \qquad \qquad \text{almost surely.}
\end{equation}
As we describe in Section \ref{sec:impossibility-of-bounds-for-continuous-data} of the \AppendixName, bounds of this kind can be obtained directly from Theorem \ref{thm:causal-bounds} if $\X_{0:\tx}(\ax_{1:\tx})$ is discrete, or by a simple modification of the proof of Theorem \ref{thm:causal-bounds} if $\X_{1:\tx}(\ax_{1:\tx})$ is discrete (but $\X_0$ is possibly continuous).
However, somewhat surprisingly, in general we cannot obtain nontrivial bounds of this kind without further assumptions in the continuous case.
To make this precise, we will say that a given $\lo{\gx}$ and $\up{\gx}$ are \emph{permissible} if \eqref{eq:almost-sure-bound-desired-result} holds when $(\X_{0:\T}(\ax_{1:\tx}'), \Y(\ax_{1:\tx}'), \A_{1:\T} : \ax_{1:\T}' \in \Aspace_{1:\T})$ are replaced by any potential outcomes $(\tilde{\X}_{0:\T}(\ax_{1:\T}'), \tilde{\Y}(\ax_{1:\tx}'), \tilde{\A}_{1:\T} : \ax_{1:\T}' \in \Aspace_{1:\T})$  for which
\[
    \Law[\tilde{\X}_{0:\T}(\tilde{\A}_{1:\T}), \tilde{\A}_{1:\T}, \tilde{\Y}(\tilde{\A}_{1:\tx})] = \Law[\X_{0:\T}(\A_{1:\T}), \A_{1:\T}, \Y(\A_{1:\tx})],
\]
and which also satisfy \eqref{eq:worst-case-behaviour-functional-assumption} (mutatis mutandis).
Intuitively, this means that $\lo{\gx}$ and $\up{\gx}$ depend only on the information we have available, i.e.\ the observational distribution and our assumed worst-case values.
We then have the following:

\begin{theorem} \label{thm:no-causal-bounds-for-continuous-data}
    Suppose $\X_0$ is almost surely constant, $\Prob(\A_1 \neq \ax_1) > 0$, and for some $\sx \in \{1, \ldots, \tx\}$ we have $\Prob(\X_{\sx}(\A_{1:\sx}) = \xx_{\sx}) = 0$ for all $\xx_{\sx} \in \Xspace_{\sx}$.
    Then $\lo{\gx}, \up{\gx} : \Xspace_{0:\tx} \to \R$ are permissible bounds only if they are trivial, i.e.\ almost surely we have
    \[
        \lo{\gx}(\X_{0:\tx}(\ax_{1:\tx})) \leq \ylo(\X_{0:\tx}(\ax_{1:\tx})) \qquad and 
        \qquad \up{\gx}(\X_{0:\tx}(\ax_{1:\tx})) \geq \yup(\X_{0:\tx}(\ax_{1:\tx})).
    \]

\end{theorem}

Here the assumption that $\X_0$ is constant essentially means we consider a special case of our model where there are no covariates available before the first action is taken, and serves mainly to simplify the proof.
We conjecture that Theorem \ref{thm:no-causal-bounds-for-continuous-data} holds more generally, provided the other assumptions are accordingly made to be conditional on $\X_0$ also.
In any case, this result shows that general-purpose bounds on $\E[\Y(\ax_{1:\tx}) \mid \X_{0:\tx}(\ax_{1:\tx})]$ are not forthcoming, and Theorem \ref{thm:causal-bounds} is the best we can hope for in general.
We show below that this result is nevertheless powerful enough to obtain useful information in practice.

\section{Falsification Methodology} \label{sec:hypotheses-from-causal-bounds}

\subsection{Hypotheses Derived from Causal Bounds} \label{sec:hypotheses-from-causal-bounds-setup}

We now use Theorem \ref{thm:causal-bounds} to obtain a hypothesis testing procedure that can be used to falsify the twin, and that does not rely on any further assumptions than we have already provided.
To this end, we first define the hypotheses $\Hyp$ satisfying \eqref{eq:falsification-hypothesis-property} that we will consider. %
Each of these will depend on a specific choice of the following parameters:

\begin{center}
\begin{tabular}{ll}
    A timestep $\tx \in \{1, \ldots, \T\}$\qquad\qquad & A sequence of actions $\ax_{1:\tx} \in \Aspace_{1:\tx}$ \\
    A measurable function $\fx : \Xspace_{0:\tx} \to \R$ \qquad\qquad & A sequence of subsets $B_{0:t} \subseteq \Xspace_{0:t}$. 
\end{tabular}
\end{center}
In practice (and in our experiments), these parameters can be chosen using a held-out dataset via a sample splitting approach \citep{cox1975note}.
To streamline notation, in this section, we will consider these parameters to be fixed.
However, we emphasize that our construction can be instantiated for many different choices of these parameters, and indeed we will do so in our case study below.
We think of $\fx$ as expressing a specific outcome of interest at time $\tx$ in terms of the data we have assumed.
Accordingly, for each $\ax_{1:\tx}' \in \Aspace_{1:\tx}$, we define new potential outcomes
$\Y(\ax_{1:\tx}') \coloneqq \fx(\X_{0:\tx}(\ax_{1:\tx}'))$.
For example, in a medical context, if $\X_{0:\tx}(\ax_{1:\tx}')$ represents a full patient history at time $\tx$ after treatments $\ax_{1:\tx}'$, then $\Y(\ax_{1:\tx}')$ might represent the patient's heart rate after these treatments. 
Likewise, $\B_{0:\tx}$ selects a subgroup of patients of interest, e.g.\ elderly patients whose blood pressure values were above some threshold at some timesteps before $\tx$.

For $\ax_{1:\tx}' \in \Aspace_{1:\tx}$, we also denote by $\Yt(\ax_{1:\tx}') \coloneqq \fx(\X_0, \Xt_{1:\tx}(\X_0, \ax_{1:\tx}'))$ the corresponding outcome of the twin.
Supposing it holds that
\begin{equation} \label{eq:B-twin-positivity-assumption}
    \Prob(\Xt_{1:\tx}(\X_0, \ax_{1:\tx}) \in \B_{1:\tx}) > 0,
\end{equation}
we may then define
$\Qt \coloneqq \E[\Yt(\ax_{1:\tx}) \mid \X_0 \in \B_0, \Xt_{1:\tx}(\X_0, \ax_{1:\tx}) \in \B_{1:\tx}]$, which is the analogue of $\Q$ for the twin.
By Proposition \ref{prop:interventional-correctness-alternative-characterisation}, if the twin is interventionally correct, then $\Qt = \Q$.
Theorem \ref{thm:causal-bounds} therefore implies that the following hypotheses have our desired property \eqref{eq:falsification-hypothesis-property}:
\begin{gather*}
\text{$\Hlo$: If \eqref{eq:B-positivity-assumption}, \eqref{eq:Y-boundedness-assumption}, and \eqref{eq:B-twin-positivity-assumption} hold, then $\Qt \geq \Qlo$} \\
\text{$\Hup$: If \eqref{eq:B-positivity-assumption}, \eqref{eq:Y-boundedness-assumption}, and \eqref{eq:B-twin-positivity-assumption} hold, then $\Qt \leq \Qup$.}
\end{gather*}
Moreover, $\Hlo$ and $\Hup$ can in principle be determined to be true or false from the information we have assumed available, since $\Qlo$ and $\Qup$ depend only on the observational data, and $\Qt$ can be estimated by generating trajectories from the twin.

When either $\Hlo$ or $\Hup$ is falsified, it immediately follows that the twin is not interventionally correct.
However, even more than this, a falsification describes a concrete failure mode with various potential implications downstream, which is considerably more useful information about the twin in practice.
For example, if $\Hlo$ is false (i.e.\ if $\Qt < \Qlo$), it follows that, among those trajectories for which $(\X_0, \Xt_{1:\tx}(\X_0, \ax_{1:\tx})) \in \B_{0:\tx}$, the mean of $\Yt(\ax_{1:\tx})$ is too small.
(Higher moments could also be considered by choosing $\fx$ appropriately.) %
In light of this, a user might choose not to rely on outputs of the twin produced under these circumstances, while a developer seeking to improve the twin could focus their attention on the specific parts of its implementation that give rise to this behaviour.

\subsection{Exact Testing Procedure} \label{sec:statistical-methodology}

We now present a procedure for testing our hypotheses using a finite dataset that obtains exact control over type I error without relying on additional assumptions or asymptotic approximations.
For simplicity, we specifically describe how to obtain a p-value for $\Hlo$ given a fixed choice of parameters $(\tx, \fx, \ax_{1:\tx}, \B_{0:\tx})$.
Our procedure for $\Hup$ is symmetrical, or may be regarded as a special case of testing $\Hlo$ by replacing $\fx$ with $-\fx$.
Multiple hypotheses may then be handled via standard techniques such as the method of \citet{holm1979simple} (which we use in our case study) or of \citet{benjamini2001control}, both of which apply without additional assumptions.

\runinhead{Dataset}

As above, we assume access to a finite dataset $\D$ of i.i.d.\ copies of \eqref{eq:observed-data-trajectory}.
We also assume access to a dataset $\Dt(\ax_{1:\tx})$ of i.i.d.\ copies of
\begin{equation} \label{eq:twin-trajecs}
    \X_0, \Xt_1(\X_0, \ax_1), \ldots, \Xt_\tx(\X_0, \ax_{1:\tx}).
\end{equation}
In practice, these copies can be obtained by initialising the twin with some value $\X_0$ taken from $\D$ without replacement and supplying inputs $\ax_{1:\tx}$.
If each $\X_0$ in $\D$ is used to initialize the twin at most once, then the resulting trajectories in $\Dt(\ax_{1:\tx})$ are guaranteed to be i.i.d., since we assumed in Section \ref{sec:causal-formulation} that the potential outcomes $\Xt_\tx(\xx_0, \ax_{1:\tx})$ produced by the twin are independent across runs.
We adopt this approach in our case study.

\runinhead{Antecedent conditions}

Observe that $\Hlo$ is false only if \eqref{eq:B-positivity-assumption}, \eqref{eq:Y-boundedness-assumption}, and \eqref{eq:B-twin-positivity-assumption} all hold.
We account for this in our testing procedure as follows.
First, \eqref{eq:Y-boundedness-assumption} immediately follows if 
\begin{equation} \label{eq:fx-boundedness-assumption}
    \ylo \leq \fx(\xx_{0:\tx}) \leq \yup \qquad \text{for all $\xx_{0:\tx} \in \B_{0:\tx}$.}
\end{equation}
This holds automatically in certain cases, such as for binary outcomes (e.g.\ patient survival), or otherwise can be enforced simply by clipping the value of $\fx$ to live within $[\ylo, \yup]$.
We describe a practical means for choosing $\fx$ in this way in Section \ref{sec:case-study}.

To account for \eqref{eq:B-positivity-assumption} and \eqref{eq:B-twin-positivity-assumption}, we simply check whether there exists some trajectory in $\D$ such that $\A_{1:\tx} = \ax_{1:\tx}$ and $\X_{0:\tx}(\A_{1:\tx}) \in \B_{0:\tx}$, and some trajectory in $\Dt(\ax_{1:\tx})$ such that $(\X_0, \Xt_{1:\tx}(\X_0, \ax_{1:\tx})) \in \B_{0:\tx}$.
If there are not, then we refuse to reject $\Hlo$ at any significance level; otherwise, we proceed to test $\Qt \geq \Qlo$ as described next.
It now easily follows that there is zero probability we will reject $\Hlo$ if \eqref{eq:B-positivity-assumption} and \eqref{eq:B-twin-positivity-assumption} do not in fact hold, and so our overall type I error is controlled at the desired level.

\runinhead{Overall testing procedure}

To test $\Qt \geq \Qlo$, we begin by constructing a one-sided lower confidence interval for $\Qlo$, and a one-sided upper confidence interval for $\Qt$.
In detail, for each significance level $\alpha \in (0, 1)$, we define random variables $\qlo{\alpha}$ and $\qt{\alpha}$ in terms of $\D$ and $\Dt(\ax_{1:\tx})$ such that
\begin{equation}
    \Prob(\Qlo \geq \qlo{\alpha}) \geq 1 - \frac{\alpha}{2} \qquad\,\, \Prob(\Qt \leq \qt{\alpha}) \geq 1 - \frac{\alpha}{2}. \label{eq:q-confidence-interval-new}
\end{equation}
We will also ensure that these are nested, i.e.\ $\qlo{\alpha} \leq \qlo{\alpha'}$ and $\qt{\alpha'} \leq \qt{\alpha}$ if $\alpha \leq \alpha'$.
We describe two methods for obtaining $\qlo{\alpha}$ and $\qt{\alpha}$ satisfying these conditions below.

From these confidence intervals, we obtain a test for the hypothesis $\Qt \geq \Qlo$ that rejects when $\qt{\alpha} < \qlo{\alpha}$.
A straightforward argument given in Section \ref{sec:hyp-testing-supplement} of the \AppendixName shows that this controls the type I error at the desired level $\alpha$.
Nestedness also yields a $p$-value obtained as the smallest value of $\alpha$ for which this test rejects, i.e.\
$\plo \coloneqq \inf\{\alpha \in (0, 1) \mid \text{$\qt{\alpha} < \qlo{\alpha}$}\}$,
or $1$ if the test does not reject at any level.

\runinhead{Obtaining confidence intervals}

We now consider how to obtain confidence intervals for $\Qlo$ and $\Qt$ satisfying the desired conditions above.
To this end, observe that both quantities are (conditional) expectations involving random variables that can be computed from $\D$ or $\Dt(\ax_{1:\tx})$.
This allows both to be estimated unbiasedly, which in turn can be used to derive confidence intervals via standard techniques.
For example, consider the subset of trajectories in $\Dt(\ax_{1:\tx})$ with $(\X_0, \Xt_{\tx}(\X_0, \ax_{1:\tx})) \in \B_{0:\tx}$.
For each such trajectory, we obtain a corresponding value of $\Yt(\ax_{1:\tx})$ that is i.i.d.\ and has expectation $\Qt$.
Similarly, for $\Qlo$, we extract the subset of trajectories in $\D$ for which $\X_{0:\N}(\A_{1:\N}) \in \B_{0:\N}$ holds. %
The values of $\Ylo$ obtained from each such trajectory are then i.i.d.\ and have expectation $\Qlo$.

At this point, our problem now reduces to that of constructing a confidence interval for the expectation of a random variable using i.i.d.\ copies of it.
Various techniques exist for this, and we consider two possibilities in our case study.
The first leverages the fact that $\Yt(\ax_{1:\tx})$ and $\Ylo$ are bounded in $[\ylo, \yup]$, which gives rise to $\qlo{\alpha}$ and $\qt{\alpha}$ via an application of Hoeffding's inequality.
This approach has the appealing property that \eqref{eq:q-confidence-interval-new} holds exactly, although at the expense of conservativeness.
In practice, this could be mitigated by instead obtaining confidence intervals via (for example) the bootstrap \citep{efron1979bootstrap}, although at the expense of requiring (often mild) asymptotic assumptions.
Section \ref{sec:confidence-intervals-methodology-supplement} of the \AppendixName describes both methods in greater detail.
Our empirical results reported in the next section all use Hoeffding's inequality, and are hence exact, but we also provide some results using bootstrapping in Section \ref{sec:boostrapping-details-supplement} of the \AppendixName.

\section{Case Study: Pulse Physiology Engine} \label{sec:case-study}

We applied our assessment methodology to the Pulse Physiology Engine \citep{pulse}, an open-source model for human physiology simulation.
Pulse simulates trajectories of various physiological metrics for patients with conditions like sepsis, COPD, and ARDS.
We describe the main steps of our experimental procedure and results below, with full details given in Section \ref{sec:experiments-supplement} of the \AppendixName. 
Code for replication is available at \mbox{\url{https://github.com/faaizT/CausalTwinAssessment}}.

\subsection{Experimental Setup} \label{sec:experimental-setup}

\runinhead{Real-world data}

We utilized the MIMIC-III dataset \citep{mimic}, a comprehensive collection of longitudinal health data from critical care patients at the Beth Israel Deaconess Medical Center (2001-2012).
We focused on patients meeting the sepsis-3 criteria \citep{sepsis-criteria}, following the methodology of \citet{ai-clinician} for selecting these.
This yielded 11,677 sepsis patient trajectories.
We randomly selected 5\% of these (583 trajectories, denoted as $\D_0$) to use for choosing the parameters of our hypotheses via a sample splitting approach \citep{cox1975note}, with the remaining 95\% (11,094 trajectories, denoted as $\D$) reserved for the actual testing.

We considered hourly observations of each patient over the first four hours of their ICU stay, i.e.\ $\T = 4$.
We defined the observation spaces $\Xspace_{0:\T}$ using a total of 17 features included in our extracted MIMIC trajectories for this time period, including static demographic quantities and patient vitals.
Following \citet{ai-clinician}, the actions we considered involved the administration of intravenous fluids and vasopressors, which both play a primary role in the treatment of sepsis in clinical practice.
Since these are recorded in MIMIC as continuous doses, we discretised their values via the same procedure as \citet{ai-clinician}, obtaining finite action spaces $\Aspace_1 = \cdots = \Aspace_4$, each with $25$ actions.

\runinhead{Hypothesis parameters}

We defined a collection of hypothesis parameters $(\tx, \fx, \ax_{1:\tx}, \B_{0:\tx})$, each of which we then used to define an $\Hlo$ and $\Hup$ to test.
For this, we chose 14 different physiological quantities of interest to assess, including heart rate, skin temperature, and respiration rate (see Table \ref{tab:hypotheses_hoeffding_full} in the \AppendixName for a complete list).
We used $\D_0$ to define a range of possible choices for each $\B_t$ via a binning procedure described in Section \ref{sec:hypothesis-parameters-supplement} of the \AppendixName.
We then selected all possible combinations of $\tx$, $\ax_{1:\tx}$, and $\B_{0:\tx}$ that were observed for at least one patient trajectory in $\D_0$.
We took \(\ylo\) and \(\yup\) to be the .2 and .8 quantiles of the same physiological quantity as was recorded in $\D_0$, and defined $\fx$ as the function that extracts this quantity from \(\Xspace_\tx\) and clips its value between \(\ylo\) and \(\yup\), so that \eqref{eq:fx-boundedness-assumption} holds.
We also investigated the sensitivity of our procedure to the choice of $\ylo$ and $\yup$ and found it to be relatively stable: see Section \ref{sec:sensitity-analysis-appendix} of the \AppendixName.
We obtained 721 unique parameter choices, each of which produced two hypotheses $\Hlo$ and $\Hup$, leading to 1,442 hypotheses in total.

\runinhead{Twin trajectories}

We generated data from Pulse to test the chosen hypotheses.
For each $\ax_{1:\tx}$ occurring in any of our hypotheses, we obtained the dataset $\Dt(\ax_{1:\tx})$ as described in Section \ref{sec:statistical-methodology}.
Specifically, we sampled $\X_0$ without replacement from $\D$, and used this to initialize a twin trajectory.
Then, at each hour $\tx' \in \{1, \ldots, 4\}$ in the simulation, we administered a dose of intravenous fluids and vasopressors corresponding to $\ax_{\tx'}$ and recorded the resulting patient features generated by Pulse. 
We describe this procedure in full in Section \ref{sec:pulse-trajectories-supplement} in the \AppendixName.
This produced a total of 26,115 simulated trajectories.

\subsection{Hypothesis Rejections} 

We tested the hypotheses just described using our methodology from Section \ref{sec:statistical-methodology}.
Here we report the results when using Hoeffding's inequality to obtain confidence intervals for $\Qlo$, $\Qup$, and $\Qt$.
We also tried confidence intervals obtained via bootstrapping, and obtained similar if less conservative results (see Section \ref{sec:experiments-supplement} of the \AppendixName).
We used the Holm-Bonferroni method to adjust for multiple tests, with family-wise error rate of 0.05.

\begin{table}%
    \centering
\begin{footnotesize}
\begin{tabular}{lcccc}
\toprule
& \multicolumn{2}{c}{Ours} & \multicolumn{2}{c}{Manski} \\
                   Physiological quantity &  Rejs. &  Hyps. &  Rejs. &  Hyps. \\
\midrule
  Chloride Blood Concentration (Chloride) &            24 &            94 & 1 & 46  \\
      Sodium Blood Concentration (Sodium) &            21 &            94 & 9 & 46  \\
Potassium Blood Concentration (Potassium) &            13 &            94 & 0 & 46  \\
                  Skin Temperature (Temp) &            10 &            86 & 9 & 46  \\
    Calcium Blood Concentration (Calcium) &             5 &            88 & 0 & 46  \\
    Glucose Blood Concentration (Glucose) &             5 &            96 & 1 & 46  \\
      Arterial CO$_2$ Pressure (paCO$_2$) &             3 &            70 & 0 & 46  \\
Bicarbonate Blood Concentration (HCO$_3$) &             2 &            90 & 1 & 46  \\
       Systolic Arterial Pressure (SysBP) &             2 &           154 & 0 & 46  \\
\bottomrule
\end{tabular}
\end{footnotesize}
    \caption{Total hypotheses (Hyps.) and rejections (Rejs.) per physiological quantity using our causal bounds, as well as those of \citet{manski}} \label{tab:hypotheses}
\end{table}

We obtained rejections for hypotheses corresponding to 10 different physiological quantities shown in Table \ref{tab:hypotheses}.
(Table \ref{tab:hypotheses_hoeffding_full} in the \AppendixName shows all hypotheses we tested, including those not rejected.)
We may therefore infer that, at a high level, Pulse does not simulate these quantities accurately for the population of sepsis patients we consider.
This appears of interest in a variety of downstream settings: for example, a developer could use this information when considering how to improve the accuracy of Pulse, while a practitioner using Pulse may wish to rely less on these outputs as a result.

To assess the relative performance of our bounds from Theorem \ref{thm:causal-bounds} compared with the unconditional bounds of \citet{manski}, we also reran this analysis with each $\tx$, $\fx$, and $\ax_{1:\tx}$ chosen as before, but with each $\B_{0:\tx}$ now set to the whole space $\Xspace_{0:\tx}$.
This in turn led to fewer hypotheses, namely 690 in total, which were evenly divided between hypotheses of the form $\E[\Yt(\ax_{1:\tx})] \geq \E[\Ylo]$ and those of the form $\E[\Yt(\ax_{1:\tx})] \leq \E[\Yup]$.
We kept all other aspects of our procedure the same as just described, including our method for obtaining confidence intervals and controlling for multiple testing.
The number of rejections that we obtained in this case is also shown in Table \ref{tab:hypotheses}.
As anticipated by our discussion in Section \ref{sec:informativeness-of-our-bounds}, the conservativeness of Manski's bounds led to considerably fewer rejections than our more general result given in Theorem \ref{thm:causal-bounds}, even when considered as a proportion of the total hypotheses we tested.

\subsection{p-Value Plots}

\begin{figure}[t]
    \centering
    \includegraphics[height=4.5cm]{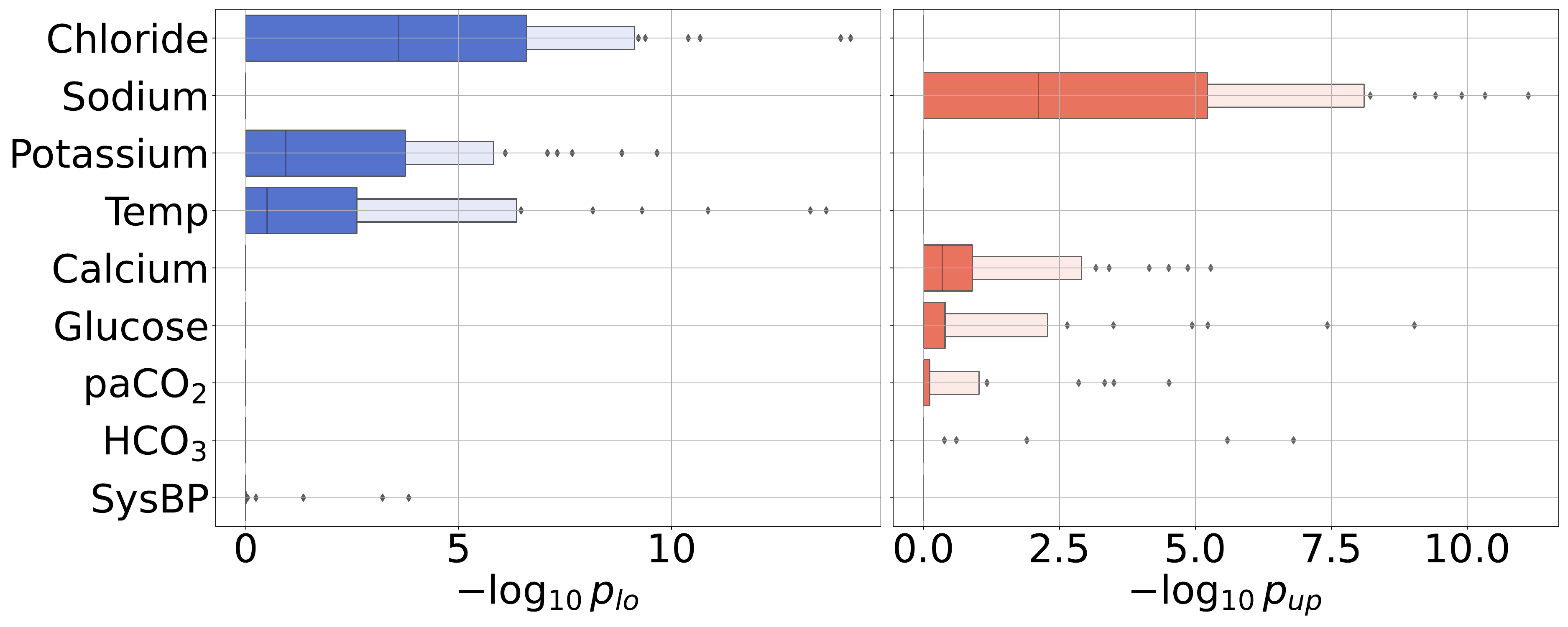}
    \caption{Distributions of $-\log_{10}{p_\textup{lo}}$ and $-\log_{10}{p_\textup{up}}$ across hypotheses, grouped by physiological quantity. Higher values indicate greater evidence in favour of rejection.}
    \label{fig:p_values}
\end{figure}

To obtain more granular information about the failure modes of the twin just identified, we examined the $p$-values obtained for each hypothesis $\Hlo$ and $\Hup$ tested using our causal bounds, which we denote here by $\plo$ and $\pup$.
Figure \ref{fig:p_values} shows the distributions of $-\log_{10}{\plo}$ and $-\log_{10}{\pup}$ that we obtained for all physiological quantities for which some hypothesis was rejected.
(The remaining $p$-values are shown in Figure \ref{fig:p_values_hoeff_complete} in the \AppendixName.)
Notably, in each row, one distribution is always tightly concentrated at $-\log_{10} p = 0$ (i.e.\ $p = 1$).
This means that, for all physiological outcomes of interest, there was either very little evidence in favour of rejecting any $\Hlo$, or very little in favour of rejecting any $\Hup$.
In other words, across configurations of $(\tx, \fx, \ax_{1:\tx}, \B_{0:\tx})$ that were rejected, the twin consistently either underestimated or overestimated each quantity on average.
For example, Pulse consistently underestimated chloride blood concentration and skin temperature, while it consistently overestimated sodium and glucose blood concentration levels.
Like Table \ref{tab:hypotheses}, this information appears of interest and actionable in a variety of downstream tasks.

\subsection{Pitfalls of Naive Assessment}

\begin{figure}%
    \centering
    \includegraphics[height=4cm]{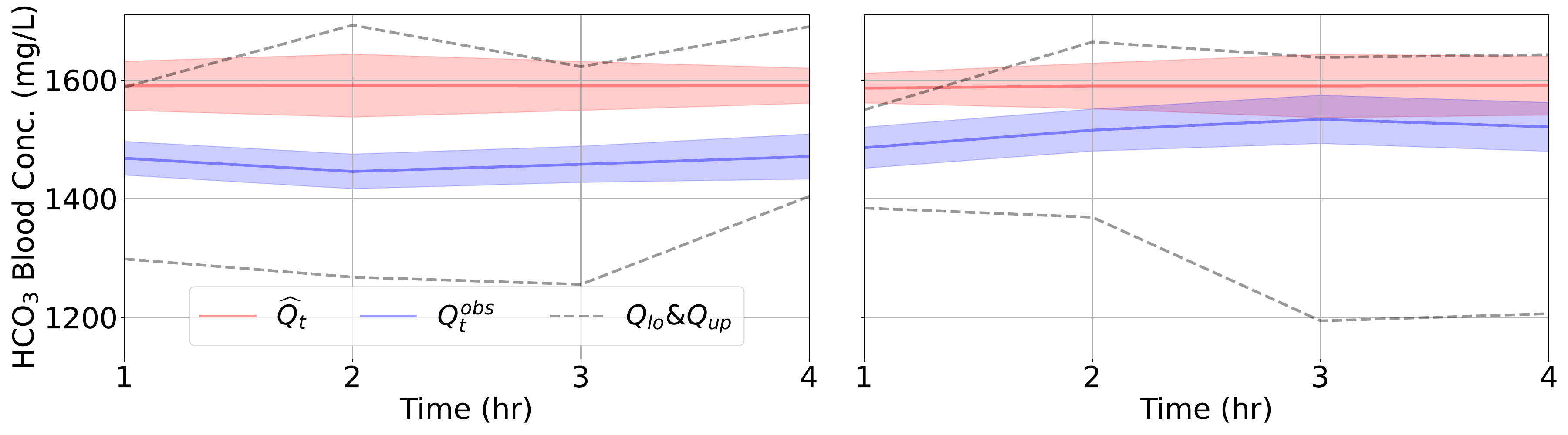}
    \caption{Estimates and 95\% confidence intervals for $\Qt_\tx$ and $\Q^{\textup{obs}}_\tx$ at each $1\leq \tx \leq 4$ for two choices of $(\B_{0:4}, \ax_{1:4})$, where $\Yt(\ax_{1:\tx})$ and $\Y(\ax_{1:\tx})$ correspond to HCO$_3$ concentration.
    The dashed lines indicate lower and upper 95\% confidence intervals for $\Qlo, \Qup$ respectively.  }
    \label{fig:longitudinal_plots}
\end{figure}

A naive approach to twin assessment involves simply comparing the output of the twin with the observational data directly, without accounting for causal considerations.
We now show that, unlike our methodology, the results produced in this way can be potentially misleading.
In Figure \ref{fig:longitudinal_plots}, for two different choices of $(\ax_{1:4}, \B_{1:4})$, we plot estimates of $\Qt_\tx$ and $\Q^{\textup{obs}}_\tx$ for $\tx \in \{1, \ldots, 4\}$, where
\begin{align*}
    \Qt_\tx &\coloneqq \E[\Yt(\ax_{1:\tx})\mid \X_0 \in \B_0, \Xt_{1:\tx}(\X_0, \ax_{1:\tx}) \in \B_{1:\tx}] \\
    \Q^{\textup{obs}}_\tx &\coloneqq \E[\Y(\A_{1:\tx})\mid \X_{0:\tx}(\A_{1:\tx})\in \B_{0:\tx}, \A_{1:\tx}=\ax_{1:\tx}].
\end{align*}
Here $\Qt_\tx$ is just $\Qt$ as defined above with its dependence on $\tx$ made explicit.
Each plot also shows one-sided 95\% confidence intervals on $\Qlo$ and $\Qup$ at each $\tx \in \{1, \ldots, 4\}$ obtained from Hoeffding's inequality. 
Directly comparing the estimates of $\Qt_\tx$ and $\Q^{\textup{obs}}_\tx$ would suggest that the twin is comparatively more accurate for the right-hand plot, as these estimates are closer to one another in that case.
However, the output of the twin in the right-hand plot is falsified at $\tx = 1$, as can be seen from the fact that confidence interval for $\Qt_1$ lies entirely above the one-sided confidence interval for $\Qup$ at that timestep.
On the other hand, the output of the twin in the left-hand plot is not falsified at any of the timesteps shown, so that the twin may in fact be accurate for these $(\ax_{1:4}, \B_{1:4})$, contrary to what a naive assessment strategy would suggest.
Our methodology provides a principled means for twin assessment that avoids drawing potentially misleading inferences like this.

\begin{figure}%
    \centering
    \begin{subfigure}[b]{0.5\textwidth}
    \centering
    \includegraphics[height=5cm]{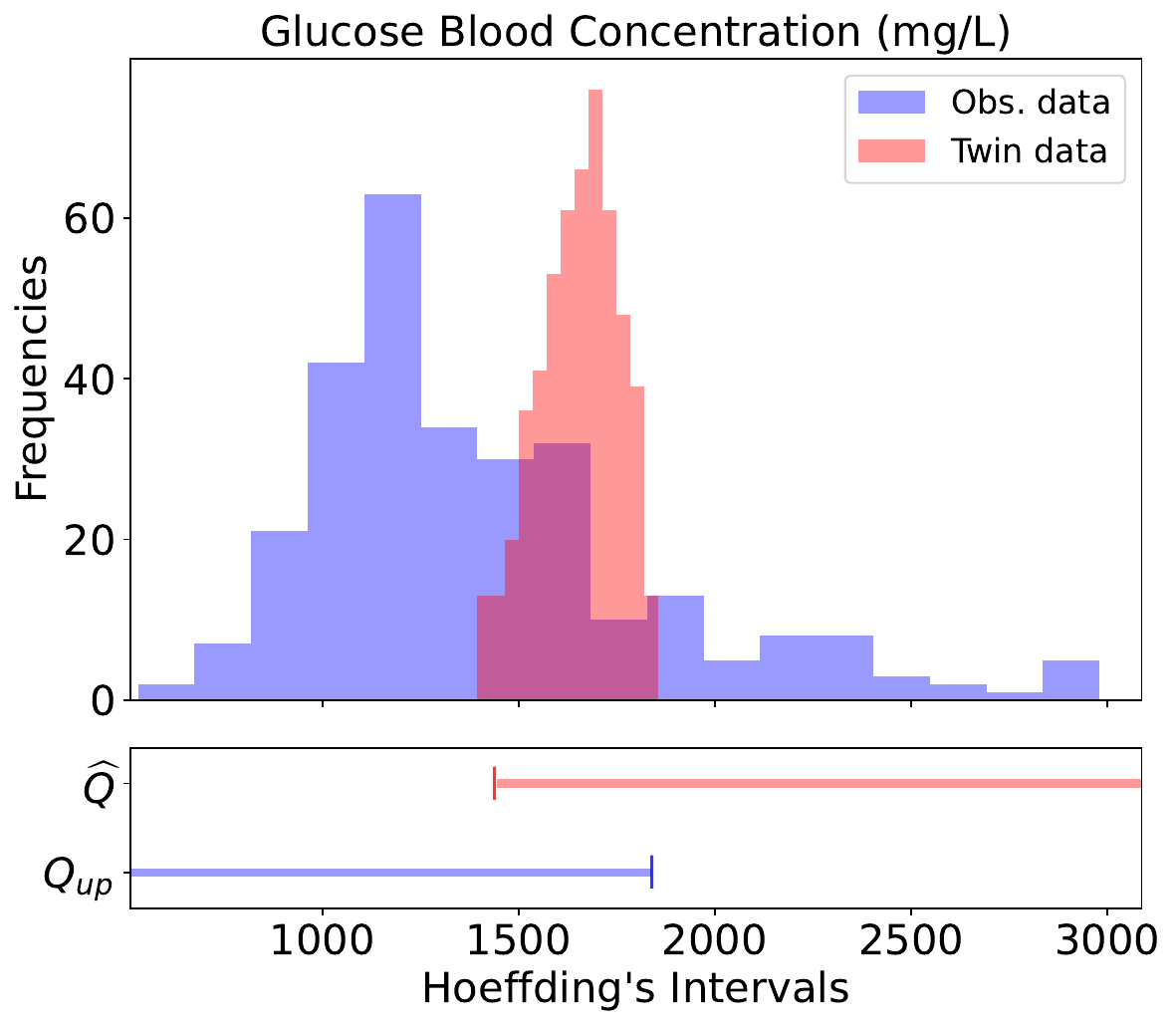}
    \subcaption{Not rejected}
    \label{fig:glucosea}
    \end{subfigure}%
    \begin{subfigure}[b]{0.5\textwidth}
    \centering
    \includegraphics[height=5cm]{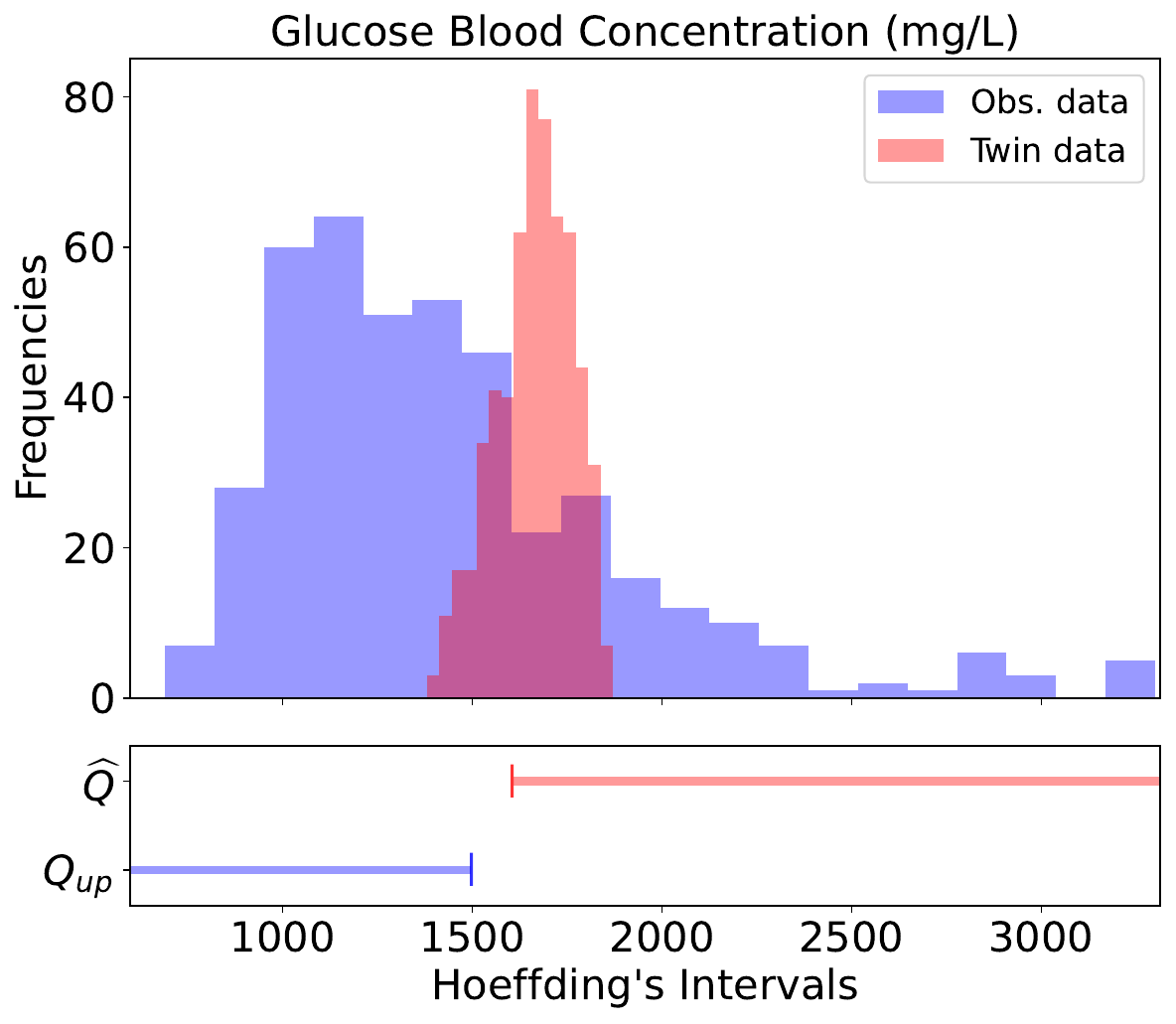}
    \subcaption{Rejected}
    \label{fig:glucoseb}
    \end{subfigure}
    \caption{Raw glucose values from the observational data and twin for two choices of $(\B_{0:\tx}, \ax_{1:\tx})$, with confidence intervals for $\Qt$ and $\Qup$ shown below. The horizontal axes are truncated to the .025 and .975 quantiles of the observational data for clarity. Untruncated plots are shown in Figure \ref{fig:histograms-supplement} of the \AppendixName.}
    \label{fig:histograms}
\end{figure}

A similar phenomenon appears in Figure \ref{fig:histograms}, which for two choices of $\B_{0:\tx}$ and $\ax_{1:\tx}$ shows histograms of raw glucose values obtained from the observational data conditional on $\A_{1:\tx}=\ax_{1:\tx}$ and $\X_{0:\tx}(\A_{1:\tx})\in \B_{0:\tx}$, and from the twin conditional on $\Xt_{0:\tx}(\ax_{1:\tx})\in \B_{0:\tx}$.
Below each histogram we also show 95\% confidence intervals for $\Qup$ and $\Qt$ obtained from Hoeffding's inequality.
While Figures \ref{fig:glucosea} and \ref{fig:glucoseb} appear visually very similar, the inferences produced by our testing procedure are different: the hypothesis corresponding to the right-hand plot is rejected, since there is no overlap between the confidence intervals underneath, while the hypothesis corresponding to the left-hand plot is not.
This was not an isolated case and several other examples of this phenomenon are shown in Figure \ref{fig:histograms-supplement} in the \AppendixName. 
This demonstrates that the inferences obtained from our procedure do not depend only on the distribution of observed outcomes (which is essentially the same for both cases).
Instead, as discussed in Section \ref{sec:causal-bounds-statement}, these also account for the worst-case effects of unmeasured confounding that may exist in the observational data.

\section{Discussion}

We have advocated for a causal approach to digital twin assessment, and have presented a statistical procedure for doing so that obtains rigorous theoretical guarantees under minimal assumptions.
We now highlight the key limitations of our approach.
Importantly, our methodology implicitly assumes that there is no distribution shift between testing and deployment time.
If the conditional distribution of $\X_{1:\T}(\ax_{1:\T})$ given $\X_0$ changes at deployment time, then so too does the set of twins that are interventionally correct, and if this change is significant enough, our assessment procedure may yield misleading results.
Distribution shift in this sense is a separate issue to unobserved confounding, and arises in a wide variety of statistical problems beyond ours.

Additionally, the procedure we used in our case study to choose the hypothesis parameters $\B_{0:\tx}$ (described in Section \ref{sec:hypothesis-parameters-supplement} of the \AppendixName) was ad hoc.
For scalability, it would likely be necessary to obtain $\B_{0:\tx}$ via a more automated procedure.
It may also be desirable to choose $\B_{0:\tx}$ dynamically in light of previous hypotheses tested, zooming in to regions containing possible failure modes to obtain increasingly granular information about the twin.
We see opportunities here for using machine learning techniques, but leave this to future work.

Various other extensions and improvements also appear possible.
For example, one can leverage ideas from the literature on partial identification \citep{manski2003partial} to obtain greater statistical efficiency, for example by building on the line of work initiated by \citet{imbens2004confidence} for obtaining more informative confidence intervals.
Beyond this, it may sometimes be useful to consider additional assumptions that lead to less conservative assessment results.
For example, various methods for \emph{sensitivity analysis} have been proposed that model the \emph{degree} to which the actions of the behavioural agent are confounded \citep{rosenbaum2002observational,tan2006distributional,yadlowsky2022bounds}.
This can yield tighter bounds on $\Q$ than Theorem \ref{thm:causal-bounds}, albeit at the expense of less robustness if these assumptions are violated.

\acks{The authors are highly appreciative of troubleshooting and development assistance provided by the Pulse team.
RC was supported by the National Research Foundation, Singapore, under a National Research Foundation Fellowship in Artificial Intelligence (Award No.: NRFFIAI1-2024-0014).
In addition, RC, AD, and CH were supported by the Engineering and Physical Sciences Research Council (EPSRC) through the Bayes4Health programme [Grant number EP/R018561/1].
MFT was funded by Google DeepMind.
The authors declare there are no competing interests.} 

\newpage
\appendix 
\section{Notation} \label{sec:notation}

\begin{tabularx}{\linewidth}{l X}
$z_{\tx:\tx'}$ & The sequence of elements $(z_\tx, \ldots, z_{\tx'})$ (or the empty sequence when $\tx > \tx'$) \\
$\mathcal{Z}_{\tx:\tx'}$ (where each $\mathcal{Z}_{i}$ is a set) & The cartesian product $\mathcal{Z}_{\tx} \times \cdots \times \mathcal{Z}_{\tx'}$ (or the set containing just the empty sequence when $\tx > \tx'$) \\
$Z_{\tx:\tx'}(\ax_{1:\tx'})$ & The sequence of potential outcomes $Z_\tx(\ax_{1:\tx}), \ldots, Z_{\tx'}(\ax_{1:\tx'})$ (or the empty sequence when $\tx > \tx'$) \\
$\Law[Z]$ & The distribution of the random variable $Z$ \\
$\Law[Z \mid M]$ & The conditional distribution of $Z$ given $M$, where $M$ is either an event or a random variable \\
$Z \eqas Z'$ & The random variables $Z$ and $Z'$ are almost surely equal, i.e.\ $\Prob(Z = Z') = 1$ \\
$Z \ci Z'$ & The random variables $Z$ and $Z'$ are independent \\
$Z \ci Z' \mid Z''$ & The random variables $Z$ and $Z'$ are conditionally independent given the random variable $Z''$ \\
$\ind(E)$ & Indicator function of some event $E$ %
\end{tabularx}

\section{Proof of Proposition \ref{prop:interventional-correctness-alternative-characterisation} (Unconditional Form of Interventional Correctness)} \label{sec:unconditional-interventional-correctness-proof}

\begin{proof}
Fix any choice of $\ax_{1:\T} \in \Aspace_{1:\T}$.
By disintegrating $\Law[\X_0, \Xt_{1:\T}(\X_0, \ax_{1:\T})]$ and $\Law[\X_{0:\T}(\ax_{1:\T})]$ along their common $\Xspace_0$-marginal (which is namely $\Law[\X_0]$), it holds that
\begin{equation} \label{eq:interventional-correctness-alternative}
    \Law[\X_0, \Xt_{1:\T}(\X_0, \ax_{1:\T})] = \Law[\X_{0:\T}(\ax_{1:\T})]
\end{equation}
if and only if
\begin{equation} \label{eq:joint-interventional-correctness-proof-intermediate-step}
    \Law[\Xt_{1:\T}(\X_0, \ax_{1:\T}) \mid \X_0 = \xx_0] = \Law[\X_{1:\T}(\ax_{1:\T}) \mid \X_0 = \xx_0]
\end{equation}
for $\Law[\X_0]$-almost all $\xx_0 \in \Xspace_0$.
But now, our definition of $\Xt_{1:\T}(\xx_0, \ax_{1:\T})$ in terms of $\twinfunction_\tx$ and $\twinnoise_{1:\tx}$ means we can write $\Xt_{1:\T}(\X_0, \ax_{1:\T}) = \boldsymbol{\twinfunction}(\X_0, \ax_{1:\T}, \twinnoise_{1:\T})$,
where
\[
    \boldsymbol{\twinfunction}(\xx_0, \ax_{1:\T}, \ux_{1:\T}) \coloneqq (\twinfunction_1(\xx_0, \ax_1, \ux_1), \ldots, \twinfunction_\T(\xx_0, \ax_{1:\T}, \ux_{1:\T})).
\]
For all $\xx_0 \in \Xspace_0$ and measurable $\B_{1:\T} \subseteq \Xspace_{1:\T}$, we then have
\begin{align*}
    \Law[\Xt_{1:\T}(\xx_0, \ax_{1:\T})](\B_{1:\T}) &= \E[\ind(\boldsymbol{\twinfunction}(\xx_0, \ax_{1:\T}, \twinnoise_{1:\T}) \in \B_{1:\T})] \\
    &= \int \ind(\boldsymbol{\twinfunction}(\xx_0, \ax_{1:\T}, \ux_{1:\T}) \in \B_{1:\T}) \, \Law[\twinnoise_{1:\T}](\dee \ux_{1:\T}).
\end{align*}
It is standard to show that the right-hand side is a Markov kernel in $\xx_0$ and $\B_{1:\T}$.
Moreover, for any measurable $\B_0 \subseteq \Xspace_0$, we have
\begin{align*}
    &\int_{\B_0} \Law[\Xt_{1:\T}(\xx_0, \ax_{1:\T})](\B_{1:\T}) \, \Law[\X_0](\dee \xx_0) \\
        &\qquad= \int_{\B_0} \left[\int \ind(\boldsymbol{\twinfunction}(\xx_0, \ax_{1:\T}, \ux_{1:\T}) \in \B_{1:\T}) \, \Law[\twinnoise_{1:\T}](\dee \ux_{1:\T}) \right] \, \Law[\X_0](\dee \xx_0) \\
        &\qquad= \int \ind(\xx_0 \in \B_0, \boldsymbol{\twinfunction}(\xx_0, \ax_{1:\T}, \ux_{1:\T}) \in \B_{1:\T}) \, \Law[\X_0, \twinnoise_{1:\T}](\dee \xx_0, \dee \ux_{1:\T}) \\
        &\qquad= \Law[\X_0, \Xt_{1:\T}(\X_0, \ax_{1:\T})](\B_{0:\T}),
\end{align*}
where the second step follows because $\X_0 \ci \twinnoise_{1:\T}$.
It therefore follows that $(\xx_0, \B_{1:\T}) \mapsto \Law[\Xt_{1:\T}(\xx_0, \ax_{1:\T})](\B_{1:\T})$ is a regular conditional distribution of $\Xt_{1:\T}(\X_0, \ax_{1:\T})$ given $\X_0$, i.e.\
\[
    \Law[\Xt_{1:\T}(\xx_0, \ax_{1:\T})] = \Law[\Xt_{1:\T}(\X_0, \ax_{1:\T}) \mid \X_0 = \xx_0] \qquad \text{for $\Law[\X_0]$-almost all $\xx_0 \in \Xspace_0$.}
\]
Substituting this into \eqref{eq:joint-interventional-correctness-proof-intermediate-step}, we see that \eqref{eq:interventional-correctness-alternative} holds if and only if
\[
    \Law[\Xt_{1:\T}(\xx_0, \ax_{1:\T})] = \Law[\X_{1:\T}(\ax_{1:\T}) \mid \X_0 = \xx_0]
\]
for $\Law[\X_0]$-almost all $\xx_0 \in \Xspace_0$.
The result now follows since $\ax_{1:\T}$ was arbitrary.
\end{proof}

\section{Online Prediction} \label{sec:online-prediction}

\subsection{Correctness in the Online Setting}

A distinguishing feature of many digital twins is their ability to integrate real-time information obtained from sensors in their environment \citep{barricelli2019survey}.
It is therefore relevant to consider a setting in which a twin is used repeatedly to make a sequence of predictions over time, each time taking all previous information into account.
One way to formalize this is to instantiate our model for the twin at each timestep.
For example, we could represent the predictions made by the twin at $\tx = 0$ after observing initial covariates $\xx_0$ as potential outcomes $(\Xt^1_{1:\T}(\xx_0, \ax_{1:\T}) : \ax_{1:\T} \in \Aspace_{1:\T})$, similar to what we did in the main text.
We could then represent the predictions made by the twin after some action $\ax_1$ is taken and an additional observation $\xx_1$ is made via potential outcomes $(\Xt^2_{2:\T}(\xx_{0:1}, \ax_{1:\T}) : \ax_{2:\T} \in \Aspace_{2:\T})$.
More generally, for $\tx \in \{1, \ldots, \T\}$, we could introduce potential outcomes $(\Xt^{\tx}_{\tx:\T}(\xx_{0:\tx-1}, \ax_{1:\T}) : \ax_{\tx:\T} \in \Aspace_{\tx:\T})$ to represent the predictions that the twin would make at time $\tx$ after the observations $\xx_{0:\tx-1}$ are made and the actions $\ax_{1:\tx-1}$ are taken.

This extended model requires a new definition of correctness than our Definition \ref{eq:interventional-correctness} from the main text.
A natural approach is to say that the twin is correct in this new setting if
\begin{equation}
    \Law[\Xt_{\tx:\T}^\tx(\xx_{0:\tx-1}, \ax_{1:\T})]
        = \Law[\X_{\tx:\T}(\ax_{1:\T}) \mid \X_{0:\tx-1}(\ax_{1:\tx-1}) = \xx_{0:\tx-1}] \label{eq:online-interventional-correctness-def}
\end{equation}
for all $\tx \in \{1, \ldots, \T\}$, $\ax_{1:\T} \in \Aspace_{1:\T}$, and $\Law[\X_{0:\tx-1}(\ax_{1:\tx-1})]$-almost all $\xx_{0:\tx-1} \in \Xspace_{0:\tx-1}$.
A twin with this property would at each step be able to accurately simulate the future in light of previous information, use this to choose a next action to take, observe the result of doing so, and then repeat.
It is possible to show that \eqref{eq:online-interventional-correctness-def} holds if and only if we have
\begin{align*}
    \Law[\Xt^1_{1:\T}(\xx_0, \ax_{1:\T})] &= \Law[\X_{1:\T}(\ax_{1:\T}) \mid \X_{0}=\xx_0] \\
    \Law[\Xt^\tx_{\tx:\T}(\xx_{0:\tx-1}, \ax_{1:\T})]
        &= \Law[\Xt^1_{\tx:\T}(\xx_0, \ax_{1:\T}) \mid \Xt_{1:\tx-1}^1(\xx_0, \ax_{1:\tx-1}) = \xx_{1:\tx-1}]
\end{align*}
for all $\tx \in \{1, \ldots, \T\}$, $\ax_{1:\T} \in \Aspace_{1:\T}$, $\Law[\X_0]$-almost all $\xx_0 \in \Xspace_0$, and $\Law[\Xt_{1:\tx-1}^1(\xx_0, \ax_{1:\tx-1})]$-almost all $\xx_{1:\tx-1} \in \Xspace_{1:\tx-1}$.
The first condition here says that $\Xt^1_{1:\T}(\xx_0, \ax_{1:\T})$ must be interventionally correct in the sense of Definition \ref{eq:interventional-correctness} from the main text.
The second condition says that the predictions made by the twin across different timesteps must be internally consistent with each other insofar as their conditional distributions must align.
This holds automatically in many circumstances, such as if the predictions of the twin are obtained from a Bayesian model (for example), and otherwise could be checked numerically given the ability to run simulations from the twin, without the need to obtain data or refer to the real-world process in any way.
As such, the problem of assessing the correctness of the twin in this new sense primarily reduces to the problem of assessing the correctness of $\Xt^1_{1:\T}(\xx_0, \ax_{1:\T})$ in the sense of Definition \ref{eq:interventional-correctness} in the main text, which motivates our focus on that condition.

\subsection{Alternative Notions of Online Correctness} \label{sec:online-correctness-alternative-notion-supp}

An important and interesting subtlety arises in this context that is worth noting.
In general it does not follow that a twin correct in the sense of \eqref{eq:online-interventional-correctness-def} satisfies
\begin{equation}
    \Law[\Xt_{\tx:\T}^\tx(\xx_{0:\tx-1}, \ax_{1:\T})] \
        = \Law[\X_{\tx:\T}(\ax_{1:\T}) \mid \X_{0:\tx-1}(\ax_{1:\tx-1}) = \xx_{0:\tx-1}, \A_{1:\tx-1} = \ax_{1:\tx-1}] \label{eq:online-interventional-correctness-alt}
\end{equation}
for all $\ax_{1:\T} \in \Aspace_{1:\T}$, and $\Law[\X_{0:\tx-1}(\ax_{1:\tx-1}) \mid \A_{1:\tx-1} = \ax_{1:\tx-1}]$-almost all $\xx_{0:\tx-1} \in \Xspace_{0:\tx-1}$, 
since in general it does not hold that 
\begin{multline*}
    \Law[\X_{\tx:\T}(\ax_{1:\T}) \mid \X_{0:\tx-1}(\ax_{1:\tx-1}) = \xx_{0:\tx-1}] \\
        = \Law[\X_{\tx:\T}(\ax_{1:\T}) \mid \X_{0:\tx-1}(\ax_{1:\tx-1}) = \xx_{0:\tx-1}, \A_{1:\tx-1} = \ax_{1:\tx-1}].
\end{multline*}
for all $\ax_{1:\T} \in \Aspace_{1:\T}$ and $\Law[\X_{0:\tx-1}(\ax_{1:\tx-1}) \mid \A_{1:\tx-1} = \ax_{1:\tx-1}]$-almost all $\xx_{0:\tx-1} \in \Xspace_{0:\tx-1}$ unless the actions $\A_{1:\tx-1}$ are unconfounded.
(Here as usual $\A_{1:\T}$ denotes the actions of a behavioural agent; see Section \ref{sec:data-driven-twin-assessment} of the main text.)
In other words, a twin that is correct in the sense of \eqref{eq:online-interventional-correctness-def} will make accurate predictions at time $\tx$ when every action taken before time $\tx$ was unconfounded (as occurs for example when the twin is directly in control of the decision-making process), but in general not when certain taken actions before time $\tx$ were chosen by a behavioural agent with access to more context than is available to the twin. %
However, should it be desirable, our framework could be extended to encompass the alternative condition in \eqref{eq:online-interventional-correctness-alt} by relabelling the observed history $(\X_{0:\tx-1}(\A_{1:\tx-1}), \A_{1:\tx-1})$ as $\X_0$, and then assessing the correctness of the potential outcomes $\Xt_{\tx:\T}^\tx(\xx_{0:\tx-1}, \ax_{1:\T})$ in the sense of Definition \ref{eq:interventional-correctness} from the main text.

Overall, the ``right'' notion of correctness in this online setting is to some extent a design choice.
We believe our causal approach to twin assessment provides a useful framework for formulating and reasoning about these possibilities, and consider the investigation of assessment strategies for additional usage regimes to be an interesting direction for future work.

\section{Proof of Theorem \ref{prop:nonidentifiability} (Interventional Distributions Are Not Identifiable)} \label{sec:non-identifiability-result-proof-supp}

It is well-known in the causal inference literature that the interventional behaviour of the real-world process cannot be uniquely identified from observational data.
For completeness, we now provide a self-contained proof of this result in our notation.
Our statement here is lengthier than Theorem \ref{prop:nonidentifiability} in the main text in order to clarify what is meant by ``uniquely identified'': intuitively, the idea is that there always exist distinct families of potential outcomes whose interventional behaviours differ and yet give rise to the same observational data.
\begin{theorem}
    Suppose we have $\ax_{1:\T} \in \Aspace_{1:\T}$ such that $\Prob(\A_{1:\T} \neq \ax_{1:\T}) > 0$.
    Then there exist potential outcomes $(\tilde{\X}_{0:\T}(\ax_{1:\T}') : \ax_{1:\T}' \in \Aspace_{1:\T})$ such that %
    \begin{equation} \label{eq:nonidentifiability-proof-almost-sure-equality}
        (\tilde{\X}_{0:\T}(\A_{1:\T}), \A_{1:\T}) \eqas (\X_{0:\T}(\A_{1:\T}), \A_{1:\T}).
    \end{equation}
    but for which $\Law[\tilde{\X}_{0:\T}(\ax_{1:\tx})] \neq \Law[\X_{0:\T}(\ax_{1:\tx})]$.
\end{theorem}

\begin{proof}
    Our assumption that $\Prob(\A_{1:\T} \neq \ax_{1:\T}) > 0$ means there must exist some $\tx \in \{1, \ldots, \T\}$ such that $\Prob(\A_{1:\tx} \neq \ax_{1:\tx}) > 0$.
    Since $\Xspace_\tx = \R^{\Xspacedim_\tx}$, we may also choose some $\xx_\tx \in \Xspace_\tx$ with $\Prob(\X_\tx(\ax_{1:\tx}) = \xx_\tx \mid \A_{1:\tx} \neq \ax_{1:\tx}) \neq 1$.
    Then, for each $\sx \in \{0, \ldots, \T\}$ and $\ax_{1:\sx}' \in \Aspace_{1:\sx}$, define
    \[
         \tilde{\X}_{\sx}(\ax_{1:\sx}') \coloneqq \begin{cases}
            \ind(\A_{1:\tx} = \ax_{1:\tx}) \, \X_{\tx}(\ax_{1:\tx}) + \ind(\A_{1:\tx} \neq \ax_{1:\tx}) \, \xx_\tx & \text{if $\sx = \tx$ and $\ax_{1:\sx}' = \ax_{1:\tx}$} \\
            \X_{\sx}(\ax_{1:\sx}') & \text{otherwise},
         \end{cases}
    \]
    It is then easily checked that \eqref{eq:nonidentifiability-proof-almost-sure-equality} holds, but
    \begin{align*}
        \MoveEqLeft \Law[\tilde{\X}_{\tx}(\ax_{1:\tx})] \\
        &= \Law[\tilde{\X}_{\tx}(\ax_{1:\tx}) \mid \A_{1:\tx} = \ax_{1:\tx}] \, \Prob(\A_{1:\tx} = \ax_{1:\tx}) + \Law[\tilde{\X}_{\tx}(\ax_{1:\tx}) \mid \A_{1:\tx} \neq \ax_{1:\tx}] \, \Prob(\A_{1:\tx} \neq \ax_{1:\tx}) \\
        &= \Law[X_{\tx}(\ax_{1:\tx}) \mid \A_{1:\tx} = \ax_{1:\tx}] \, \Prob(\A_{1:\tx} = \ax_{1:\tx}) + \mathrm{Dirac}(\xx_\tx) \, \Prob(\A_{1:\tx} \neq \ax_{1:\tx}) \\
        &\neq \Law[X_{\tx}(\ax_{1:\tx}) \mid \A_{1:\tx} = \ax_{1:\tx}] \, \Prob(\A_{1:\tx} = \ax_{1:\tx}) + \Law[\X_{\tx}(\ax_{1:\tx}) \mid \A_{1:\tx} \neq \ax_{1:\tx}] \, \Prob(\A_{1:\tx} \neq \ax_{1:\tx}) \\
        &= \Law[X_{\tx}(\ax_{1:\tx})],
    \end{align*}
    from which the result follows.
\end{proof}

\section{Deterministic Potential Outcomes Are Unconfounded} \label{eq:deterministic-potential-outcomes-are-unconfounded}

In this section we expand on our earlier claim that, if the real-world process is deterministic, then the observational data is unconfounded.
We first make this claim precise.
By ``deterministic'', we mean that there exist measurable functions $\gx_\tx$ for $\tx \in \{1, \ldots, \T\}$ such that
\begin{equation} \label{eq:potential-outcomes-are-deterministic}
    \X_\tx(\ax_{1:\tx}) \eqas \gx_\tx(\X_{0:\tx-1}(\ax_{1:\tx-1}), \ax_{1:\tx}) \qquad \text{for all $\tx \in \{1, \ldots, \T\}$ and $\ax_{1:\tx} \in \Aspace_{1:\tx}$.}
\end{equation}
By ``unconfounded'', we mean that the \emph{sequential randomisation assumption (SRA)} introduced by \citet{robins1986new} holds, i.e.\
\begin{equation} \label{eq:actions-are-unconfounded}
    (\X_{\sx}(\ax_{1:\sx}) : \sx \in \{1, \ldots, \T\}, \ax_{1:\sx} \in \Aspace_{1:\sx}) \ci \A_\tx \mid \X_{0:\tx-1}(\A_{1:\tx-1}), \A_{1:\tx-1} \quad \text{for all $\tx \in \{1, \ldots, \T\}$},
\end{equation}
where $\ci$ denotes conditional independence.
Intuitively, this says that, apart from the historical observations $(\X_{0:\tx-1}(\A_{1:\tx-1}), \A_{1:\tx-1})$, any additional factors that influence the agent's choice of action $\A_\tx$ are independent of the behaviour of the real-world process.
The SRA provides a standard formulation of the notion of unconfoundedness in longitudinal settings such as ours (see \citealt[Chapter 5]{tsiatis2019dynamic} for a review).

It is now a standard exercise to show that \eqref{eq:potential-outcomes-are-deterministic} implies \eqref{eq:actions-are-unconfounded}.
We include a proof below for completeness.
Key to this is the following straightforward Lemma.

\begin{lemma}\label{lem:determinism_conditional_independence}
Suppose $U$ and $V$ are random variables such that, for some measurable function $g$, it holds that $U \eqas g(V)$.
Then, for any other random variable $W$, we have
\[
    U \ci W \mid V.
\]
\end{lemma}

\begin{proof}
By standard properties of conditional expectations, for any measurable sets $S_1$ and $S_2$, we have almost surely
\begin{align*}
    \Prob(U \in S_1, W \in S_2 \mid V) &= \E[\ind(g(V) \in S_1) \, \ind(W \in S_2) \mid V] \\
    &= \ind(g(V) \in S_1) \, \E[\ind(W \in S_2) \mid V] \\
    &= \E[\ind(U \in S_1) \mid V] \, \Prob(W \in S_2 \mid V) \\
    &= \Prob(U \in S_1 \mid V) \, \Prob(W \in S_2 \mid V),
\end{align*}
which gives the result.
\end{proof}

It is now easy to see that \eqref{eq:potential-outcomes-are-deterministic} implies \eqref{eq:actions-are-unconfounded}.
Indeed, by recursive substitution, it is straightforward to show that there exist measurable functions $\tilde{g}_\tx$ for $\tx \in \{1, \ldots, \T\}$ such that
\[
    \X_\tx(\ax_{1:\tx}) \eqas \tilde{g}_\tx(\X_{0}, \ax_{1:\tx}) \qquad \text{for all $\tx \in \{1, \ldots, \T\}$ and $\ax_{1:\tx} \in \Aspace_{1:\tx}$},
\]
and so
\[
    (\X_{\sx}(\ax_{1:\sx}) : \sx \in \{1, \ldots, \T\}, \ax_{1:\sx} \in \Aspace_{1:\sx})
        = (\tilde{g}_\tx(\X_{0}, \ax_{1:\sx}) : \sx \in \{1, \ldots, \T\}, \ax_{1:\sx} \in \Aspace_{1:\sx}).
\]
The right-hand side is now seen to be a measurable function of $\X_0$ and hence certainly of $(\X_{0:\tx-1}(\A_{1:\tx-1}), \A_{1:\tx-1})$, so that the result follows by Lemma \ref{lem:determinism_conditional_independence}.

\section{Motivating Toy Example} \label{sec:motivating-example}

We provide here a toy scenario that illustrates intuitively the pitfalls that may arise when assessing twins using observational data without properly accounting for unmeasured confounding.
Suppose a digital twin has been designed for a particular make of car, e.g.\ to facilitate autonomous driving \citep{allamaa2022sim2real}.
The twin simulates how quantities such as the velocity and fuel consumption of the car respond as certain inputs are applied to it, such as braking, acceleration, steering, etc.
We wish to assess the accuracy of this twin using a dataset obtained from a fleet of the same make.
The braking performance of these vehicles is significantly affected by the age of their brake pads: if these are fairly new, then an aggressive braking strategy will stop the car, while if these are old, then the same aggressive strategy will send the car into a skid that will reduce braking efficacy.
Brake pad age is not recorded in the data we have obtained, but \emph{was} known to the drivers who operated these vehicles (e.g.\ perhaps they were aware of how recently their car was serviced), and so the drivers of cars with old brake pads tended to avoid braking aggressively out of safety concerns.

\begin{figure}
    \centering
    \includegraphics[height=5cm]{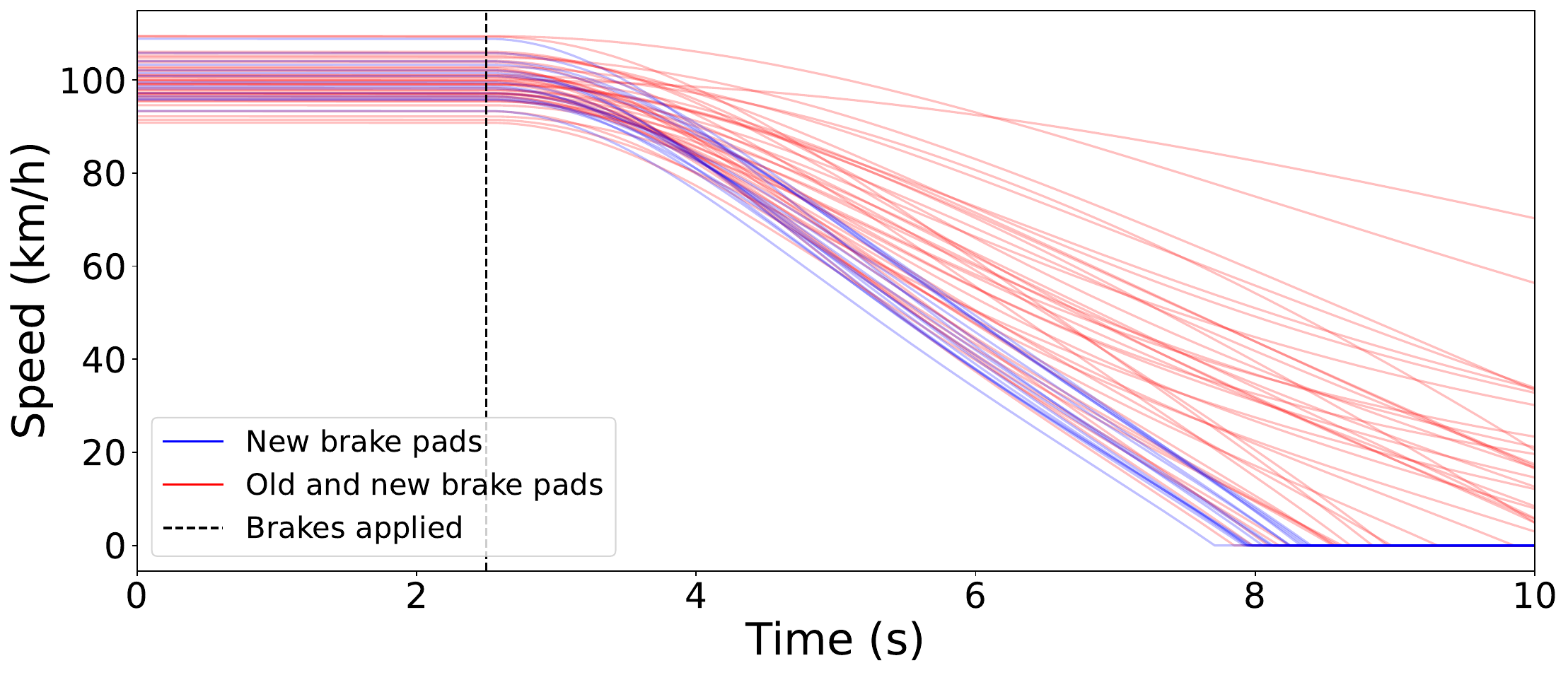}

    \caption{The discrepancy between observational data and interventional behavior. The data only show the effect of aggressive braking on cars with new brake pads (blue). This differs from what \emph{would} be observed if aggressive braking were applied to the entire fleet of cars, encompassing both those with old and new brake pads (red).}

    \label{fig:syn_ex}
\end{figure}

A naive approach to twin assessment in this situation would directly compare the outputs of the twin with the data and conclude the twin is accurate if these match closely.
However, in this scenario, the data contains a spurious relationship between braking strategy and the performance of the car: since aggressive braking is only observed for cars with new brake pads, the data appears to show that aggressive braking is effective at stopping the car, while in fact this is not the case for cars with older brake pads.
As such, the naive assessment approach would yield misleading information about the twin: a twin that captures only the behaviour of cars with newer brake pads would appear to be correct, while a twin that captures the full range of possibilities (i.e.\ regardless of brake pad age) would deviate from the observational data and appear therefore less accurate.
Figure \ref{fig:syn_ex} illustrates this pictorially under a toy model for this scenario.

More generally, whenever an \emph{unmeasured confounder} such as brake pad age is present, a potential discrepancy arises between how the real-world process was observed to behave in the dataset and how it \emph{would} behave under certain interventions.
An obvious approach towards mitigating this possibility is to measure additional quantities that may affect the outcome of interest.
For example, if brake pad age were included in the data in the scenario above, then it would be possible to adjust for its effect on braking performance. 
However, in many cases, gathering additional data may be costly or impractical.
Moreover, even if this strategy is pursued, it is rarely possible to rule out the possibility of unmeasured confounding altogether, especially for complicated real-world problems \citep{tsiatis2019dynamic}. 
For example, in the scenario above, it is very conceivable that some other factor such as weather conditions could play a similar confounding role as brake pad age, and so would need also to be included in the data, and so on.
Analogous scenarios are also easily forthcoming for other application domains such as medicine and economics \citep{manski1995identification,tsiatis2019dynamic,hernan2020causal}.
As such, rather than attempting to sidestep the issue of unmeasured confounding, we instead propose a methodology for assessing twins using data that is robust to its presence. %

\section{Causal Bounds} \label{sec:causal-bounds-proofs}

\subsection{Proof of Lemma \ref{lem:causal-bounds}}

\begin{proof}
We prove the lower bound, i.e.\ \eqref{eq:causal-bounds-lower}; the upper bound is analogous.
To streamline notation, we will denote
\begin{align*}
    p &\coloneqq \Prob(\A_{1:\tx} = \ax_{1:\tx} \mid \X_{0:\tx}(\ax_{1:\tx}) \in \B_{0:\tx}) \\
    q &\coloneqq \Prob(\A_{1:\tx} = \ax_{1:\tx} \mid \X_{0:\N}(\A_{1:\N}) \in \B_{0:\N}).
\end{align*}
Both quantities are always well-defined since we have
\[
    0 < \Prob(\X_{0:\tx}(\ax_{1:\tx}) \in \B_{0:\tx}) \leq \Prob(\X_{0:\N}(\A_{1:\N}) \in \B_{0:\N}),
\]
where the first inequality holds by assumption, and the second since $\N \leq \tx$.

Now, from the definition of $\Ylo$, we can write the left-hand side of \eqref{eq:causal-bounds-lower} as follows:
\begin{align*}
    \E[\Ylo \mid \X_{0:\tx}(\ax_{1:\tx}) \in \B_{0:\tx}]
        &= p \, \E[\Y(\A_{1:\tx}) \mid \X_{0:\tx}(\ax_{1:\tx}) \in \B_{0:\tx}, \A_{1:\tx} = \ax_{1:\tx}] \, + (1 - p) \, \ylo \\
        &= p \, \E[\Y(\A_{1:\tx}) \mid \X_{0:\N}(\A_{1:\N}) \in \B_{0:\N}, \A_{1:\tx} = \ax_{1:\tx}] \, + (1 - p) \, \ylo.
\end{align*}
Here the first step uses the law of total probability, and the second step uses the fact that $\N = \tx$ on the event $\{\A_{1:\tx} = \ax_{1:\tx}\}$, since $\N = \max \{0 \leq \sx \leq \tx \mid \A_{1:\sx} = \ax_{1:\sx}\}$.
Similarly, the right-hand side of \eqref{eq:causal-bounds-lower} can be written as
\[
    \E[\Ylo \mid \X_{0:\N}(\A_{1:\N}) \in \B_{0:\N}]
        = q \, \E[\Y(\A_{1:\tx}) \mid \X_{0:\N}(\A_{1:\N}) \in \B_{0:\N}, \A_{1:\tx} = \ax_{1:\tx}] \, + (1 - q) \, \ylo,
\]
again by the law of total probability.

Now, suppose we can show that $p \geq q$.
It is then a general fact about convex combinations that for any real numbers $u \geq v$, we have
\[
    p u + (1-p) v \geq q u + (1-q) v.
\]
Intuitively, moving from the left to the right-hand side shifts mass from the larger value $u$ to the smaller value $v$, which can only decrease the value of the convex combination overall.
Applying this with $u \coloneqq \E[\Y(\A_{1:\tx}) \mid \X_{0:\N}(\A_{1:\N}) \in \B_{0:\N}, \A_{1:\tx} = \ax_{1:\tx}]$ and $v \coloneqq \ylo$ then gives the result \eqref{eq:causal-bounds-lower}, since our assumption $\Prob(\ylo \leq \Y(\ax_{1:\tx}) \leq \yup \mid \X_{0:\tx}(\ax_{1:\tx}) \in \B_{0:\tx}) = 1$ straightforwardly implies that $u \geq v$.

We therefore wish to show that $q \leq p$ indeed holds.
To this end, observe the following:
\begin{align*}
    \Prob(\A_{1:\tx} = \ax_{1:\tx} \mid \X_{0:\tx}(\ax_{1:\tx}) \in \B_{0:\tx})
        &= \frac{\Prob(\X_{0:\tx}(\ax_{1:\tx}) \in \B_{0:\tx}, \A_{1:\tx} = \ax_{1:\tx})}{\Prob(\X_{0:\tx}(\ax_{1:\tx}) \in \B_{0:\tx})} \notag \\
        &\geq \frac{\Prob(\X_{0:\tx}(\ax_{1:\tx}) \in \B_{0:\tx}, \A_{1:\tx} = \ax_{1:\tx})}{\Prob(\X_{0:\N}(\ax_{1:\N}) \in \B_{0:\N})} \notag \\
        &= \frac{\Prob(\X_{0:\N}(\A_{1:\N}) \in \B_{0:\N}, \A_{1:\tx} = \ax_{1:\tx})}{\Prob(\X_{0:\N}(\ax_{1:\N}) \in \B_{0:\N})} \notag \\
        &= \frac{\Prob(\X_{0:\N}(\A_{1:\N}) \in \B_{0:\N}, \A_{1:\tx} = \ax_{1:\tx})}{\Prob(\X_{0:\N}(\A_{1:\N}) \in \B_{0:\N})} \notag \\
        &= \Prob(\A_{1:\tx} = \ax_{1:\tx} \mid \X_{0:\N}(\A_{1:\N}) \in \B_{0:\N}).
\end{align*}
Here the second step follows because $\N \leq \tx$, and hence
\[
    \Prob(\X_{0:\tx}(\ax_{1:\tx}) \in \B_{0:\tx}) \leq \Prob(\X_{0:\N}(\ax_{1:\N}) \in \B_{0:\N}).
\]
The third step holds because on the event $\{\A_{1:\tx} = \ax_{1:\tx}\}$, we have $\N = \tx$ and hence $\X_{0:\tx}(\ax_{1:\tx}) = \X_{0:\N}(\A_{1:\N})$.
The fourth step then follows because by definition of $\N = \max \{0 \leq \sx \leq \tx \mid \A_{1:\sx} = \ax_{1:\sx}\}$, we have $\ax_{1:\N} = \A_{1:\N}$ and hence $\X_{0:\N}(\ax_{1:\N}) = \X_{0:\N}(\A_{1:\N})$.
The result now follows.
\end{proof}

\subsection{Proof of Proposition \ref{prop:our-bounds-vs-manskis}}

\begin{proof}
From the definition of $\Yup$, we have straightforwardly
\begin{multline*}
    \Qup = \E[\Y(\A_{1:\tx}) \mid \X_{0:\N}(\A_{1:\N}) \in \B_{0:\N}, \A_{1:\tx} = \ax_{1:\tx}] \, \Prob(\A_{1:\tx} = \ax_{1:\tx} \mid \X_{0:\N}(\A_{1:\N}) \in \B_{0:\N}) \\
            +  \yup \, \Prob(\A_{1:\tx} \neq \ax_{1:\tx} \mid \X_{0:\N}(\A_{1:\N}) \in \B_{0:\N}).
\end{multline*}
A similar expression holds for $\Qlo$.
Subtracting these two expressions yields
\[
    \Qup - \Qlo = (\yup - \ylo) \, (1 - \Prob(\A_{1:\tx}=\ax_{1:\tx} \mid \X_{0:\N}(\A_{1:\N}) \in \B_{0:\N})). 
\]
Similar manipulations show that
\[
    \E[\Yup \mid \X_0 \in \B_0] - \E[\Ylo \mid \X_0 \in \B_0] = (\yup - \ylo) \, (1 - \Prob(\A_{1:\tx}=\ax_{1:\tx} \mid \X_0 \in \B_0)),
\]
and the result now follows.
\end{proof}

\subsection{Proof of Proposition \ref{prop:sharpness-of-bounds}} \label{sec:sharpness-of-bounds-supplement}

\begin{proof}
    We consider the case of the lower bound; the case of the upper bound is analogous.
    Choose $\xx_{1:\T} \in \B_{1:\T}$ arbitrarily.
    (Certainly some choice is always possible, since each $\B_{\sx}$ has positive measure and is therefore nonempty.)
    Define
    \begin{align*}
        \tilde{\X}_0 &\coloneqq \X_0 \\
        \tilde{\X}_{\sx}(\ax_{1:\sx}') &\coloneqq \ind(\A_{1:\sx} = \ax_{1:\sx}') \, \X_{\sx}(\ax_{1:\sx}') + \ind(\A_{1:\sx} \neq \ax_{1:\sx}') \, \xx_{\sx}
    \end{align*}
    for each $\sx \in \{0, \ldots, \T\}$ and $\ax_{1:\sx}' \in \Aspace_{1:\sx}$, and similarly let
    \[
        \tilde{\Y}(\ax_{1:\tx}') = \ind(\A_{1:\tx} = \ax_{1:\tx}') \, \Y(\ax_{1:\tx}') + \ind(\A_{1:\tx} \neq \ax_{1:\tx}')\, \ylo \qquad \text{for all $\ax_{1:\tx}' \in \Aspace_{1:\tx}$}.
    \]
    It is easy to check that $(\tilde{\X}_{0:\T}(\A_{1:\T}), \tilde{\Y}(\A_{1:\tx}), \A_{1:\T}) \eqas (\X_{0:\T}(\A_{1:\T}), \Y(\A_{1:\tx}), \A_{1:\T})$.
    But now we have directly $\tilde{\Y}(\ax_{1:\tx}) = \Ylo$.
    Moreover, it is easily checked from the definition of $\N$ and $\tilde{\X}_{0:\tx}(\ax_{1:\tx})$ that
    \[
        \tilde{\X}_{0:\tx}(\ax_{1:\tx}) \eqas (\X_{0:\N}(\A_{1:\N}), \xx_{\N+1:\tx}),
    \]
    so that
    \begin{align*}
        \ind(\tilde{\X}_{0:\tx}(\ax_{1:\tx}) \in \B_{0:\tx})
        &\eqas \ind(\tilde{\X}_{0:\N}(\ax_{1:\N}) \in \B_{0:\N}, \xx_{\N+1:\tx} \in \B_{\N+1:\tx}) \\
        &\eqas \ind(\X_{0:\N}(\A_{1:\N}) \in \B_{0:\N})
    \end{align*}
    since each $\xx_{\sx} \in \B_\sx$.
    Consequently,
    \begin{align*}
        \E[\tilde{\Y}(\ax_{1:\tx}) \mid \tilde{\X}_{0:\tx}(\ax_{1:\tx}) \in \B_{0:\tx}]
        &= \E[\Ylo \mid \tilde{\X}_{0:\tx}(\ax_{1:\tx}) \in \B_{0:\tx}] \\
        &= \E[\Ylo \mid \X_{0:\N}(\A_{1:\N}) \in \B_{0:\N}],
    \end{align*}
    which gives the result.
\end{proof}

\subsection{Bounds on the Conditional Expectation Given Specific Covariate Values} \label{sec:impossibility-of-bounds-for-continuous-data}

Theorem \ref{thm:causal-bounds} provides a bound on $\E[\Y(\ax_{1:\tx}) \mid \X_{0:\tx}(\ax_{1:\tx}) \in \B_{0:\tx}]$, i.e.\ the conditional expectation given the \emph{event} $\{\X_{0:\tx}(\ax_{1:\tx}) \in \B_{0:\tx}\}$, which is assumed to have positive probability.
We consider here the prospect of obtaining bounds on $\E[\Y(\ax_{1:\tx}) \mid \X_{0:\tx}(\ax_{1:\tx})]$, i.e.\ the conditional expectation given the \emph{value} of $\X_{0:\tx}(\ax_{1:\tx})$.
For falsification purposes, this would provide a means for determining that twin is incorrect when it outputs specific values of $\Xt_{0:\tx}(\ax_{1:\tx})$, rather than just that it is incorrect on average across all runs that output values $\Xt_{0:\tx}(\ax_{1:\tx}) \in \B_{0:\tx}$.

When $\X_{0:\tx}(\ax_{1:\tx})$ is discrete, Theorem \ref{thm:causal-bounds} yields measurable functions $\lo{\gx}, \up{\gx} : \Xspace_{0:\tx} \to \R$ such that
\begin{equation} \label{eq:psi-lo-up-defining-property}
    \lo{\gx}(\X_{0:\tx}(\ax_{1:\tx}))
        \leq \E[\Y(\ax_{1:\tx}) \mid \X_{0:\tx}(\ax_{1:\tx})] \leq \up{\gx}(\X_{0:\tx}(\ax_{1:\tx})) \qquad \text{almost surely}.
\end{equation}
In particular, $\lo{\gx}(\xx_{0:\tx})$ is obtained as the value of $\E[\Ylo \mid \X_{0:\N}(\A_{1:\N}) \in \B_{0:\N}]$ for $\B_{0:\tx} \coloneqq \{\xx_{0:\tx}\}$, and similarly for $\up{\gx}(\xx_{0:\tx})$.
Moreover, since the constants $\ylo, \yup \in \R$ in Theorem \ref{thm:causal-bounds} were allowed to depend on $\B_{0:\tx}$, and hence here on each choice of $\xx_{0:\tx} \in \Xspace_{0:\tx}$, we may think of these now as measurable functions $\ylo, \yup : \Xspace_{0:\tx} \to \R$ satisfying
\begin{equation} \label{eq:y-boundedness-functional-assumption}
    \ylo(\X_{0:\tx}(\ax_{1:\tx})) \leq \Y(\ax_{1:\tx}) \leq \yup(\X_{0:\tx}(\ax_{1:\tx})) \qquad \text{almost surely}.
\end{equation}
In other words, when $\X_{0:\tx}(\ax_{1:\tx})$ is discrete, Theorem \ref{thm:causal-bounds} provides bounds on the conditional expectation of $\Y(\ax_{1:\tx})$ given the value of $\X_{0:\tx}(\ax_{1:\tx})$ whenever we have $\ylo$ and $\yup$ such that \eqref{eq:y-boundedness-functional-assumption} holds.

When $\Prob(\X_{1:\tx}(\ax_{1:\tx}) \in \B_{1:\tx}) > 0$, we can also obtain bounds of the following form:
\begin{multline}
    \E[\Ylo \mid \X_0, \X_{1:\N}(\A_{1:\N}) \in \B_{1:\N}] 
        \leq \E[\Y(\ax_{1:\tx}) \mid \X_0, \X_{1:\tx}(\ax_{1:\tx}) \in \B_{1:\tx}] \\
        \leq \E[\Yup \mid \X_0, \X_{1:\N}(\A_{1:\N}) \in \B_{1:\N}] \qquad \text{almost surely}. \label{eq:bareinboim-style-bounds-with-continuous-initial-covariates}
\end{multline}
This can be seen by straightforwardly modifying the proof of Theorem \ref{thm:causal-bounds} to make all relevant quantities conditional on $\X_0$ (rather than $\X_0 \in \B_0$).
In particular, this holds regardless of whether or not $\X_0$ is discrete.
In turn, if $\X_{1:\tx}(\ax_{1:\tx})$ is discrete, then by a similar argument as was given in the previous subsection, this yields almost sure bounds on $\E[\Y(\ax_{1:\tx}) \mid \X_{0:\tx}(\ax_{1:\tx})]$ of the form in \eqref{eq:psi-lo-up-defining-property}, provided \eqref{eq:y-boundedness-functional-assumption} holds.
Alternatively, by taking $\B_{1:\tx} \coloneqq \Xspace_{1:\tx}$, \eqref{eq:bareinboim-style-bounds-with-continuous-initial-covariates} yields bounds of the form
\[
    \E[\Ylo \mid \X_0] \leq \E[\Y(\ax_{1:\tx}) \mid \X_0] \leq \E[\Yup \mid \X_0].
\]
If the action sequence $\ax_{1:\tx}$ is thought of as a single choice of an action from the extended action space $\Aspace_{1:\tx}$, then this recovers the bounds originally proposed by \citet{manski}, which allowed conditioning on potentially continuous pre-treatment covariates corresponding to our $\X_0$.

\subsection{Proof of Theorem \ref{thm:no-causal-bounds-for-continuous-data} (and Discussion)}

\begin{proof}
    Suppose we have a permissible $\lo{\gx}$.
    (The case of $\up{\gx}$ is analogous).
    Choose $\xx_{1:\T} \in \Xspace_{1:\T}$ arbitrarily, and define new potential outcomes
    \begin{align*}
        \tilde{\X}_0 &\coloneqq \X_0 \\
        \tilde{\X}_r(\ax_{1:r}') &\coloneqq %
            \ind(\A_{1:r} = \ax_{1:r}') \, \X_{r}(\ax_{1:r}') + \ind(\A_{1:r} \neq \ax_{1:r}') \, \xx_{r} \quad \text{for $r \in \{1, \ldots, \T\}$ and $\ax_{1:r}' \in \Aspace_{1:r}$}.
    \end{align*}
    Similarly, define
    \begin{align*}
        \tilde{\Y}(\ax_{1:\tx}') &\coloneqq \ind(\A_{1:\tx} = \ax_{1:\tx}') \, \Y(\ax_{1:\tx}') + \ind(\A_{1:\tx} \neq \ax_{1:\tx}') \, \ylo(\tilde{\X}_{0:\tx}(\ax_{1:\tx}')) \qquad \text{for all $\ax_{1:\tx}' \in \Aspace_{1:\tx}$}.
    \end{align*}
    It immediately follows that
    \[
        (\tilde{\X}_{0:\T}(\A_{1:\T}), \tilde{\Y}(\A_{1:\tx}), \A_{1:\T}) \eqas (\X_{0:\T}(\A_{1:\T}), \Y(\A_{1:\tx}), \A_{1:\T}).
    \]
    Moreover, it is easily checked that
    \[
        \ylo(\tilde{\X}_{0:\tx}(\ax_{1:\tx})) \leq \tilde{\Y}(\ax_{1:\tx}) \leq \yup(\tilde{\X}_{0:\tx}(\ax_{1:\tx})) \qquad \text{almost surely}.
    \]
    As such, since $\lo{\gx}$ is permissible, we must have, almost surely,
    \begin{align}
        \lo{\gx}(\tilde{\X}_{0:\tx}(\ax_{1:\tx})) &\leq \E[\tilde{\Y}(\ax_{1:\tx}) \mid \tilde{\X}_{0:\tx}(\ax_{1:\tx})] \notag \\
         &= \begin{multlined}[t]
            \E[\tilde{\Y}(\A_{1:\tx}) \mid \tilde{\X}_{0:\tx}(\ax_{1:\tx}), \A_{1:\tx} = \ax_{1:\tx}] \, \Prob(\A_{1:\tx} = \ax_{1:\tx} \mid \tilde{\X}_{0:\tx}(\ax_{1:\tx})) \\
                + \underbrace{\E[\tilde{\Y}(\ax_{1:\tx}) \mid \tilde{\X}_{0:\tx}(\ax_{1:\tx}), \A_{1:\tx} \neq \ax_{1:\tx}]}_{=\ylo(\tilde{\X}_{0:\tx}(\ax_{1:\tx}))} \, \Prob(\A_{1:\tx} \neq \ax_{1:\tx} \mid \tilde{\X}_{0:\tx}(\ax_{1:\tx})).
        \end{multlined} \label{eq:no-continuous-bounds-proof-convex-combination}
    \end{align}
    Now, by our definition of $\tilde{\X}_{0:\tx}(\ax_{1:\tx})$, we have almost surely
    \begin{align*}
        \MoveEqLeft \ind(\A_{1} \neq \ax_{1}) \, \Prob(\A_{1:\tx} = \ax_{1:\tx} \mid \tilde{\X}_{0:\tx}(\ax_{1:\tx})) \\
            &= \ind(\A_{1} \neq \ax_{1}, \tilde{\X}_{\sx}(\ax_{1:\sx}) = \xx_\sx) \, \Prob(\A_{1:\tx} = \ax_{1:\tx} \mid \tilde{\X}_{0:\tx}(\ax_{1:\tx})) \\
            &= \ind(\A_{1} \neq \ax_{1}) \, \E[\ind(\A_{1:\tx} = \ax_{1:\tx}, \tilde{\X}_{\sx}(\ax_{1:\sx}) = \xx_\sx) \mid \tilde{\X}_{0:\tx}(\ax_{1:\tx})] \\
            &= \ind(\A_{1} \neq \ax_{1}) \, \E[\ind(\A_{1:\tx} = \ax_{1:\tx}, \X_{\sx}(\A_{1:\sx}) = \xx_\sx) \mid \tilde{\X}_{0:\tx}(\ax_{1:\tx})] \\
            &= 0,
    \end{align*}
    where the last step follows by our assumption that $\Prob(\X_{\sx}(\A_{1:\sx}) = \xx_{\sx}) = 0$.
    Combining this with \eqref{eq:no-continuous-bounds-proof-convex-combination}, we get, almost surely,
    \begin{align}
        \ind(\A_{1} \neq \ax_{1}) \, \lo{\gx}(\X_{0}, \xx_{1:\tx}) &=  \ind(\A_{1} \neq \ax_{1}) \, \lo{\gx}(\tilde{\X}_{0:\tx}(\ax_{1:\tx})) \notag \\
            &\leq \ind(\A_{1} \neq \ax_{1}) \, \ylo(\tilde{\X}_{0:\tx}(\ax_{1:\tx})) \notag \\
            &= \ind(\A_{1} \neq \ax_{1}) \, \ylo(\X_{0}, \xx_{1:\tx}). \label{eq:no-continuous-bounds-intermediate-step}
    \end{align}
    Now let $\xx_0 \in \Xspace_0$ be the value such that $\Prob(\X_0 = \xx_0) = 1$.
    Using our assumption that $\Prob(\A_1 \neq \ax_1) > 0$ and the fact that $\xx_{1:\tx}$ was arbitrary, we obtain
    \[
        \lo{\gx}(\xx_{0:\tx}) \leq \ylo(\xx_{0:\tx}) \qquad \text{for all $\xx_{1:\tx} \in \Xspace_{1:\tx}$}.
    \]
    The result now follows.
\end{proof}

To gain intuition for the phenomenon underlying Theorem \ref{thm:no-causal-bounds-for-continuous-data}, consider a simplified model consisting of $\Xspace$-valued potential outcomes $(\X(\ax') : \ax' \in \Aspace)$, real-valued potential outcomes $(\Y(\ax') : \ax' \in \Aspace)$, and an $\Aspace$-valued random variable $\A$ representing the choice of action.
(This constitutes a special case of our setup with $\T = 1$ and $\Xspace_0$ taken to be a singleton set.)
Suppose moreover that the following conditions hold:
\begin{align*}
    \Prob(\X(\A) = \xx) &= 0 \qquad \text{for all $\xx \in \Xspace$} \\
    \Prob(\A = \ax) &< 1.
\end{align*}
We then have
\begin{multline} \label{eq:no-continuous-bounds-toy-example}
    \E[\Y(\ax) \mid \X(\ax)] \\
        \eqas \E[\Y(\A) \mid \X(\A), \A = \ax] \, \Prob(\A = \ax \mid \X(\ax)) + \E[\Y(\ax) \mid \X(\ax), \A \neq \ax] \, \Prob(\A \neq \ax \mid \X(\ax)).
\end{multline}
But now, since the behaviour of $\X(\ax)$ is only observed on $\{\A = \ax\}$, for any given value of $\xx \in \Xspace$, we cannot rule out the possibility that
\[
    \X(\ax) = \ind(\A = \ax) \, \X(\A) + \ind(\A \neq \ax) \, \xx \qquad \text{almost surely}.
\]
In turn, since $\Prob(\A = \ax) > 0$, this would imply $\Prob(\X(\ax) = \xx) > 0$, and, since $\Prob(\X(\A) = \xx) = 0$, that $\Prob(\A = \ax \mid \X(\ax) = \xx) = 0$.
From \eqref{eq:no-continuous-bounds-toy-example}, this would yield
\[
    \E[\Y(\ax) \mid \X(\ax) = \xx] = \E[\Y(\ax) \mid \X(\ax) = \xx, \A \neq \ax].
\]
But now, since the behaviour of $\Y(\ax)$ is unobserved on $\{\A \neq \ax\}$, intuitively speaking, the observational distribution does not provide any information about the value of the right-hand side, and therefore about the behaviour of $\E[\Y(\ax) \mid \X(\ax)]$ more generally since $\xx \in \Xspace$ was arbitrary.

\section{Hypothesis Testing Methodology} 

\subsection{Validity of Testing Procedure} \label{sec:hyp-testing-supplement}

We show here that our procedure for testing $\Qt \geq \Qlo$ based on the one-sided confidence intervals $\qlo{\alpha}$ and $\qt{\alpha}$ has the correct probability of type I error, provided $\qlo{\alpha}$ and $\qt{\alpha}$ have the correct coverage probabilities.
In particular, the result below (which applies a standard union bound argument) shows that if $\Qt \geq \Qlo$, then our test rejects (i.e.\ $\qt{\alpha} < \qlo{\alpha}$) with probability at most $\alpha$.
An analogous result is easily proven for testing $\Qt \leq \Qup$ also, with $\qlo{\alpha}$ replaced by a one-sided upper $(1-\alpha/2)$-confidence interval for $\Qup$, and $\qt{\alpha}$ replaced by a one-sided lower $(1 - \alpha/2)$-confidence interval for $\Qt$.

\begin{proposition}
Suppose that for some $\alpha \in (0, 1)$ we have random variables $\qt{\alpha}$ and $\qlo{\alpha}$ satisfying
\begin{align}
    \Prob(\Qlo \geq \qlo{\alpha}) &\geq 1 - \frac{\alpha}{2} \label{eq:qlo-confidence-interval-guarantee} \\
    \Prob(\Qt \leq \qt{\alpha}) &\geq 1 - \frac{\alpha}{2} \label{eq:qt-confidence-interval-guarantee}.
\end{align}
If $\Qt \geq \Qlo$, then $\Prob(\qt{\alpha} < \qlo{\alpha}) \leq \alpha$.
\end{proposition}

\begin{proof}
If $\Qt \geq \Qlo$, then we have
\[
    \{\qt{\alpha} < \qlo{\alpha}\} \subseteq \{\Qt > \qt{\alpha}\} \cup \{\Qlo < \qlo{\alpha}\}.
\]
To see this, note that
\begin{align*}
    (\{\Qt > \qt{\alpha}\} \cup \{\Qlo < \qlo{\alpha}\})^c
        &= \{\Qt > \qt{\alpha}\}^c \cap \{\Qlo < \qlo{\alpha}\}^c \\
        &= \{\Qt \leq \qt{\alpha}\} \cap \{\Qlo \geq \qlo{\alpha}\} \\
        &\subseteq \{\qlo{\alpha} \leq \qt{\alpha}\}.
\end{align*}
As such,
\begin{align*}
    \Prob(\qt{\alpha} < \qlo{\alpha}) &\leq \Prob(\{\Qt > \qt{\alpha}\} \cup \{\Qlo < \qlo{\alpha}\}) \\
        &\leq \Prob(\Qt > \qt{\alpha}) + \Prob(\Qlo < \qlo{\alpha}) \\
        &\leq \alpha/2 + \alpha/2 \\
        &= \alpha.
\end{align*}
\end{proof}

\subsection{Unbiased Sample Mean Estimates of $\Qlo$, $\Qt$, and $\Qup$} \label{sec:confidence-intervals-methodology-supplement}

We use our data to obtain one-sided confidence intervals $\qlo{\alpha}$ and $\qt{\alpha}$ satisfying \eqref{eq:qlo-confidence-interval-guarantee} and \eqref{eq:qt-confidence-interval-guarantee} as required by our procedure for testing $\Qt \geq \Qlo$.
We use an analogous procedure to obtain confidence intervals for testing $\Qt \leq \Qup$.
We tried two techniques for this: an exact method based on Hoeffding's inequality, and an approximate method based on bootstrapping.
Conceptually, both are based on obtaining unbiased sample mean estimates of $\Qlo$ and $\Qt$, which we describe now, before giving the particulars of each method in the next two subsections.

We begin with our sample mean estimator of $\Qlo$.
Recall that we assume access to a dataset $\D$ consisting of i.i.d.\ copies of observational trajectories of the form
\[
    \X_0, \A_1, \X_1(\A_1), \ldots, \A_\T, \X_\T(\A_{1:\T}).
\]
Let $\D(\ax_{1:\tx}, \B_{0:\tx})$ be the subset of trajectories in $\D$ for which $\X_{0:\N}(\A_{1:\N})\in\B_{0:\N}$.
Obtaining $\D(\ax_{1:\tx}, \B_{0:\tx})$ is possible since the only random quantity that $\N = \max\{0 \leq \sx \leq \tx \mid \A_{1:\sx} = \ax_{1:\sx}\}$ depends on is $\A_{1:\tx}$, which is included in the data.
We denote the cardinality of $\D(\ax_{1:\tx}, \B_{0:\tx})$ by $\nx \coloneqq \abs{\D(\ax_{1:\tx}, \B_{0:\tx})}$.
We then denote by $\Ylo^{(i)}$ for $i \in \{1, \ldots, \nx\}$ the corresponding values of
\begin{align*}
    \Ylo &= \ind(\A_{1:\tx} = \ax_{1:\tx}) \, \fx(\X_{0:\tx}(\A_{1:\tx})) + \ind(\A_{1:\tx} \neq \ax_{1:\tx}) \, \ylo
\end{align*}
obtained from each trajectory in $\D(\ax_{1:\tx}, \B_{0:\tx})$.
This is again possible since both terms only depend on the observational quantities $(\X_{0:\tx}(\A_{1:\tx}), \A_{1:\tx})$.
It is easily seen that the values of $\Ylo^{(i)}$ are i.i.d.\ and satisfy $\E[\Ylo^{(i)}] = \Qlo$.
As a result, the sample mean
\begin{equation} \label{eq:YClmean-definition-supplement}
    \YClmean \coloneqq \frac{1}{\nx} \sum_{i=1}^{\nx} \Ylo^{(i)} 
\end{equation}
is an unbiased estimator of $\Qlo$.

We obtain an unbiased sample mean estimate of $\Qt$ in a similar fashion as for $\Qlo$.
Recall that we assume access to a dataset $\Dt(\ax_{1:\tx})$ consisting of i.i.d.\ copies of
\[
    \X_0, \Xt_1(\X_0, \ax_1), \ldots, \Xt_t(\X_0, \ax_{1:\tx}).
\]
Let $\Dt(\ax_{1:\tx}, \B_{0:\tx})$ denote the subset of trajectories in $\Dt(\ax_{1:\tx})$ with $(\X_0, \Xt_{\tx}(\X_0, \ax_{1:\tx})) \in \B_{0:\tx}$, and denote its cardinality by $\widehat{\nx} \coloneqq \abs{\Dt(\ax_{1:\tx}, \B_{0:\tx})}$.
Then denote by $\Yt^{(i)}$ for $i \in \{1 \ldots, \widehat{\nx}\}$ the corresponding values of
\[
    \Yt = \fx(\X_0, \Xt_{1:\tx}(\X_0, \ax_{1:\tx}))
\]
obtained from each trajectory in $\Dt(\ax_{1:\tx}, \B_{0:\tx})$.
It is easily seen that the values $\Yt^{(i)}$ are i.i.d.\ (since the entries of $\Dt(\ax_{1:\tx})$ are) and satisfy $\E[\Yt^{(i)}] = \Qt$.
As a result, the sample mean
\[
    \Ytmean \coloneqq \frac{1}{\widehat{\nx}} \sum_{i=1}^{\widehat{\nx}} \Yt^{(i)}
\]
is an unbiased estimator of $\Qt$.

\subsection{Exact Confidence Intervals via Hoeffding's Inequality}

Recall that we assume in Section \ref{sec:hypotheses-from-causal-bounds-setup} that $\Y(\ax_{1:\tx})$ has the form $\Y(\ax_{1:\tx}) = \fx(\X_{0:\tx}(\ax_{1:\tx}))$, and that moreover
\begin{equation} \label{eq:f-boundedness-assumption-hoeffding-proof}
    \ylo \leq \fx(\xx_{0:\tx}) \leq \yup \qquad \text{for all $\xx_{0:\tx} \in \B_{0:\tx}$.}
\end{equation}
This means $\Yt^{(i)}$ is almost surely bounded in $[\ylo, \yup]$, and so $\Ytmean$ gives rise to one-sided confidence intervals via an application of Hoeffding's inequality.
The exact form of these confidence intervals is as follows:

\begin{proposition} \label{prop:hoeffding-confidence-bounds-supp}
If \eqref{eq:f-boundedness-assumption-hoeffding-proof} holds, then for each $\alpha \in (0, 1)$, letting
\[
    \CIlen \coloneqq (\yup - \ylo) \, \sqrt{\frac{1}{2 \nx} \, \log \frac{2}{\alpha}}  \qquad \text{and} \qquad \widehat{\CIlen} \coloneqq (\yup - \ylo) \, \sqrt{\frac{1}{2 \widehat{\nx}} \, \log \frac{2}{\alpha}},
\]
and similarly
\begin{align*}
    \qlo{\alpha} \coloneqq \YClmean - \CIlen  \qquad \text{and} \qquad
    \qt{\alpha} \coloneqq \Ytmean + \widehat{\CIlen},
\end{align*}
it follows that
\begin{align*}
    \Prob(\Qlo \geq \qlo{\alpha}) \geq 1 - \frac{\alpha}{2} \qquad \text{and} \qquad \Prob(\Qt \leq \qt{\alpha}) \geq 1 - \frac{\alpha}{2}.
\end{align*}
\end{proposition}

\begin{proof}
We only prove the result for $\qlo{\alpha}$; the other statement can be proved analogously.
Recall that $\YClmean$ is the empirical mean of i.i.d.\ samples $\Ylo^{(i)}$ for $i\in \{1, \ldots, \nx\}$ with $\E[\Ylo^{(i)}]=\Qlo$ (see Equation~\ref{eq:YClmean-definition-supplement}).
Moreover, by \eqref{eq:f-boundedness-assumption-hoeffding-proof}, $\Ylo^{(i)}$ is almost surely bounded in $[\ylo, \yup]$.
Hoeffding's inequality then implies that
\begin{align*}
    \Prob(\YClmean - \Qlo > \CIlen) &\leq \exp\left(- \frac{2 \nx \CIlen^2}{(\yup - \ylo)^2 } \right).
\end{align*}
In turn, some basic manipulations yield
\begin{align*}
    \Prob(\Qlo \geq \qlo{\alpha}) &= 1 - \Prob(\Qlo < \YClmean - \CIlen) \\
    &\geq 1 - \exp\left(- \frac{2 \nx \CIlen^2}{(\yup - \ylo)^2 } \right) \\
    &= 1- \frac{\alpha}{2}.
\end{align*}
\end{proof}

\subsection{Approximate Confidence Intervals via Bootstrapping} \label{subsec:bootstrapping}

While Hoeffding's inequality yields the probability guarantees in \eqref{eq:qlo-confidence-interval-guarantee} and \eqref{eq:qt-confidence-interval-guarantee} exactly, the confidence intervals obtained can be conservative.
Consequently, our testing procedure may have lower probability of falsifying certain twins that in fact do not satisfy the causal bounds.
To address this, we also consider an approximate approach based on bootstrapping \citep{efron1979bootstrap} that can produce tighter confidence intervals.
While other schemes are possible, bootstrapping provides a general-purpose approach that is straightforward to implement and works well in practice.

At a high level, our approach here is again to construct one-sided level $1 - \alpha/2$ confidence intervals via bootstrapping on $\Qlo$ and $\Qt$.
Many bootstrapping procedures for obtaining confidence intervals have been proposed in the literature \citep{tibshirani1993introduction,davison1997bootstrap,hesterberg2015what}. 
Our results reported below were obtained via the \emph{reverse percentile} bootstrap (see \citealt{hesterberg2015what} for an overview).
(We also tried the \emph{percentile} bootstrap method, which obtained nearly indistinguishable results.)
In particular, this method takes
\[
    \qlo{\alpha} \coloneqq 2 \YClmean - \Delta_{\mathrm{b}}
    \qquad \qt{\alpha} \coloneqq 2 \widehat{\mu} - \widehat{\Delta}_{\mathrm{b}},
\]
where $\Delta_{\mathrm{b}}$ and $\widehat{\Delta}_{\mathrm{b}}$ correspond to the approximate $1 - \alpha / 2$ and $\alpha / 2$ quantiles of the distributions of 
\[
    \frac{1}{\nx} \sum_{i=1}^{\nx} \Ylo^{(i^\ast)} 
    \qquad\text{and}\qquad
    \frac{1}{\widehat{\nx}} \sum_{i=1}^{\widehat{\nx}} \Yt^{(i^\ast)},
\]
where each $\Ylo^{(i^\ast)}$ and $\Y^{(i^\ast)}$ is obtained by sampling uniformly with replacement from among the values of $\Ylo^{(i)}$ and $\Y^{(i)}$.
In our case study, as is typically done in practice, we approximated $\Delta_{\mathrm{b}}$ and $\widehat{\Delta}_{\mathrm{b}}$ via Monte Carlo sampling.
It can be shown that the confidence intervals produced in this way obtain a coverage level that approaches the desired level of $1 - \alpha/2$ as $\nx$ and $\widehat{\nx}$ grow to infinity under mild assumptions \citep{hall1988theoretical}.

\section{Experimental Details} \label{sec:experiments-supplement}

\subsection{MIMIC Preprocessing} \label{sec:mimic-preprocessing-supp}

For data extraction and preprocessing, we used the same procedure as \citet{ai-clinician} with minor modifications.
For completeness, we describe the pre-processing steps applied in \citet{ai-clinician} and subsequently outline our modifications to these.
First, we extracted adult patients fulfilling the sepsis-3 criteria \citep{sepsis-criteria}. 
Sepsis was defined as a suspected infection (as indicated by prescription of antibiotics and sampling of bodily fluids for microbiological culture) combined with evidence of organ dysfunction, defined by a SOFA score $\geq 2$ \citep{sepsis-criteria, seymour2016assessment}.  
Then we excluded patients for whom any of the following was true: their age was less than $18$ years old at the time of ICU admission; their mortality not documented; their IV fluid/vasopressors intake was not documented; their treatment was withdrawn.

We made the following modifications to the preprocessing code of \citet{ai-clinician} for our experiment.
First, instead of extracting physiological quantities (e.g.\ heart rate) every 4 hours, we extracted these every hour.
Additionally, we excluded patients with any missing hourly vitals during the first 4 hours of their ICU stay.

We then extracted a total of 19 quantities of interest listed in Table \ref{tab:mimic-features}.
Of these, 17 were physiological quantities associated with the patient, including static demographic quantities (e.g.\ age), patient vital signs (e.g.\ heart rate), and patient lab values (e.g.\ potassium blood concentration).
All of these were continuous values, apart from sex.
These were chosen since they were the subset of physiological quantities extracted from MIMIC by \citet{ai-clinician} that are also modelled by Pulse, and were used to define our observation spaces $\Xspace_\tx$ as described next.
The remaining 2 quantities (intravenous fluids and vasopressor doses) were chosen since they correspond to treatments that the patient received, and were used to define our action spaces $\Aspace_\tx$ as described below.

\subsection{Sample Splitting}

Before proceeding further, we randomly selected 5\% of our extracted trajectories (583 trajectories, denoted as $\D_0$) to use for preliminary tasks such as choosing the parameters of our hypotheses.
We reserved the remaining 95\% (11,094 trajectories, denoted as $\D$) for the actual testing.
By a standard sample splitting argument \citep{cox1975note}, the statistical guarantees of our testing procedure established above continue to apply even when our hypotheses are defined in this data-dependent way.

\begin{table}[t]%
\centering
\begin{footnotesize}
\begin{tabular}{lll}
\toprule
Category &  Physiological quantity \\
\midrule
Demographic & Age \\
& Sex  \\
& Weight \\
\midrule
Vital Signs & Heart rate (HR)  \\
& Systolic blood pressure (SysBP)  \\
& Diastolic blood pressure (DiaBP)  \\
& Mean blood pressure (MeanBP) \\
& Respiratory Rate (RR) \\
& Skin Temperature (Temp) \\
\midrule
Lab Values & Potassium Blood Concentration (Potassium)  \\
& Sodium Blood Concentration (Sodium)  \\
& Chloride Blood Concentration (Chloride)  \\
& Glucose Blood Concentration (Glucose)  \\
& Calcium Blood Concentration (Calcium)  \\
& Bicarbonate Blood Concentration ($\textup{HCO}_3$)  \\
& Arterial $\textup{O}_2$ Pressure ($\textup{PaO}_2$)  \\
& Arterial $\textup{CO}_2$ Pressure ($\textup{PaCO}_2$) \\
\midrule
Treatments & Intravenous fluid (IV) dose \\
& Vasopressor dose \\
\bottomrule
\end{tabular}
\end{footnotesize}
\caption{Physiological quantities and treatments extracted from MIMIC}\label{tab:mimic-features}
\end{table}

\subsection{Observation Spaces} \label{sec:observation-space-definition-supp}

Our $\Xspace_0$ consisted of the following features: age, sex, weight, heart rate, systolic blood pressure, diastolic blood pressure and respiration rate.
We chose $\Xspace_0$ in this way because, out of the 17 physiological quantities we extracted from MIMIC, these were the quantities that can be initialised to user-provided values before starting a simulation in version 4 of Pulse, which we used. %
(In contrast, Pulse initialises the other 10 features to default values.)
For the remaining observation spaces, we used the full collection of the 17 physiological quantities we extracted to define $\Xspace_1 = \cdots = \Xspace_4$.
We encoded all features in $\Xspace_\tx$ numerically, i.e.\ $\Xspace_0 = \R^7$, and $\Xspace_\tx = \R^{17}$ for $\tx \in \{1, 2, 3, 4\}$.

\subsection{Action Spaces} \label{sec:action-space-definition-supp}

Following \citet{ai-clinician}, we constructed our action space using 2 features obtained from MIMIC, namely intravenous fluid (IV) and vasopressor doses.
To obtain discrete action spaces suitable for our framework, we used the same discretization procedure for these quantities as was used by \citet{ai-clinician}.
Specifically, we divided the hourly doses of intravenous fluids and vasopressors into 5 bins each, with the first bin corresponding to zero drug dosage, and the remaining 4 bins based on the quartiles of the non-zero drug dosages in our held-out observational dataset $\D_0$.
From this we obtained action spaces $\Aspace_1 = \cdots = \Aspace_{4}$ with $5 \times 5 = 25$ elements. 
Table \ref{tab:act_space} shows the dosage bins constructed in this way, as well as the frequency of each bin's occurrence in the observational data.

\begin{table}[t]%
    \centering
    \begin{footnotesize}
\begin{tabular}{ll|ccccc}
\multicolumn{1}{c}{} & \multicolumn{1}{c}{} & \multicolumn{5}{c}{Vasopressor dose ($\mu$g/kg/min)}\\
\multicolumn{1}{c}{} & \multicolumn{1}{c}{} &      0 &  0.0 - 0.061 &  0.061 - 0.15 &  0.15 - 0.313 &  $>0.313$ \\
\midrule
\multirow{5}{*}{IV dose (mL/h)} & 0        &  16659 &          329 &           256 &           152 &      145 \\
& 0 - 20   &   5840 &          428 &           351 &           244 &      145 \\
& 20 - 75  &   6330 &          297 &           378 &           383 &      309 \\
& 75 - 214 &   6232 &          176 &           175 &           197 &      273 \\
& $>214$    &   5283 &          347 &           488 &           544 &      747 \\
\bottomrule
\end{tabular}
    \end{footnotesize}
\caption{Action space with frequency of occurrence in observational data} \label{tab:act_space}
\end{table}

\subsection{Hypothesis Parameters} \label{sec:hypothesis-parameters-supplement}

We used our held-out observational dataset $\D_0$ to obtain a collection of hypothesis parameters $(\tx, \fx, \ax_{1:\tx}, \B_{0:\tx})$.
Specifically, for each physiological quantity of interest (e.g.\ heart rate) in the list of Vital Signs and Lab Values given in Table \ref{tab:mimic-features}, we did the following.
First, for each $\tx \in \{0, \ldots, 4\}$, we obtained 16 choices of $\B_{\tx}$ by discretizing the patient space $\Xspace_{\tx}$ into 16 subsets based on the values of certain features as follows: 2 bins corresponding to sex; 4 bins corresponding to the quartiles of the ages of patients in $\D_0$; 2 bins corresponding to whether or not the value of the chosen physiological quantity of interest at time $\tx$ was above or below its median value in $\D_0$.

Next, for each $\tx \in \{1, \ldots, 4\}$, $\ax_{1:\tx} \in \Aspace_{1:\tx}$, and sequence $\B_{0:\tx}$ with each $\B_{\tx'}$ as defined in the previous step, let $\D_0(\tx, \ax_{1:\tx}, \B_{0:\tx})$ denote the subset of $\D_0$ corresponding to $(\tx, \ax_{1:\tx}, \B_{0:\tx})$, i.e.\
\begin{multline*}
    \D_0(\tx, \ax_{1:\tx}, \B_{0:\tx}) \\
        \coloneqq \{\X_{0:\tx}(\A_{1:\tx}) \mid \text{$(\X_{0:\T}(\A_{1:\T}), \A_{1:\T}) \in \D_0$ with $\A_{1:\tx} = \ax_{1:\tx}$ and $\X_{0:\tx}(\A_{1:\T}) \in \B_{0:\tx}$}\}.
\end{multline*}
We then selected the set of all triples $(\tx, \ax_{1:\tx}, \B_{0:\tx})$ such that $\D_0(\tx, \ax_{1:\tx}, \B_{0:\tx})$ contained at least one trajectory.
This meant the number of combinations of hypotheses parameters that we considered was limited to a tractable quantity, which had benefits both computationally, and also by ensuring that we did not sacrifice too much power when adjusting for multiple testing.

Finally, for each selected triple $(\tx, \ax_{1:\tx}, \B_{0:\tx})$, we chose a corresponding $\fx$ as follows.
First, we let $i \in \{1, \ldots, \Xspacedim_\tx\}$ denote the index of the physiological quantity of interest in $\Xspace_\tx = \R^{\Xspacedim_\tx}$.
We then set $\ylo, \yup$ to be the .2 and the .8 quantiles of the values in
\[
    \{(\X_\tx(\A_{1:\tx}))_i \mid \X_{0:\tx}(\A_{1:\tx}) \in \D_0(\tx, \ax_{1:\tx}, \B_{0:\tx})\}.
\]
We then obtained $\fx : \Xspace_{0:\tx} \to \R$ as the function that extracts the physiological quantity of interest from $\Xspace_\tx$ and clips its value to lie between the values $\ylo$ and $\yup$. Explicitly, $\fx(\xx_{0:\tx}) \coloneqq \min(\max((\xx_{\tx})_{i}, \ylo), \yup)$.

Overall, accounting for all physiological quantities of interest, we obtained 721 distinct choices of $(\tx, \fx, \ax_{1:\tx}, \B_{0:\tx})$ in this way.
Figure \ref{fig:n_histograms} shows the amount of non-held out observational and twin data that we subsequently used for testing each hypothesis, i.e.\ the values of $n$ and $\widehat{n}$ as defined in Section \ref{sec:confidence-intervals-methodology-supplement} above.
We describe how we generated our dataset of twin trajectories next.

\subsection{Generating Twin Trajectories Using the Pulse Physiology Engine}\label{sec:pulse-trajectories-supplement}

The Pulse Physiology Engine is a comprehensive open-source human physiology simulator that has been used in medical education, research, and training.
Pulse allows users to initialize patient trajectories with given age, sex, weight, heart rate, systolic blood pressure, diastolic blood pressure and respiration rate and medical conditions such as sepsis, COPD, ARDS, etc. Once initialised, users have the ability to advance patient trajectories by a given time step (one hour in our case), and administer actions (e.g. administer a given dose of IV fluids or vasopressors).
Detailed documentation is available at: \href{https://pulse.kitware.com/}{pulse.kitware.com}.

\begin{figure}[t]
    \centering
    \includegraphics[height=20.2cm]{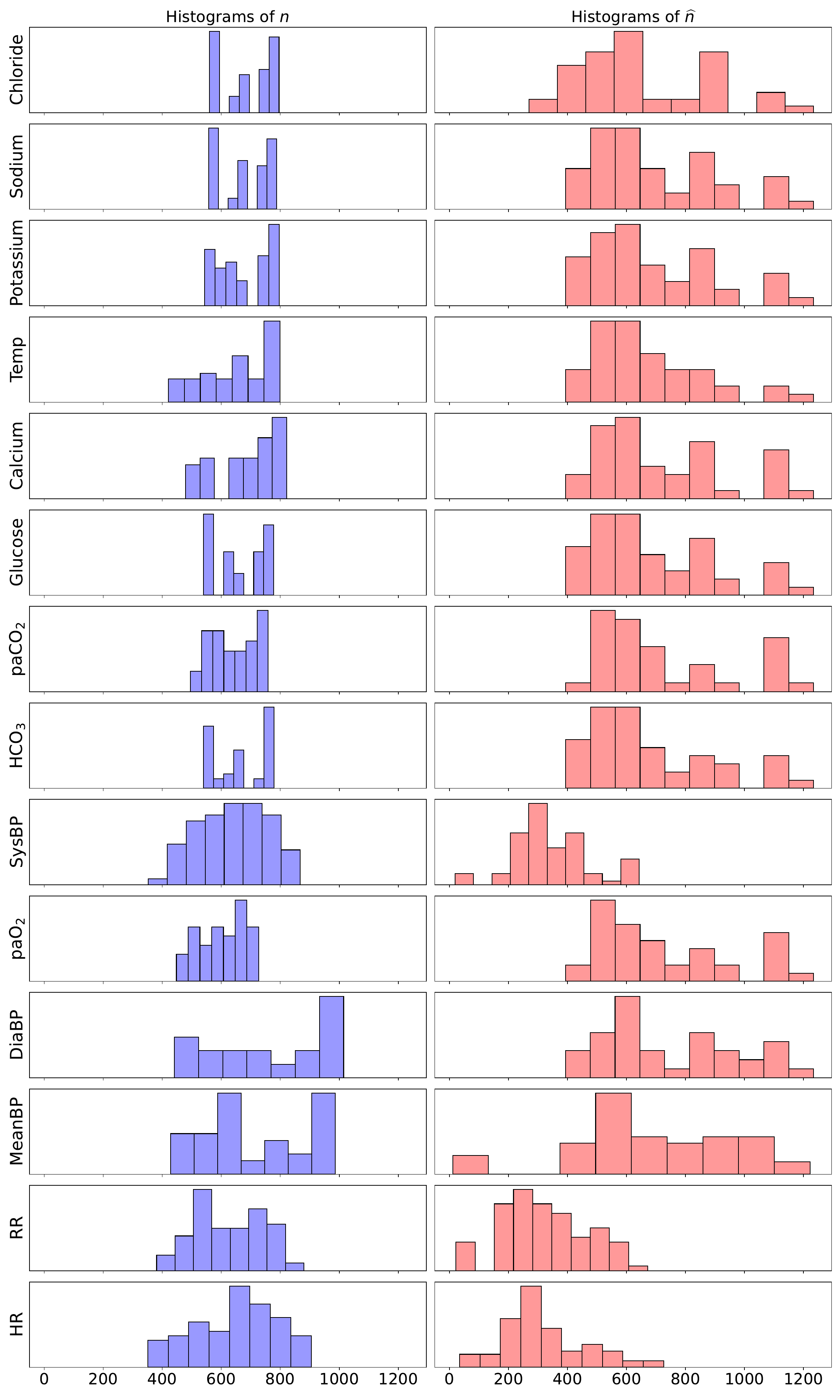}
    \caption{Histograms of $n$ and $\widehat{n}$ (as defined in Section \ref{sec:confidence-intervals-methodology-supplement}) across all hypothesis parameters corresponding to each physiological quantity of interest.}
    \label{fig:n_histograms}
\end{figure}

In Algorithm \ref{algo:twin-data-generation} we describe how we generated the twin data to test the chosen hypotheses. Note that we sampled $\X_0$ without replacement as it ensures that each $\X_0$ is chosen at most once and consequently twin trajectories in $\Dt(\ax_{1:\tx})$ are i.i.d. 
Additionally, Algorithm \ref{algo:twin-data-generation} can be easily parallelised to improve efficiency.
Figure \ref{fig:n_histograms} shows histograms of the number of twin trajectories $\widehat{\nx}$ (as defined in Section \ref{sec:confidence-intervals-methodology-supplement} above) obtained in this way across all hypotheses.

\begin{algorithm}
\SetAlgoLined
\textbf{Inputs:} Action sequence $\ax_{1:\tx}$; Observational dataset $\D$.\\
\textbf{Output:} Twin data $\Dt(\ax_{1:\tx})$ of size $m$.\\
\For{$i = 1, \dots, m$}{
Sample $\X_0$ without replacement from $\D$\;
Initialize the Pulse trajectory with the information of $\X_0$\;
\For{$\sx = 1, \dots, \tx$}{
Administer the median doses of IV fluids and vasopressors in action bin $\ax_{\tx'}$\;
\If{$\sx \equiv 0$ \textup{(mod 3)}}{
Virtual patient in Pulse consumes nutrients and water, and urinates\;
}
Advance the twin trajectory by one hour\;
}
Add the trajectory to $\Dt(\ax_{1:\tx})$\;
}
\textbf{Return} $\Dt(\ax_{1:\tx})$
\caption{Generating Twin data $\Dt(\ax_{1:\tx})$.}
\label{algo:twin-data-generation}
\end{algorithm}

\begin{figure}[t]%
    \centering
    \includegraphics[height=8cm]{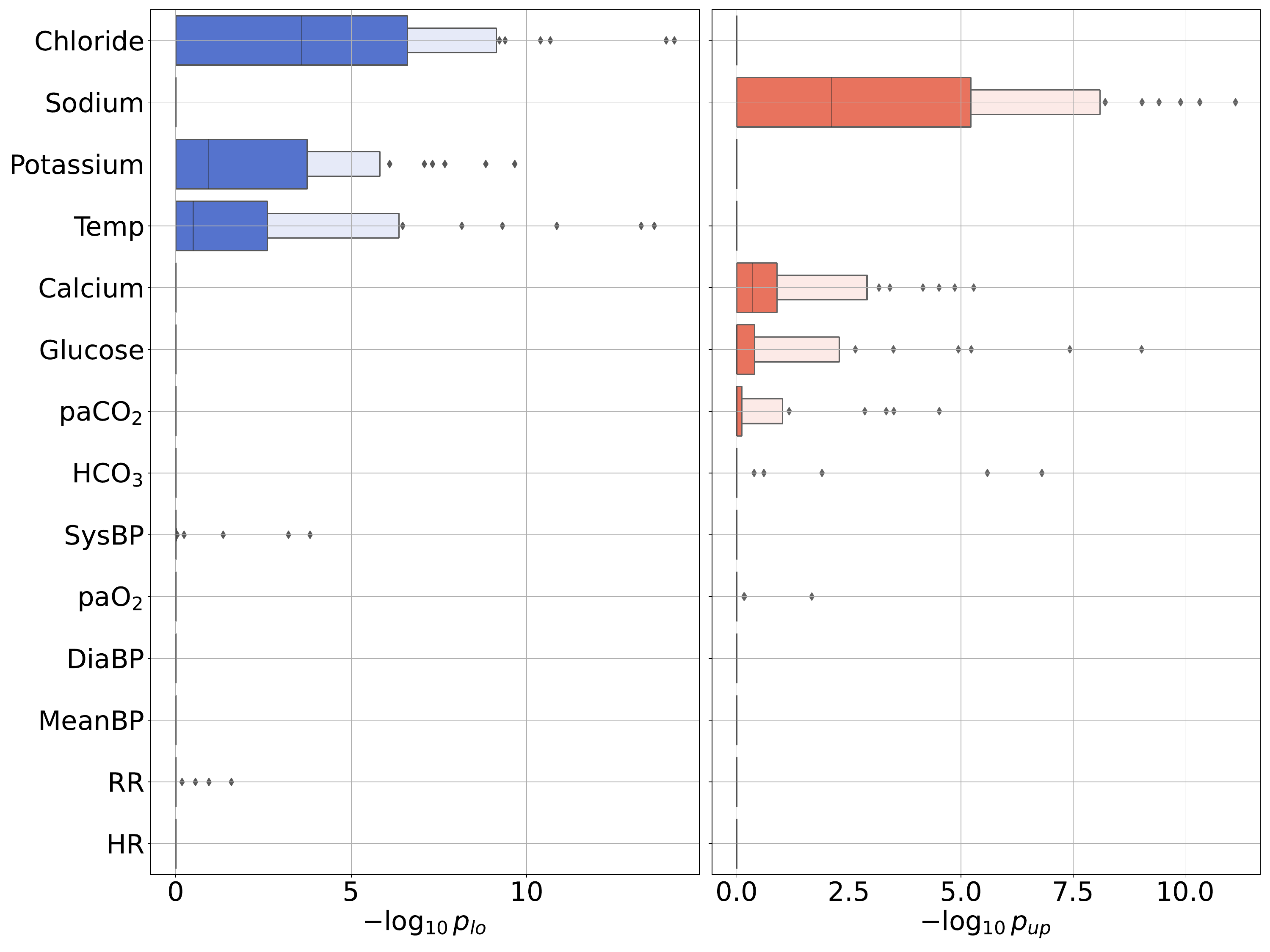}
    \caption{Boxenplots showing distributions of $-\log_{10}{p_\textup{lo}}$ and $-\log_{10}{p_\textup{up}}$ for different physiological quantities obtained via Hoeffding's inequality. Higher values indicate greater evidence in favour of rejection.}
    \label{fig:p_values_hoeff_complete}
\end{figure}

\begin{table}%
    \centering
\begin{footnotesize}
\begin{tabular}{lcccc}
\toprule
& \multicolumn{2}{c}{Ours} & \multicolumn{2}{c}{Manski} \\
                   Physiological quantity &  Rejs. &  Hyps. &  Rejs. &  Hyps. \\
\midrule
  Chloride Blood Concentration (Chloride) &            24 &            94 & 1 & 46 \\
      Sodium Blood Concentration (Sodium) &            21 &            94 & 9 & 46 \\
Potassium Blood Concentration (Potassium) &            13 &            94 & 0 & 46 \\
                  Skin Temperature (Temp) &            10 &            86 & 9 & 46 \\
    Calcium Blood Concentration (Calcium) &             5 &            88 & 0 & 46 \\
    Glucose Blood Concentration (Glucose) &             5 &            96 & 1 & 46 \\
      Arterial CO$_2$ Pressure (paCO$_2$) &             3 &            70 & 0 & 46 \\
Bicarbonate Blood Concentration (HCO$_3$) &             2 &            90 & 1 & 46 \\
       Systolic Arterial Pressure (SysBP) &             2 &           154 & 0 & 46 \\
        Arterial O$_2$ Pressure (paO$_2$) &             0 &            78 & 1 & 46 \\
                Arterial pH (Arterial\_pH) &             0 &            80 & 0 & 46 \\
      Diastolic Arterial Pressure (DiaBP) &             0 &            72 & 0 & 46 \\
          Mean Arterial Pressure (MeanBP) &             0 &            92 & 0 & 46 \\
                    Respiration Rate (RR) &             0 &           172 & 0 & 46 \\
                          Heart Rate (HR) &             0 &           162 & 0 & 46 \\
\bottomrule
\end{tabular}
\end{footnotesize}
    \caption{Total hypotheses (Hyps.) and rejections (Rejs.) per physiological quantity obtained using Hoeffding's inequality} \label{tab:hypotheses_hoeffding_full}
\end{table}

\begin{table}[t]
\centering
\begin{footnotesize}
\begin{tabular}{lcccc}
\toprule
& \multicolumn{2}{c}{Ours} & \multicolumn{2}{c}{Manski} \\
                   Physiological quantity &  Rejs. &  Hyps. &  Rejs. &  Hyps. \\
\midrule
  Chloride Blood Concentration (Chloride) &            47 &            94 & 1 & 46 \\
      Sodium Blood Concentration (Sodium) &            46 &            94 & 12 & 46 \\
Potassium Blood Concentration (Potassium) &            33 &            94 & 0 & 46 \\
                  Skin Temperature (Temp) &            43 &            86 & 13 & 46 \\
    Calcium Blood Concentration (Calcium) &            44 &            88 & 0 & 46 \\
    Glucose Blood Concentration (Glucose) &            19 &            96 & 0 & 46 \\
      Arterial CO$_2$ Pressure (paCO$_2$) &            13 &            70 & 0 & 46 \\
Bicarbonate Blood Concentration (HCO$_3$) &             8 &            90 & 0 & 46 \\
       Systolic Arterial Pressure (SysBP) &             8 &           154 & 0 & 46 \\
        Arterial O$_2$ Pressure (paO$_2$) &             4 &            78 & 1 & 46 \\
                Arterial pH (Arterial\_pH) &             0 &            80 & 0 & 46 \\
      Diastolic Arterial Pressure (DiaBP) &             0 &            72 & 0 & 46 \\
          Mean Arterial Pressure (MeanBP) &             3 &            92 & 0 & 46 \\
                    Respiration Rate (RR) &            12 &           172 & 0 & 46 \\
                          Heart Rate (HR) &             1 &           162 & 0 & 46 \\
\bottomrule
\end{tabular}
\end{footnotesize}
\caption{Total hypotheses (Hyps.) and rejections (Rejs.) per physiological quantity obtained using the reverse percentile bootstrap}  \label{tab:hypotheses_rev_percentile}
\end{table}

\subsection{Bootstrapping Details} \label{sec:boostrapping-details-supplement}

In addition to Hoeffding's inequality, we also used reverse percentile bootstrap method (see e.g.\ \citealt{hesterberg2015what}) to obtain our confidence intervals on $\Qlo$ and $\Qup$ as described in Section \ref{sec:confidence-intervals-methodology-supplement}.
We used 100 bootstrap samples for each confidence interval.
To avoid bootstrapping on small numbers of data points, we did not reject any hypothesis where either the number of observational trajectories $n$ or twin trajectories $\widehat{n}$ was less than 100, and returned a $p$-value of 1 in each such case.

Table \ref{tab:hypotheses_rev_percentile} shows the number of rejected hypotheses for each physiological quantity using this approach.
We observed a similar trend as in our results obtained using Hoeffding's inequality (Table \ref{tab:hypotheses_hoeffding_full}).
For example, we obtained high number of rejections for Sodium, Chloride and Potassium blood concentrations but few rejections for Arterial Pressure and Heart Rate.
Overall, bootstrapping increased the number of rejected hypotheses by a factor of roughly 3.3 compared with Hoeffding's inequality (281 vs.\ 85 rejections in total).
Like we described for Hoeffding's inequality in the main text, we also ran this analysis using the unconditional bounds of \citet{manski}, and again obtained substantially fewer rejections compared with our approach based on Theorem \ref{thm:causal-bounds}.

\begin{figure}[t]
    \centering
        \begin{subfigure}[b]{0.29\textwidth}
    \includegraphics[height=3.68cm]{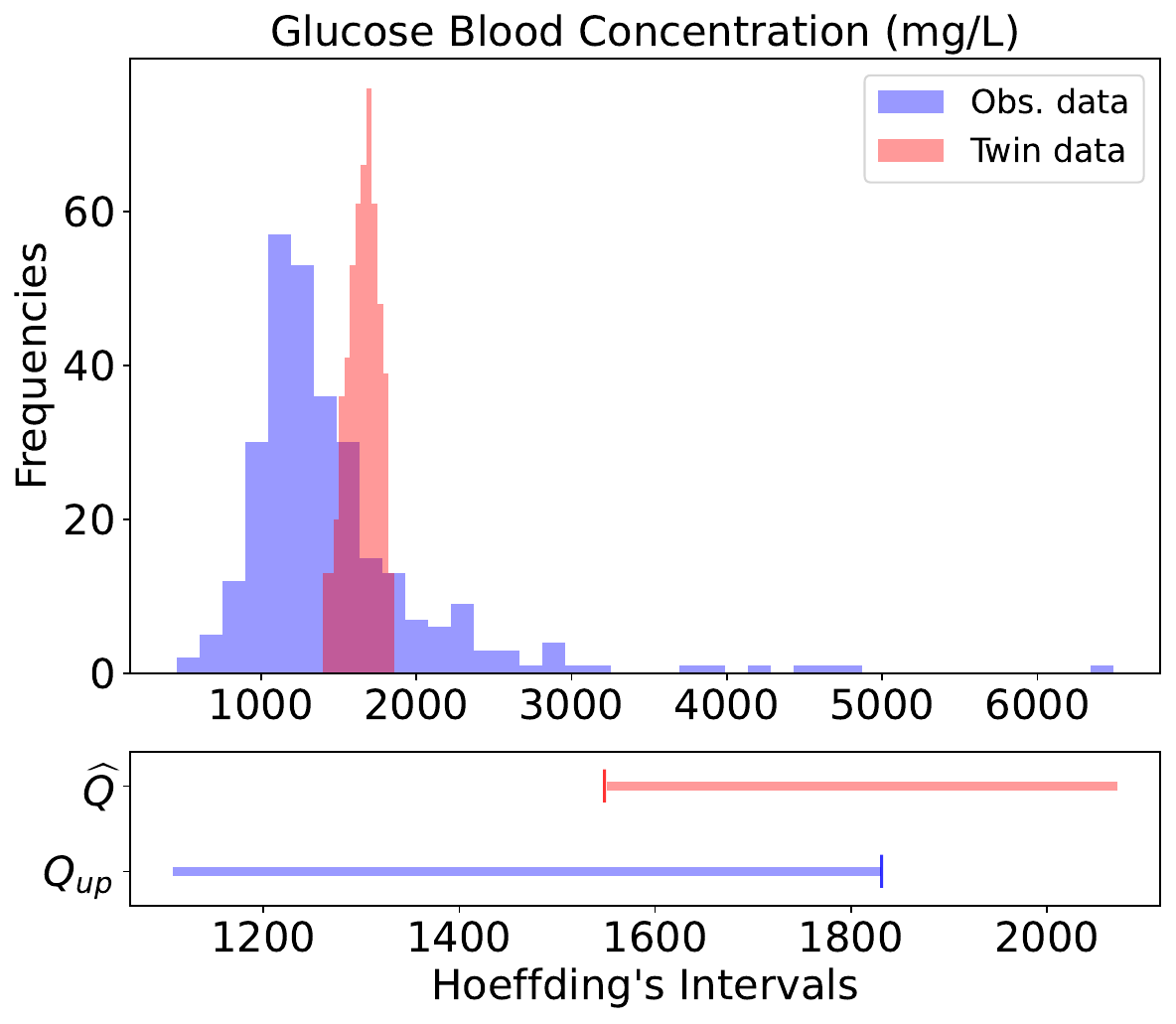}
    \subcaption{Not rejected}
    \label{fig:glucosea-supp}
    \end{subfigure}\hspace{1cm}%
    \begin{subfigure}[b]{0.29\textwidth}
    \includegraphics[height=3.68cm]{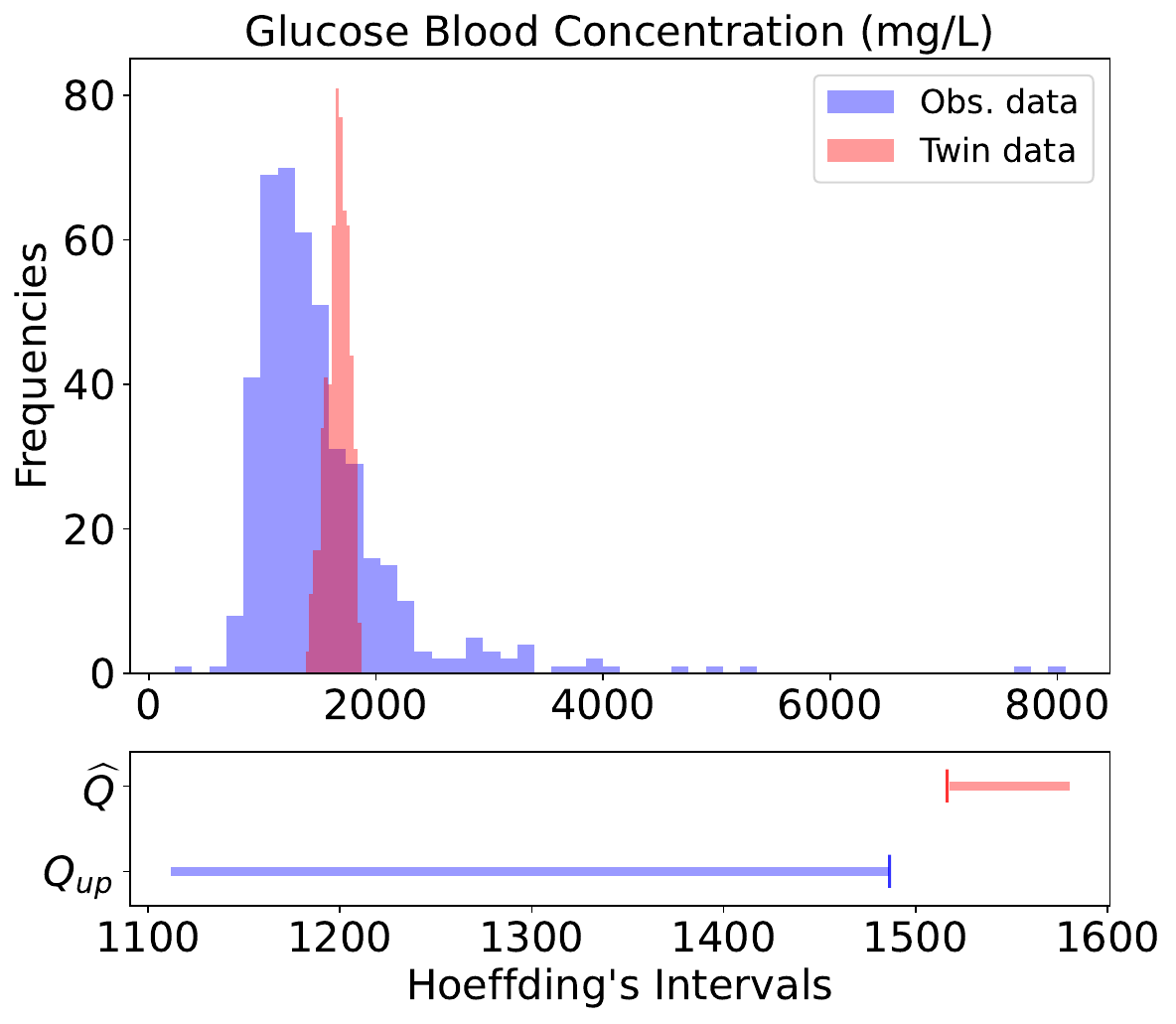}
    \subcaption{Rejected}
    \label{fig:glucoseb-supp}
    \end{subfigure}\\      
    \begin{subfigure}[b]{0.29\textwidth}
    \includegraphics[height=3.68cm]{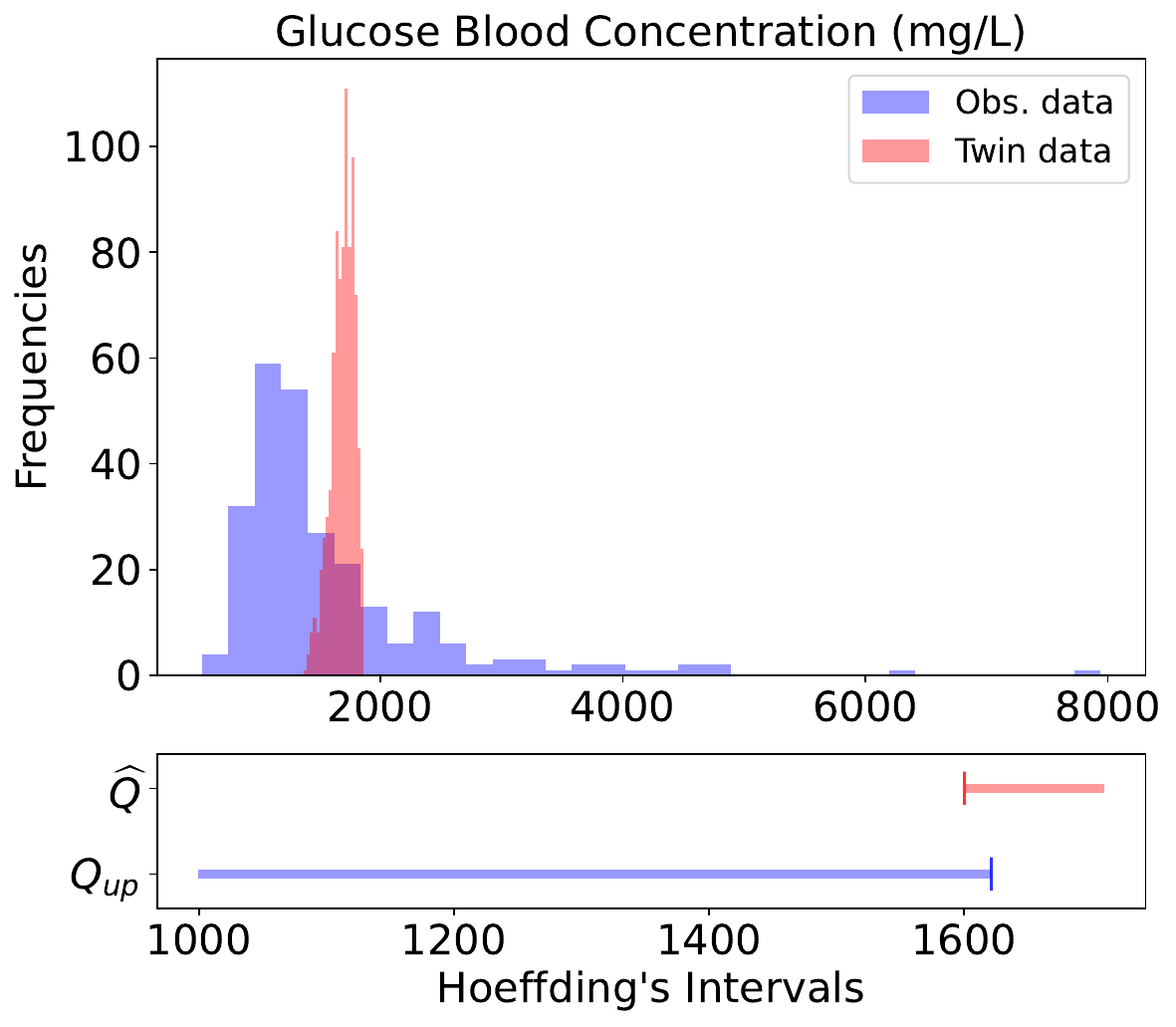}
    \subcaption{Not rejected}
    \label{fig:chloridea}
    \end{subfigure}\hspace{1cm}%
    \begin{subfigure}[b]{0.29\textwidth}
    \includegraphics[height=3.68cm]{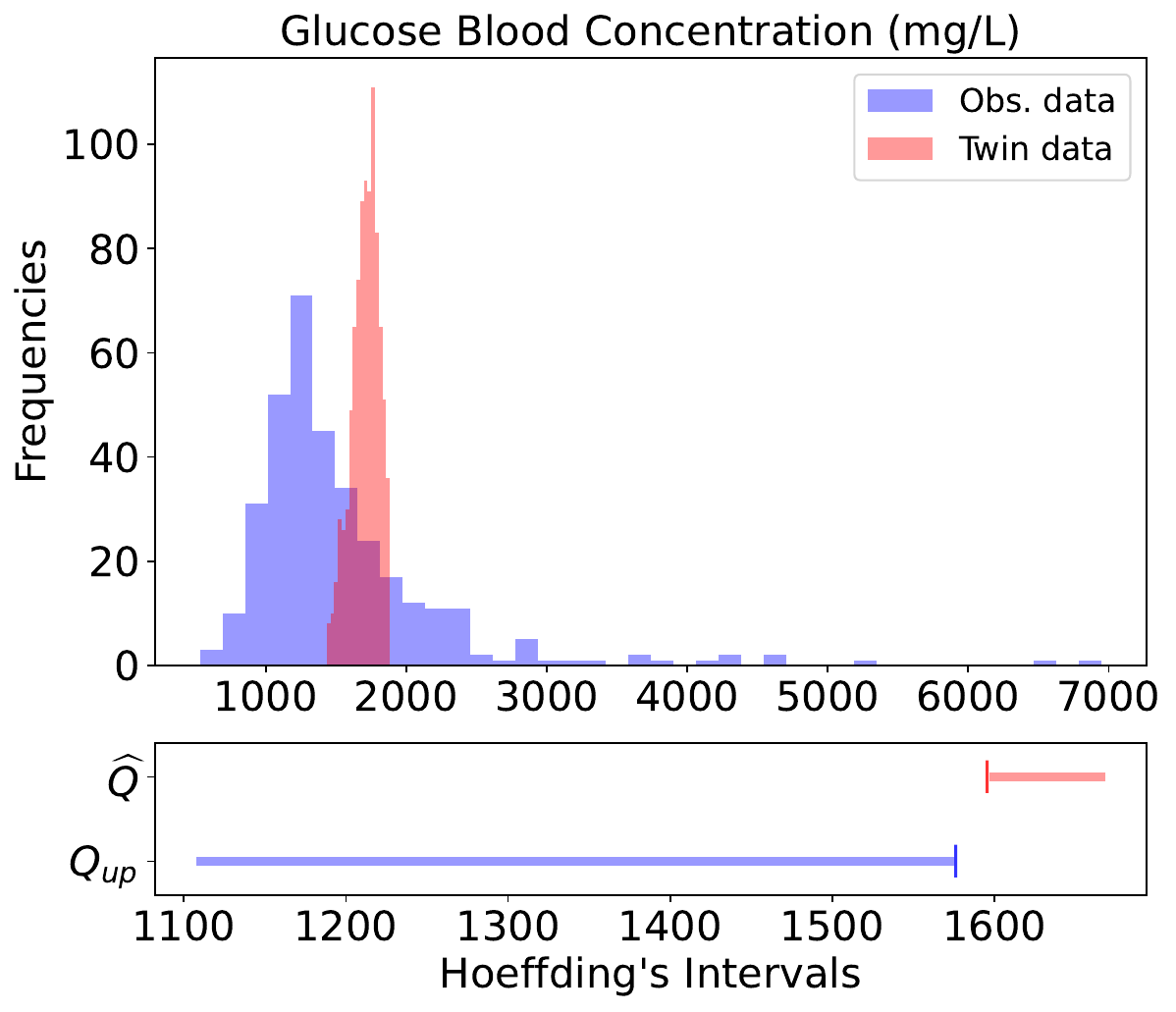}
    \subcaption{Rejected}
    \label{fig:chlorideb}
    \end{subfigure}\\
    \begin{subfigure}[b]{0.29\textwidth}
    \includegraphics[height=3.68cm]{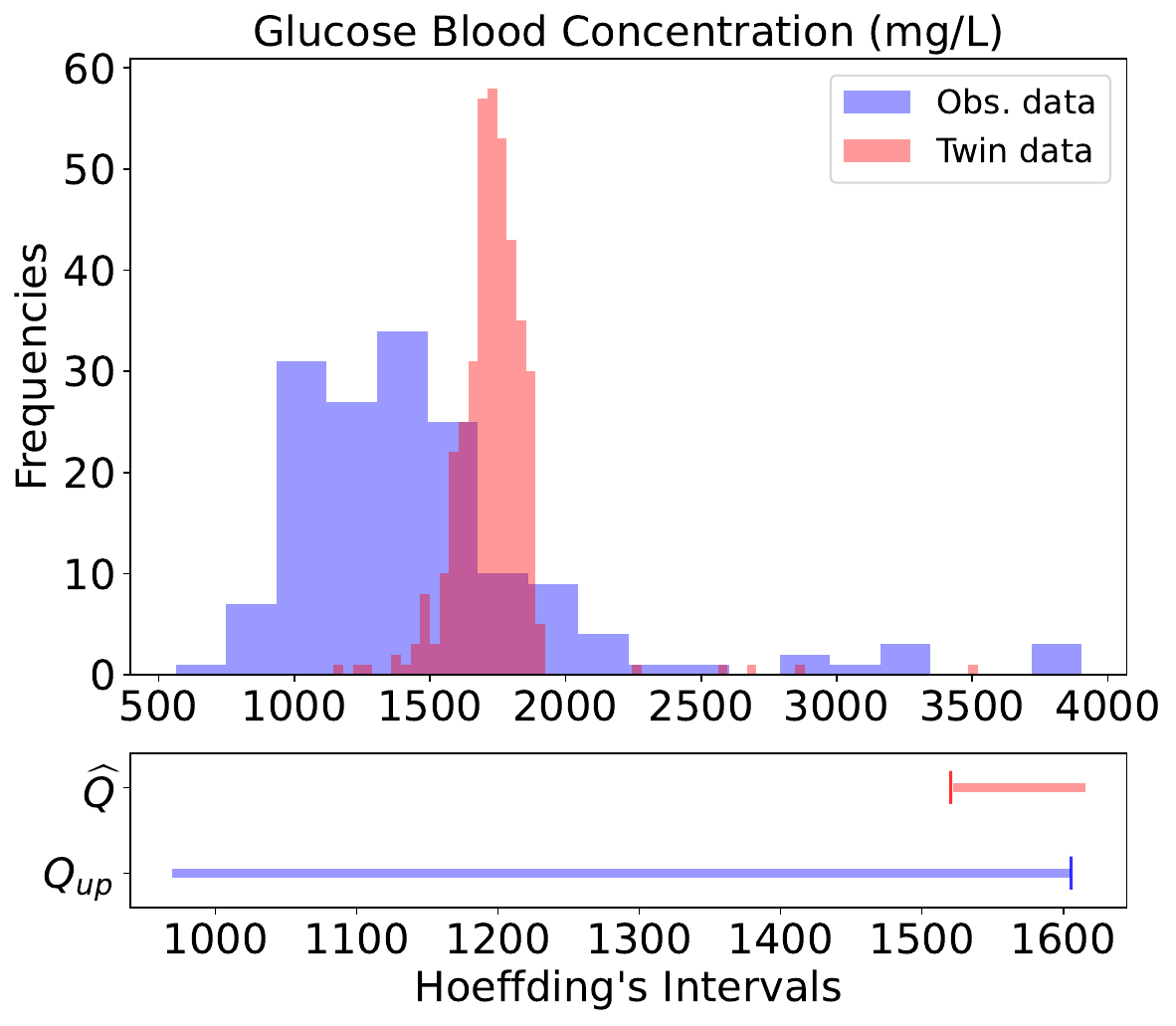}
    \subcaption{Not rejected}
    \label{fig:potassiuma}
    \end{subfigure}\hspace{1cm}%
    \begin{subfigure}[b]{0.29\textwidth}
    \includegraphics[height=3.68cm]{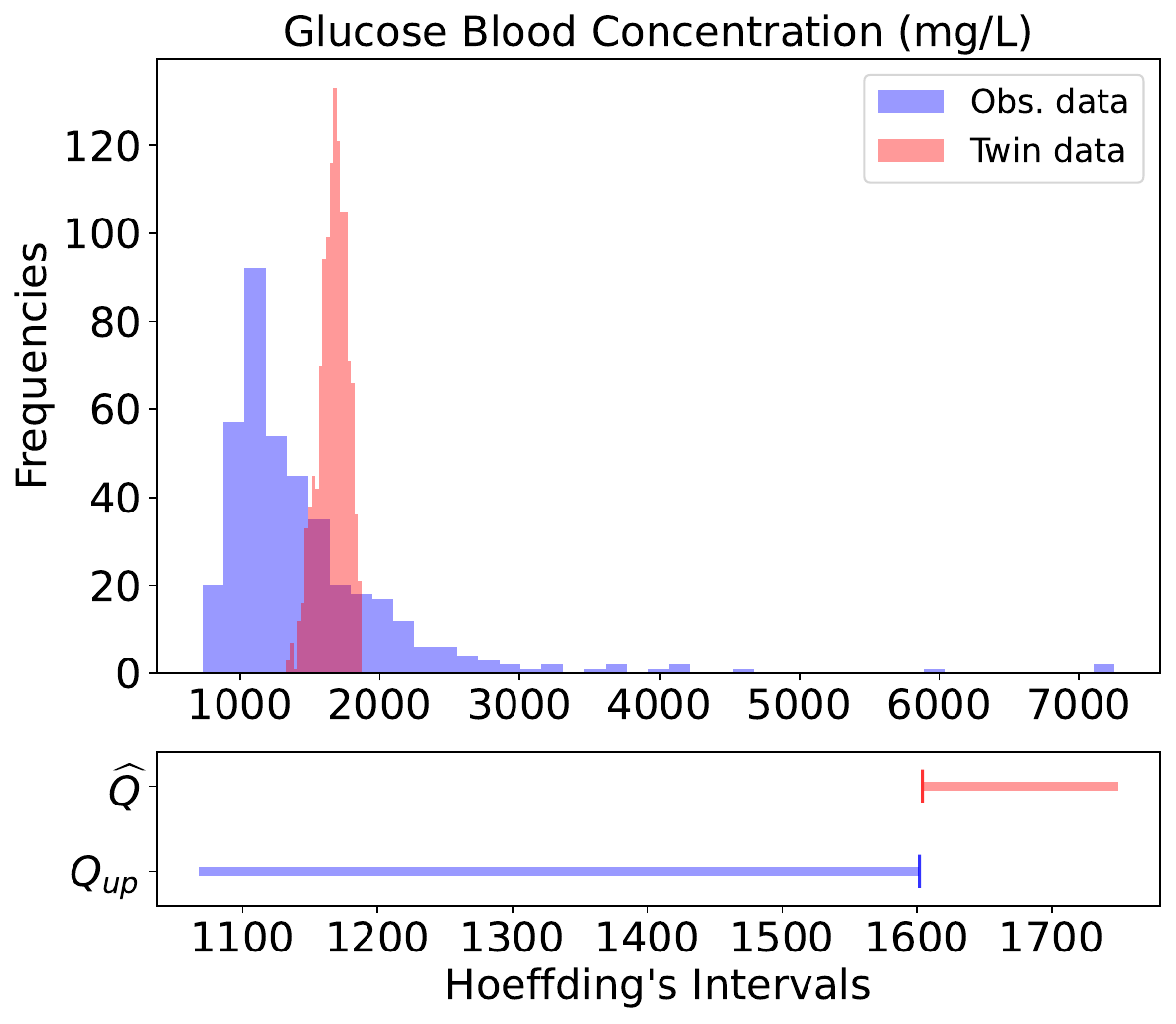}
    \subcaption{Rejected}
    \label{fig:potassiumb}
    \end{subfigure}\\
    \begin{subfigure}[b]{0.29\textwidth}
    \includegraphics[height=3.68cm]{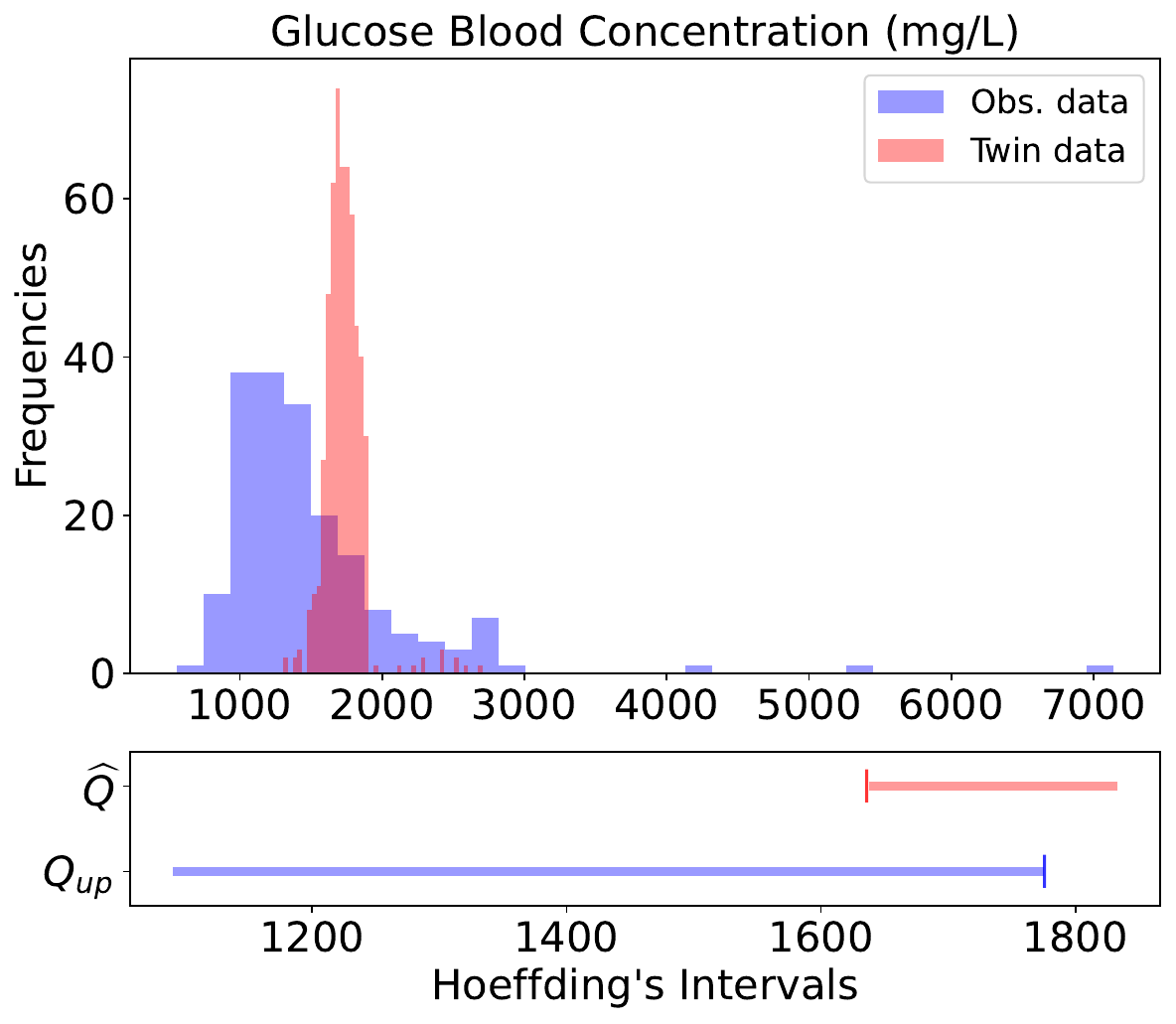}
    \subcaption{Not rejected}
    \label{fig:paco2a}
    \end{subfigure}\hspace{1cm}%
    \begin{subfigure}[b]{0.29\textwidth}
    \includegraphics[height=3.68cm]{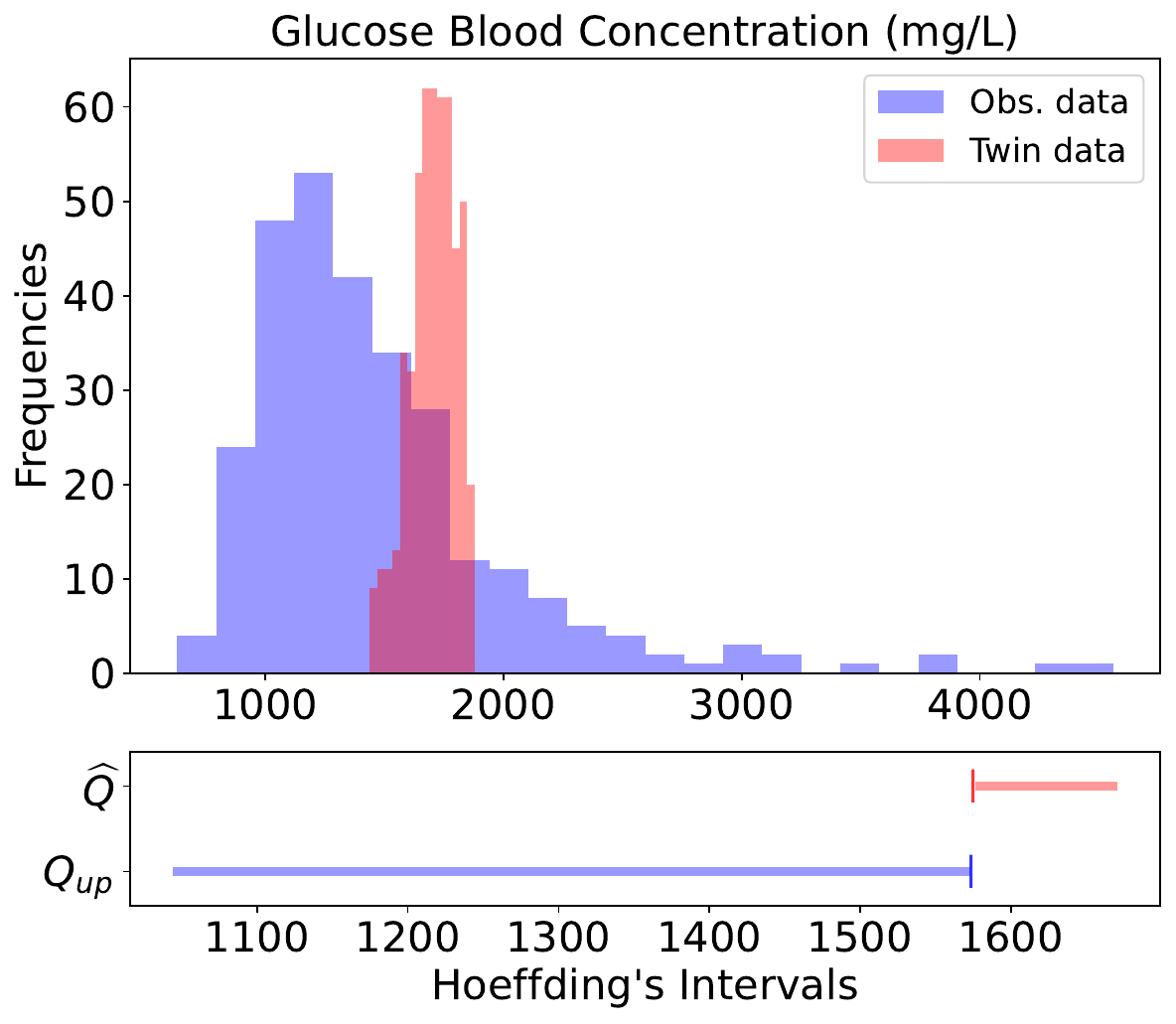}
    \subcaption{Rejected}
    \label{fig:paco2b}
    \end{subfigure}

    \caption{Raw observational data values conditional on $\A_{1:\tx}=\ax_{1:\tx}$ and $\X_{0:\tx}(\A_{1:\tx})\in \B_{0:\tx}$, and from the output of the twin conditional on $\Xt_{0:\tx}(\ax_{1:\tx})\in \B_{0:\tx}$.
    Each row shows two distinct choices of $(\B_{0:\tx}, \ax_{1:\tx})$.
    Below each figure are shown 95\% Hoeffding confidence intervals for $\Qt$ and $\Qup$.
    Unlike Figure \ref{fig:histograms} from the main text, the horizontal axes of the histograms are not truncated, and the first row is in particular an untruncated version of Figure \ref{fig:histograms} from the main text.
    Note however that the scales of the horizontal axes of the confidence intervals differ from those of the histograms, since it is visually more difficult to determine whether or not the confidence intervals overlap when fully zoomed out.} \label{fig:histograms-supplement}
\end{figure}

\subsection{Tightness of Bounds and Number of Data Points per Hypothesis}
    In this section, we show empirically how both the tightness of the bounds $[\Qlo, \Qup]$ and the number of data points per hypothesis relate to the number of falsifications obtained in our case study.
    Recall that the tightness of $[\Qlo, \Qup]$ is determined by the value of $\Prob(\A_{1:\tx} = \ax_{1:\tx} \mid \X_{0:\N}(\A_{1:\N}) \in \B_{0:\N})$, since we have
    \begin{equation} \label{eq:N-propensity-tightness-relation}
        \frac{\Qup - \Qlo}{\yup - \ylo} = 1 - \Prob(\A_{1:\tx} = \ax_{1:\tx} \mid \X_{0:\N}(\A_{1:\N}) \in \B_{0:\N}).
    \end{equation}
    Here the left-hand side is a number in $[0, 1]$ that quantifies the tightness of the bounds $[\Qlo, \Qup]$ relative to the trivial worst-case bounds $[\ylo, \yup]$, with smaller values meaning tighter bounds. The equation above shows that the higher the value of $\Prob(\A_{1:\tx} = \ax_{1:\tx} \mid \X_{0:\N}(\A_{1:\N}) \in \B_{0:\N})$, the tighter the bounds are.
    
\begin{figure}[t]
        \centering
        \includegraphics[height=6cm]{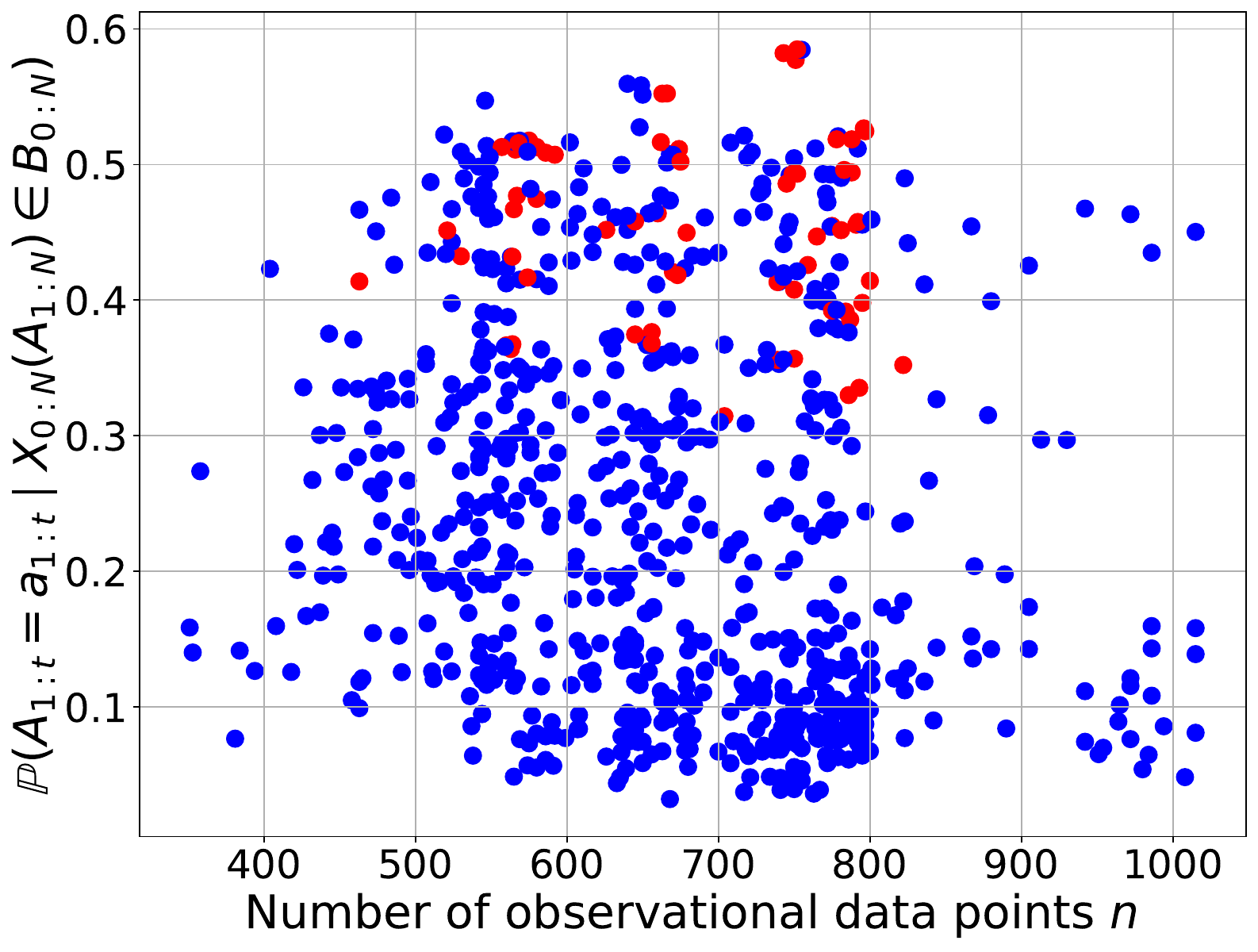}
        \caption{Sample mean estimate of $\Prob(\A_{1:\tx} = \ax_{1:\tx} \mid \X_{0:\N}(\A_{1:\N}) \in \B_{0:\N})$ for each pair of hypotheses $(\Hlo, \Hup)$ corresponding to the same set of parameters $(\tx, \fx, \ax_{1:\tx}, \B_{0:\tx})$ that we tested, along with the corresponding number of observational data points used to test each hypothesis.
        Red points indicate that either $\Hlo$ or $\Hup$ were rejected, while blue points indicate that both $\Hlo$ and $\Hup$ were not rejected.}
        \label{fig:scatter-plot}
    \end{figure}

    Figure \ref{fig:scatter-plot} shows the bounds are often informative in practice, with $\Prob(\A_{1:\tx} = \ax_{1:\tx} \mid \X_{0:\N}(\A_{1:\N}) \in \B_{0:\N})$ being reasonably large (and hence the bounds tight, by \eqref{eq:N-propensity-tightness-relation} above) for a significant number of hypotheses we consider.
    However, rejections still occur even when the bounds are reasonably loose (e.g.\ $\Prob(\A_{1:\tx} = \ax_{1:\tx} \mid \X_{0:\N}(\A_{1:\N}) \in \B_{0:\N}) \approx 0.3$), which shows our method can still yield useful information even in this case.
    We moreover observe rejections across a range of different numbers of observational data points used to test each hypothesis, which shows that our method is not strongly dependent on the size of the dataset obtained. 

    \subsection{Sensitivity to $\ylo$ and $\yup$} \label{sec:sensitity-analysis-appendix}

\begin{figure}[t]
    \centering
\includegraphics[width=0.6\textwidth]{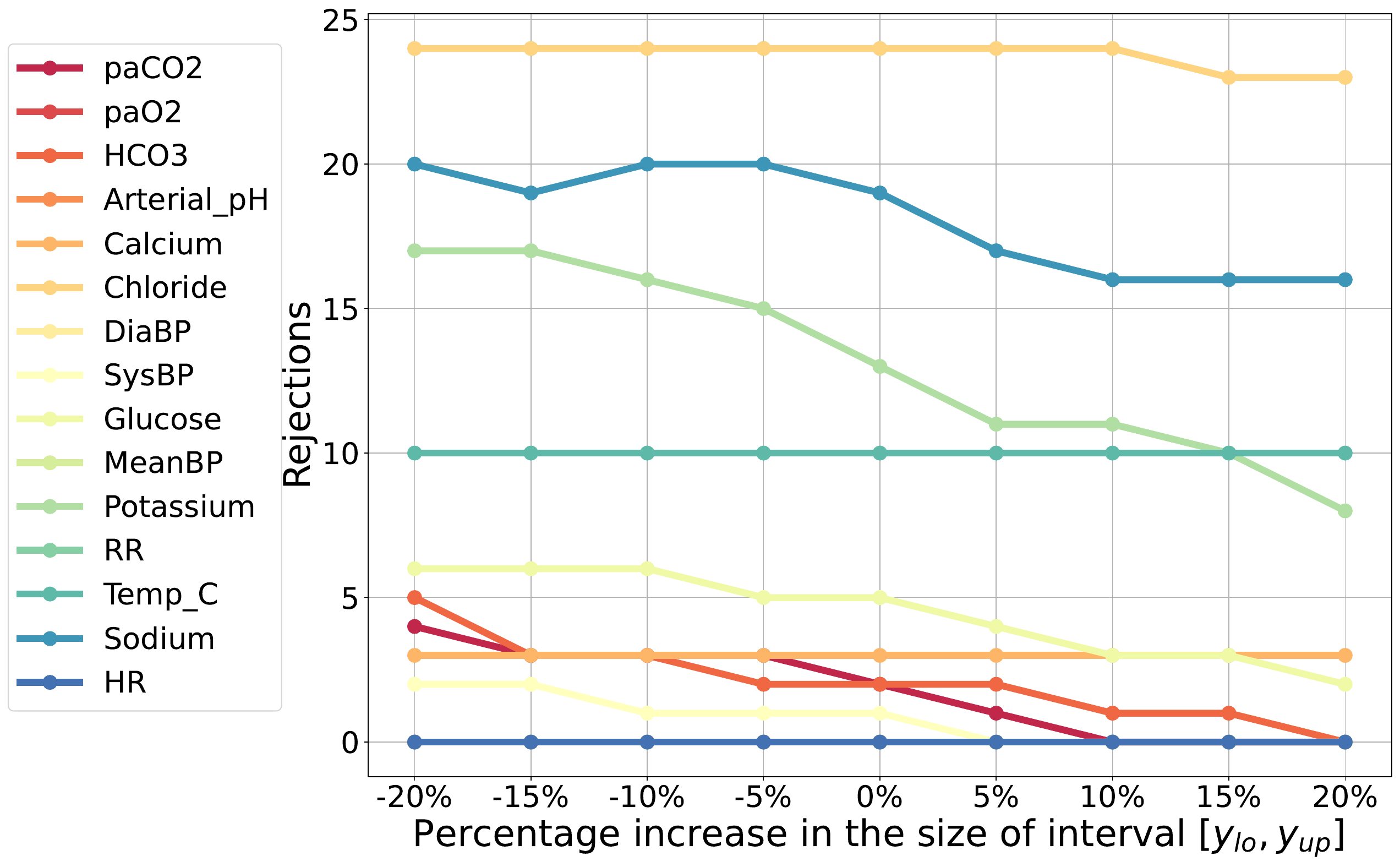}
    \caption{Rejections obtained as the width of the $[\ylo, \yup]$ interval changes. Here, the interval is increased (or decreased) symmetrically on each side.}
    \label{fig:sensitivity-plot-rejections}
\end{figure}

    We investigated the sensitivity of our methodology with respect to our choices of the values $\ylo$ and $\yup$.
    Specifically, we repeated our procedure with the intervals $[\ylo, \yup]$ replaced with $[\ylo\, (1- \Delta/2), \yup\,(1 + \Delta/2)]$ for a range of different values of $\Delta \in \R$.
    Figure \ref{fig:sensitivity-plot-rejections} plots the number of rejections for different values of $\Delta$.
    We observe that for significantly larger $[\ylo, \yup]$ intervals, we do obtain fewer rejections, although this is to be expected since the widths of the bounds $[\Qlo, \Qup]$ and our confidence intervals $\qlo{\alpha}$ and $\qup{\alpha}$ obtained using Hoeffding's inequality (see Proposition \ref{prop:hoeffding-confidence-bounds-supp}) both grow increasingly large as the width of $[\ylo, \yup]$ grows.
    However, we observe that the number of rejections per outcome is stable for a moderate range of widths of $[\ylo, \yup]$, which indicates that our method is reasonably robust to the choice of $\ylo, \yup$ parameters.

\clearpage %

\vskip 0.2in
\bibliography{main}

@article{pulse,
	Author = {Bray, Aaron and Webb, Jeffrey B. and Enquobahrie, Andinet and Vicory, Jared and Heneghan, Jerry and Hubal, Robert and TerMaath, Stephanie and Asare, Philip and Clipp, Rachel B.},
	Da = {2019/05/01},
	Doi = {10.1007/s42399-019-00053-w},
	Id = {Bray2019},
	Isbn = {2523-8973},
	Journal = {SN Comprehensive Clinical Medicine},
	Number = {5},
	Pages = {362--377},
	Title = {{Pulse Physiology Engine}: An open-source software platform for computational modeling of human medical simulation},
	Ty = {JOUR},
	Url = {https://doi.org/10.1007/s42399-019-00053-w},
	Volume = {1},
	Year = {2019}
}

@article{manski,
 ISSN = {00028282},
 URL = {http://www.jstor.org/stable/2006592},
 author = {Charles F. Manski},
 journal = {The American Economic Review},
 number = {2},
 pages = {319--323},
 publisher = {American Economic Association},
 title = {Nonparametric Bounds on Treatment Effects},
 urldate = {2022-06-24},
 volume = {80},
 year = {1990}
}

@inproceedings{bareinboim,
     author = {Zhang, Junzhe and Bareinboim, Elias},
     booktitle = {Advances in Neural Information Processing Systems},
     title = {Near-Optimal Reinforcement Learning in Dynamic Treatment Regimes},
     url = {https://proceedings.neurips.cc/paper/2019/file/8252831b9fce7a49421e622c14ce0f65-Paper.pdf},
     volume = {32},
     year = {2019}
}

@ARTICLE{sepsis-modelling,
    AUTHOR={McDaniel, Matthew and Keller, Jonathan M. and White, Steven and Baird, Austin},   
    TITLE={A whole-body mathematical model of sepsis progression and treatment designed in the {BioGears Physiology Engine}},      
    JOURNAL={Frontiers in Physiology},      
    VOLUME={10},      
    PAGES={1321},     
    YEAR={2019},      
    URL={https://www.frontiersin.org/article/10.3389/fphys.2019.01321},       
    DOI={10.3389/fphys.2019.01321},      
    ISSN={1664-042X},   
}

@article{mimic,
	Author = {Johnson, Alistair E. W. and Pollard, Tom J. and Shen, Lu and Lehman, Li-wei H. and Feng, Mengling and Ghassemi, Mohammad and Moody, Benjamin and Szolovits, Peter and Celi, Leo Anthony and Mark, Roger G.},
	Da = {2016/05/24},
	Doi = {10.1038/sdata.2016.35},
	Id = {Johnson2016},
	Isbn = {2052-4463},
	Journal = {Scientific Data},
	Number = {1},
	Pages = {160035},
	Title = {{MIMIC-III}, a freely accessible critical care database},
	Ty = {JOUR},
	Url = {https://doi.org/10.1038/sdata.2016.35},
	Volume = {3},
	Year = {2016}
 }

@article{sepsis-criteria,
    author = {Singer, Mervyn and Deutschman, Clifford S. and Seymour, Christopher W. and Shankar-Hari, Manu and Annane, Djillali and Bauer, Michael and Bellomo, Rinaldo and Bernard, Gordon R. and Chiche, Jean-Daniel and Coopersmith, Craig M. and Hotchkiss, Richard S. and Levy, Mitchell M. and Marshall, John C. and Martin, Greg S. and Opal, Steven M. and Rubenfeld, Gordon D. and van der Poll, Tom and Vincent, Jean-Louis and Angus, Derek C.},
    title = "{The Third International Consensus Definitions for Sepsis and Septic Shock (Sepsis-3)}",
    journal = {JAMA},
    volume = {315},
    number = {8},
    pages = {801--810},
    year = {2016},
    issn = {0098-7484},
    doi = {10.1001/jama.2016.0287},
    url = {https://doi.org/10.1001/jama.2016.0287},
    eprint = {https://jamanetwork.com/journals/jama/articlepdf/2492881/jsc160002.pdf},
}

@article{seymour2016assessment,
    author = {Seymour, Christopher W. and Liu, Vincent X. and Iwashyna, Theodore J. and Brunkhorst, Frank M. and Rea, Thomas D. and Scherag, André and Rubenfeld, Gordon and Kahn, Jeremy M. and Shankar-Hari, Manu and Singer, Mervyn and Deutschman, Clifford S. and Escobar, Gabriel J. and Angus, Derek C.},
    title = {Assessment of clinical criteria for sepsis: For the {Third International Consensus Definitions for Sepsis and Septic Shock (Sepsis-3)}},
    journal = {JAMA},
    volume = {315},
    number = {8},
    pages = {762--774},
    year = {2016},
    issn = {0098-7484},
    doi = {10.1001/jama.2016.0288},
    url = {https://doi.org/10.1001/jama.2016.0288},
    eprint = {https://jamanetwork.com/journals/jama/articlepdf/2492875/joi160006.pdf},
}

@article{DT-patient,
    author = {Lal, Amos and Li, Guangxi and Cubro, Edin and Chalmers, Sarah and Li, Heyi and Herasevich, Vitaly and Dong, Yue and Pickering, Brian and Kilickaya, Oguz and Gajic, Ognjen},
    year = {2021},
    pages = {611},
    title = {Development and Verification of a Digital Twin Patient Model to Predict Treatment Response in Sepsis},
    volume = {49},
    number = {1},
    journal = {Critical Care Medicine},
    doi = {10.1097/01.ccm.0000730744.82258.38}
}

@INPROCEEDINGS{biogears,
  author={McDaniel, Matthew and Baird, Austin},
  booktitle={2019 41st Annual International Conference of the IEEE Engineering in Medicine and Biology Society (EMBC)}, 
  title={A full-body model of burn pathophysiology and treatment using the {BioGears Engine}}, 
  year={2019},
  volume={},
  number={},
  pages={261--264},
  doi={10.1109/EMBC.2019.8857686}
}

@article{rubin1974estimating,
    author = {Donald B. Rubin},
    title = {Estimating causal effects of treatments in randomized and nonrandomized studies},
    journal = {Journal of Educational Psychology},
    volume = {66},
    pages = {688--701},
    year = {1974},
    doi = {https://doi.org/10.1037/h0037350},
}

@article{ai-clinician,
	Author = {Komorowski, Matthieu and Celi, Leo Anthony and Badawi, Omar and Gordon, Anthony C. and Faisal, A. Aldo},
	Da = {2018/11/01},
	Doi = {10.1038/s41591-018-0213-5},
	Id = {Komorowski2018},
	Isbn = {1546-170X},
	Journal = {Nature Medicine},
	Number = {11},
	Pages = {1716--1720},
	Title = {The Artificial Intelligence Clinician learns optimal treatment strategies for sepsis in intensive care},
	Ty = {JOUR},
	Url = {https://doi.org/10.1038/s41591-018-0213-5},
	Volume = {24},
	Year = {2018}
}

@book{tsiatis2019dynamic,
  title={Dynamic Treatment Regimes: Statistical Methods for Precision Medicine},
  author={Tsiatis, Anastasios A. and Davidian, Marie and Holloway, Shannon T. and Laber, Eric B.},
  year={2019},
  publisher={Chapman and Hall/CRC}
}

@article{murphy2003optimal,
  title={Optimal dynamic treatment regimes},
  author={Murphy, Susan A.},
  journal={Journal of the Royal Statistical Society: Series B (Statistical Methodology)},
  volume={65},
  number={2},
  pages={331--355},
  year={2003},
  publisher={Wiley Online Library}
}

@article{murphy2005experimental,
author = {Murphy, Susan A.},
title = {An experimental design for the development of adaptive treatment strategies},
journal = {Statistics in Medicine},
volume = {24},
number = {10},
pages = {1455--1481},
doi = {https://doi.org/10.1002/sim.2022},
url = {https://onlinelibrary.wiley.com/doi/abs/10.1002/sim.2022},
eprint = {https://onlinelibrary.wiley.com/doi/pdf/10.1002/sim.2022},
year = {2005}
}

@article{lavori2004dynamic,
  title={Dynamic treatment regimes: Practical design considerations},
  author={Lavori, Philip W. and Dawson, Ree},
  journal={Clinical Trials},
  volume={1},
  number={1},
  pages={9--20},
  year={2004},
  publisher={Sage Publications Sage CA: Thousand Oaks, CA}
}

@article{robins1986new,
  title={A new approach to causal inference in mortality studies with a sustained exposure period—application to control of the healthy worker survivor effect},
  author={Robins, James},
  journal={Mathematical Modelling},
  volume={7},
  number={9--12},
  pages={1393--1512},
  year={1986},
  publisher={Elsevier}
}

@book{popper2005logic,
  title={The Logic of Scientific Discovery},
  author={Popper, Karl},
  year={2005},
  publisher={Routledge}
}

@book{davison1997bootstrap, 
place={Cambridge}, 
series={Cambridge Series in Statistical and Probabilistic Mathematics}, 
title={Bootstrap Methods and Their Application}, 
DOI={10.1017/CBO9780511802843}, 
publisher={Cambridge University Press}, 
author={Davison, Anthony C. and Hinkley, David V.}, 
year={1997}, 
collection={Cambridge Series in Statistical and Probabilistic Mathematics}
}

@book{tibshirani1993introduction,
  title={An Introduction to the Bootstrap},
  author={Efron, Bradley and Tibshirani, Robert J.},
  series={Monographs on Statistics and Applied Probability},
  number={57},
  publisher={Chapman \& Hall/CRC},
  address={New York},
  year={1993}
}

@article{hesterberg2015what,
author = {Tim C. Hesterberg},
title = {What Teachers Should Know About the Bootstrap: Resampling in the Undergraduate Statistics Curriculum},
journal = {The American Statistician},
volume = {69},
number = {4},
pages = {371--386},
year  = {2015},
publisher = {Taylor & Francis},
doi = {10.1080/00031305.2015.1089789},
    note ={PMID: 27019512},
URL = {https://doi.org/10.1080/00031305.2015.1089789},
eprint = {https://doi.org/10.1080/00031305.2015.1089789}
}

@article{efron1979bootstrap,
author = {Bradley Efron},
title = {Bootstrap Methods: Another Look at the Jackknife},
volume = {7},
journal = {The Annals of Statistics},
number = {1},
publisher = {Institute of Mathematical Statistics},
pages = {1--26},
year = {1979},
doi = {10.1214/aos/1176344552},
URL = {https://doi.org/10.1214/aos/1176344552}
}

@article{hall1988theoretical,
author = {Peter Hall},
title = {Theoretical Comparison of Bootstrap Confidence Intervals},
volume = {16},
journal = {The Annals of Statistics},
number = {3},
publisher = {Institute of Mathematical Statistics},
pages = {927--953},
year = {1988},
doi = {10.1214/aos/1176350933},
URL = {https://doi.org/10.1214/aos/1176350933}
}

@article{benjamini2001control,
author = {Yoav Benjamini and Daniel Yekutieli},
title = {The control of the false discovery rate in multiple testing under dependency},
volume = {29},
journal = {The Annals of Statistics},
number = {4},
publisher = {Institute of Mathematical Statistics},
pages = {1165--1188},
year = {2001},
doi = {10.1214/aos/1013699998},
URL = {https://doi.org/10.1214/aos/1013699998}
}

@article{holm1979simple,
 ISSN = {03036898, 14679469},
 URL = {http://www.jstor.org/stable/4615733},
 author = {Sture Holm},
 journal = {Scandinavian Journal of Statistics},
 number = {2},
 pages = {65--70},
 publisher = {[Board of the Foundation of the Scandinavian Journal of Statistics, Wiley]},
 title = {A Simple Sequentially Rejective Multiple Test Procedure},
 urldate = {2022-12-13},
 volume = {6},
 year = {1979}
}

@article{rubin2005causal,
author = {Donald B Rubin},
title = {Causal Inference Using Potential Outcomes},
journal = {Journal of the American Statistical Association},
volume = {100},
number = {469},
pages = {322--331},
year  = {2005},
publisher = {Taylor \& Francis},
doi = {10.1198/016214504000001880},
URL = {https://doi.org/10.1198/016214504000001880},
eprint = {https://doi.org/10.1198/016214504000001880}
}

@article{holland1986statistics,
 ISSN = {01621459},
 URL = {http://www.jstor.org/stable/2289064},
 author = {Paul W. Holland},
 journal = {Journal of the American Statistical Association},
 number = {396},
 pages = {945--960},
 publisher = {[American Statistical Association, Taylor \& Francis, Ltd.]},
 title = {Statistics and Causal Inference},
 urldate = {2022-12-13},
 volume = {81},
 year = {1986}
}

@article{rosenbaum1983central,
 ISSN = {00063444},
 URL = {http://www.jstor.org/stable/2335942},
 author = {Paul R. Rosenbaum and Donald B. Rubin},
 journal = {Biometrika},
 number = {1},
 pages = {41--55},
 publisher = {[Oxford University Press, Biometrika Trust]},
 title = {The Central Role of the Propensity Score in Observational Studies for Causal Effects},
 urldate = {2022-12-13},
 volume = {70},
 year = {1983}
}

@book{pearl2009causality, place={Cambridge}, edition={Second}, title={Causality}, DOI={10.1017/CBO9780511803161}, publisher={Cambridge University Press}, author={Pearl, Judea}, year={2009}
}

@book{manski2009identification,
  title={Identification for Prediction and Decision},
  author={Manski, Charles F.},
  year={2009},
  publisher={Harvard University Press}
}

@book{manski2003partial,
  title={Partial Identification of Probability Distributions},
  author={Manski, Charles F.},
  year={2003},
  publisher={Springer}
}

@article{manski1989anatomy,
 ISSN = {0022166X},
 URL = {http://www.jstor.org/stable/145818},
 author = {Charles F. Manski},
 journal = {The Journal of Human Resources},
 number = {3},
 pages = {343--360},
 publisher = {[University of Wisconsin Press, Board of Regents of the University of Wisconsin System]},
 title = {Anatomy of the Selection Problem},
 urldate = {2022-12-13},
 volume = {24},
 year = {1989}
}

@article{balke1997bounds,
author = {Alexander   Balke  and  Judea   Pearl},
title = {Bounds on Treatment Effects from Studies with Imperfect Compliance},
journal = {Journal of the American Statistical Association},
volume = {92},
number = {439},
pages = {1171--1176},
year  = {1997},
publisher = {Taylor \& Francis},
doi = {10.1080/01621459.1997.10474074},
URL = {https://doi.org/10.1080/01621459.1997.10474074
        },
eprint = {https://doi.org/10.1080/01621459.1997.10474074
}
}

@article{manski2000monotone,
 ISSN = {00129682, 14680262},
 URL = {http://www.jstor.org/stable/2999533},
 author = {Charles F. Manski and John V. Pepper},
 journal = {Econometrica},
 number = {4},
 pages = {997--1010},
 publisher = {[Wiley, Econometric Society]},
 title = {Monotone Instrumental Variables: With an Application to the Returns to Schooling},
 urldate = {2022-11-24},
 volume = {68},
 year = {2000}
}

@article{manski1997monotone,
 ISSN = {00129682, 14680262},
 URL = {http://www.jstor.org/stable/2171738},
 author = {Charles F. Manski},
 journal = {Econometrica},
 number = {6},
 pages = {1311--1334},
 publisher = {[Wiley, Econometric Society]},
 title = {Monotone Treatment Response},
 urldate = {2022-11-24},
 volume = {65},
 year = {1997}
}

@article{sacks2020construction,
  title={Construction with digital twin information systems},
  author={Sacks, Rafael and Brilakis, Ioannis and Pikas, Ergo and Xie, Haiyan Sally and Girolami, Mark},
  journal={Data-Centric Engineering},
  volume={1},
  pages={e14},
  year={2020},
  publisher={Cambridge University Press}
}

@article{jans2020digital,
    title={Digital twin of an urban-integrated hydroponic farm},
    volume={1},
    DOI={10.1017/dce.2020.21},
    journal={Data-Centric Engineering},
    publisher={Cambridge University Press},
    author={Jans-Singh, Melanie and Leeming, Kathryn and Choudhary, Ruchi and Girolami, Mark},
    year={2020},
    pages={e20}
}

@article{barricelli2019survey,
  title={A Survey on Digital Twin: Definitions, Characteristics, Applications, and Design Implications},
  author={Barricelli, Barbara Rita and Casiraghi, Elena and Fogli, Daniela},
  journal={IEEE Access},
  volume={7},
  pages={167653--167671},
  year={2019},
  publisher={IEEE}
}

@article{niederer2021scaling,
  title={Scaling digital twins from the artisanal to the industrial},
  author={Niederer, Steven A. and Sacks, Michael S. and Girolami, Mark and Willcox, Karen},
  journal={Nature Computational Science},
  volume={1},
  number={5},
  pages={313--320},
  year={2021},
  publisher={Nature Publishing Group}
}

@article{jones2020characterising,
  title={Characterising the Digital Twin: A systematic literature review},
  author={Jones, David and Snider, Chris and Nassehi, Aydin and Yon, Jason and Hicks, Ben},
  journal={CIRP Journal of Manufacturing Science and Technology},
  volume={29},
  pages={36--52},
  year={2020},
  publisher={Elsevier}
}

@article{tuegel2011reengineering,
	Author = {Tuegel, Eric J. and Ingraffea, Anthony R. and Eason, Thomas G. and Spottswood, S. Michael},
	Da = {2011/10/23},
	Doi = {10.1155/2011/154798},
	Isbn = {1687-5966},
	Journal = {International Journal of Aerospace Engineering},
	Pages = {154798},
	Publisher = {Hindawi Publishing Corporation},
	Title = {Reengineering Aircraft Structural Life Prediction Using a Digital Twin},
	Ty = {JOUR},
	Url = {https://doi.org/10.1155/2011/154798},
	Volume = {2011},
	Year = {2011}
}

@article{corral2020digital,
    author = {Corral-Acero, Jorge and Margara, Francesca and Marciniak, Maciej and Rodero, Cristobal and Loncaric, Filip and Feng, Yingjing and Gilbert, Andrew and Fernandes, Joao F and Bukhari, Hassaan A and Wajdan, Ali and Martinez, Manuel Villegas and Santos, Mariana Sousa and Shamohammdi, Mehrdad and Luo, Hongxing and Westphal, Philip and Leeson, Paul and DiAchille, Paolo and Gurev, Viatcheslav and Mayr, Manuel and Geris, Liesbet and Pathmanathan, Pras and Morrison, Tina and Cornelussen, Richard and Prinzen, Frits and Delhaas, Tammo and Doltra, Ada and Sitges, Marta and Vigmond, Edward J and Zacur, Ernesto and Grau, Vicente and Rodriguez, Blanca and Remme, Espen W and Niederer, Steven and Mortier, Peter and McLeod, Kristin and Potse, Mark and Pueyo, Esther and Bueno-Orovio, Alfonso and Lamata, Pablo},
    title = {The `{Digital Twin}' to enable the vision of precision cardiology},
    journal = {European Heart Journal},
    volume = {41},
    number = {48},
    pages = {4556--4564},
    year = {2020},
    issn = {0195-668X},
    doi = {10.1093/eurheartj/ehaa159},
    url = {https://doi.org/10.1093/eurheartj/ehaa159},
    eprint = {https://academic.oup.com/eurheartj/article-pdf/41/48/4556/46616698/ehaa159.pdf},
}

@article{lu2020digital,
  title={Digital Twin-driven smart manufacturing: Connotation, reference model, applications and research issues},
  author={Lu, Yuqian and Liu, Chao and Wang, Kevin I-Kai and Huang, Huiyue and Xu, Xun},
  journal={Robotics and Computer-Integrated Manufacturing},
  volume={61},
  pages={101837},
  year={2020},
  publisher={Elsevier}
}

@article{roy2011comprehensive,
title = {A comprehensive framework for verification, validation, and uncertainty quantification in scientific computing},
journal = {Computer Methods in Applied Mechanics and Engineering},
volume = {200},
number = {25},
pages = {2131--2144},
year = {2011},
issn = {0045-7825},
doi = {https://doi.org/10.1016/j.cma.2011.03.016},
url = {https://www.sciencedirect.com/science/article/pii/S0045782511001290},
author = {Christopher J. Roy and William L. Oberkampf}
}

@article{galappaththige2022credibility,
  title={Credibility assessment of patient-specific computational modeling using patient-specific cardiac modeling as an exemplar},
  author={Galappaththige, Suran and Gray, Richard A. and Costa, Caroline Mendonca and Niederer, Steven and Pathmanathan, Pras},
  journal={PLoS Computational Biology},
  volume={18},
  number={10},
  pages={e1010541},
  year={2022},
  publisher={Public Library of Science San Francisco, CA USA}
}

@book{amse2018assessing,
  title={Assessing Credibility of Computational Modeling through Verification and Validation: Application to Medical Devices},
  author={{ASME}},
  year={2018},
  publisher={American Society of Mechanical Engineers},
  address={New York, NY},
  note={ASME V\&V 40-2018}
}

@ARTICLE{dahmen2022verification,
      author       = {Dahmen, Ulrich Richard and Osterloh, Tobias and Roßmann,
                      Heinz-Jürgen},
      title        = {Verification and validation of digital twins and virtual
                      testbeds},
      journal      = {International Journal of Advances in Applied Sciences},
      volume       = {11},
      number       = {1},
      pages        = {47--64},
      year         = {2022},
      cin          = {615210},
      cid          = {$I:(DE-82)615210_20140620$},
      typ          = {PUB:(DE-HGF)16},
      doi          = {10.11591/ijaas.v11.i1.pp47-64},
      url          = {https://publications.rwth-aachen.de/record/843535},
}

@inproceedings{khan2018digital,
  title={Digital twin for legacy systems: Simulation model testing and validation},
  author={Khan, Adnan and Dahl, Martin and Falkman, Petter and Fabian, Martin},
  booktitle={2018 IEEE 14th International Conference on Automation Science and Engineering (CASE)},
  pages={421--426},
  year={2018},
  organization={IEEE}
}

@article{hemmler2019patient,
  title={Patient-specific in silico endovascular repair of abdominal aortic aneurysms: Application and validation},
  author={Hemmler, Andr{\'e} and Lutz, Brigitta and Kalender, G{\"u}nay and Reeps, Christian and Gee, Michael W.},
  journal={Biomechanics and Modeling in Mechanobiology},
  volume={18},
  number={4},
  pages={983--1004},
  year={2019},
  publisher={Springer}
}

@article{larrabide2012fast,
	Title = {Fast virtual deployment of self-expandable stents: Method and in vitro evaluation for intracranial aneurysmal stenting},
	Author = {Larrabide, Ignacio and Kim, Minsuok and Augsburger, Luca and Villa-Uriol, Maria Cruz and Rüfenacht, Daniel and Frangi, Alejandro F.},
	DOI = {10.1016/j.media.2010.04.009},
	Number = {3},
	Volume = {16},
	Year = {2012},
	Journal = {Medical Image Analysis},
	ISSN = {1361-8415},
	Pages = {721--730},
	URL = {https://doi.org/10.1016/j.media.2010.04.009},
}

@article{coorey2022health,
  title={The health digital twin to tackle cardiovascular disease—a review of an emerging interdisciplinary field},
  author={Coorey, Genevieve and Figtree, Gemma A. and Fletcher, David F. and Snelson, Victoria J. and Vernon, Stephen Thomas and Winlaw, David and Grieve, Stuart M. and McEwan, Alistair and Yang, Jean Yee Hwa and Qian, Pierre and O'Brien, Kieran and Orchard, Jessica and Kim, Jinman and Patel, Sanjay and Redfern, Julie},
  journal={npj Digital Medicine},
  volume={5},
  number={1},
  pages={126},
  year={2022},
  publisher={Nature Publishing Group}
}

@article{kochunas2021digital,
  title={Digital Twin Concepts with Uncertainty for Nuclear Power Applications},
  author={Kochunas, Brendan and Huan, Xun},
  journal={Energies},
  volume={14},
  number={14},
  pages={4235},
  year={2021},
  publisher={MDPI}
}

@incollection{grieves2017digital,
  title={Digital Twin: Mitigating Unpredictable, Undesirable Emergent Behavior in Complex Systems},
  author={Grieves, Michael and Vickers, John},
  booktitle={Transdisciplinary Perspectives on Complex Systems},
  pages={85--113},
  year={2017},
  publisher={Springer}
}

@article{kapteyn2021probabilistic,
  title={A probabilistic graphical model foundation for enabling predictive digital twins at scale},
  author={Kapteyn, Michael G. and Pretorius, Jacob V. R. and Willcox, Karen E.},
  journal={Nature Computational Science},
  volume={1},
  number={5},
  pages={337--347},
  year={2021},
  publisher={Nature Publishing Group}
}

@article{masison2021modular,
  title={A modular computational framework for medical digital twins},
  author={Masison, J. and Beezley, J. and Mei, Y. and Ribeiro, H. A. L. and Knapp, A. C. and {Sordo Vieira}, L. and Adhikari, B. and Scindia, Y. and Grauer, M. and Helba, B. and Schroeder, W. and Mehrad, B. and Laubenbacher, R.},
  journal={Proceedings of the National Academy of Sciences},
  volume={118},
  number={20},
  pages={e2024287118},
  year={2021},
  publisher={National Acad Sciences}
}

@book{hernan2020causal,
  title={Causal Inference: What If},
  author={Hern{\'a}n, Miguel A. and Robins, James M.},
  year={2020},
  publisher={Chapman and Hall/CRC},
  address={Boca Raton}
}

@article{imbens2004confidence,
  title={Confidence intervals for partially identified parameters},
  author={Imbens, Guido W. and Manski, Charles F.},
  journal={Econometrica},
  volume={72},
  number={6},
  pages={1845--1857},
  year={2004},
  publisher={Wiley Online Library}
}

@article{chernozhukov2007estimation,
 ISSN = {00129682, 14680262},
 URL = {http://www.jstor.org/stable/4502031},
 author = {Victor Chernozhukov and Han Hong and Elie Tamer},
 journal = {Econometrica},
 number = {5},
 pages = {1243--1284},
 publisher = {[Wiley, The Econometric Society]},
 title = {Estimation and Confidence Regions for Parameter Sets in Econometric Models},
 urldate = {2022-12-09},
 volume = {75},
 year = {2007}
}

@inproceedings{namkoong2020offpolicy,
 author = {Namkoong, Hongseok and Keramati, Ramtin and Yadlowsky, Steve and Brunskill, Emma},
 booktitle = {Advances in Neural Information Processing Systems},
 pages = {18819--18831},
 title = {Off-policy Policy Evaluation For Sequential Decisions Under Unobserved Confounding},
 url = {https://proceedings.neurips.cc/paper/2020/file/da21bae82c02d1e2b8168d57cd3fbab7-Paper.pdf},
 volume = {33},
 year = {2020}
}

@book{rosenbaum2002observational,
  title={Observational Studies},
  author={Rosenbaum, Paul R.},
  year={2002},
  publisher={Springer},
  address={New York, NY}
}

@article{tan2006distributional,
  title={A distributional approach for causal inference using propensity scores},
  author={Tan, Zhiqiang},
  journal={Journal of the American Statistical Association},
  volume={101},
  number={476},
  pages={1619--1637},
  year={2006},
  publisher={Taylor \& Francis}
}

@inproceedings{hussain2022falsification,
 author = {Hussain, Zeshan M. and Oberst, Michael and Shih, Ming-Chieh and Sontag, David},
 booktitle = {Advances in Neural Information Processing Systems},
 volume = {35},
 pages = {6161--6174},
 title = {Falsification before Extrapolation in Causal Effect Estimation},
 year = {2022}
}

@article{cox1975note,
  title={A note on data-splitting for the evaluation of significance levels},
  author={Cox, David R.},
  journal={Biometrika},
  volume={62},
  number={2},
  pages={441--444},
  year={1975},
  publisher={Oxford University Press}
}

@article{yadlowsky2022bounds,
author = {Steve Yadlowsky and Hongseok Namkoong and Sanjay Basu and John Duchi and Lu Tian},
title = {{Bounds on the conditional and average treatment effect with unobserved confounding factors}},
volume = {50},
journal = {The Annals of Statistics},
number = {5},
publisher = {Institute of Mathematical Statistics},
pages = {2587--2615},
year = {2022},
doi = {10.1214/22-AOS2195},
URL = {https://doi.org/10.1214/22-AOS2195}
}

@book{manski1995identification,
  title={{Identification Problems in the Social Sciences}},
  author={Manski, Charles F.},
  year={1995},
  publisher={Harvard University Press}
}

@article{allamaa2022sim2real,
title = {Sim2real for Autonomous Vehicle Control using Executable Digital Twin},
journal = {IFAC-PapersOnLine},
volume = {55},
number = {24},
pages = {385--391},
year = {2022},
note = {10th IFAC Symposium on Advances in Automotive Control AAC 2022},
issn = {2405-8963},
doi = {https://doi.org/10.1016/j.ifacol.2022.10.314},
url = {https://www.sciencedirect.com/science/article/pii/S2405896322023461},
author = {Jean Pierre Allamaa and Panagiotis Patrinos and Herman {Van der Auweraer} and Tong Duy Son}
}

@InProceedings{zhang2022partial,
  title = 	 {Partial Counterfactual Identification from Observational and Experimental Data},
  author =       {Zhang, Junzhe and Tian, Jin and Bareinboim, Elias},
  booktitle = 	 {Proceedings of the 39th International Conference on Machine Learning},
  pages = 	 {26548--26558},
  year = 	 {2022},
  editor = 	 {Chaudhuri, Kamalika and Jegelka, Stefanie and Song, Le and Szepesvari, Csaba and Niu, Gang and Sabato, Sivan},
  volume = 	 {162},
  series = 	 {Proceedings of Machine Learning Research},
  month = 	 {17--23 Jul},
  publisher =    {PMLR},
  url = 	 {https://proceedings.mlr.press/v162/zhang22ab.html}
}

@incollection{molinari2020chapter,
title = {Microeconometrics with partial identification},
editor = {Steven N. Durlauf and Lars Peter Hansen and James J. Heckman and Rosa L. Matzkin},
series = {Handbook of Econometrics},
publisher = {Elsevier},
volume = {7},
pages = {355--486},
year = {2020},
booktitle = {Handbook of Econometrics, Volume 7A},
issn = {1573-4412},
doi = {https://doi.org/10.1016/bs.hoe.2020.05.002},
url = {https://www.sciencedirect.com/science/article/pii/S1573441220300027},
author = {Francesca Molinari}
}

@article{imbens2009recent,
 ISSN = {00220515},
 URL = {http://www.jstor.org/stable/27647134},
 author = {Guido W. Imbens and Jeffrey M. Wooldridge},
 journal = {Journal of Economic Literature},
 number = {1},
 pages = {5--86},
 publisher = {American Economic Association},
 title = {Recent Developments in the Econometrics of Program Evaluation},
 urldate = {2026-04-08},
 volume = {47},
 year = {2009}
}

@article{bugni2015specification,
title = {Specification tests for partially identified models defined by moment inequalities},
journal = {Journal of Econometrics},
volume = {185},
number = {1},
pages = {259--282},
year = {2015},
issn = {0304-4076},
doi = {https://doi.org/10.1016/j.jeconom.2014.10.013},
url = {https://www.sciencedirect.com/science/article/pii/S0304407614002577},
author = {Federico A. Bugni and Ivan A. Canay and Xiaoxia Shi}
}

@article{kitagawa2015test,
author = {Kitagawa, Toru},
title = {A Test for Instrument Validity},
journal = {Econometrica},
volume = {83},
number = {5},
pages = {2043--2063},
doi = {https://doi.org/10.3982/ECTA11974},
url = {https://onlinelibrary.wiley.com/doi/abs/10.3982/ECTA11974},
eprint = {https://onlinelibrary.wiley.com/doi/pdf/10.3982/ECTA11974},
year = {2015}
}

@article{mourifie2017testing,
    author = {Mourifié, Ismael and Wan, Yuanyuan},
    title = {Testing Local Average Treatment Effect Assumptions},
    journal = {The Review of Economics and Statistics},
    volume = {99},
    number = {2},
    pages = {305--313},
    year = {2017},
    issn = {0034-6535}
}

@article{canay2023users,
title = {A User’s guide for inference in models defined by moment inequalities},
journal = {Journal of Econometrics},
pages = {105558},
year = {2023},
issn = {0304-4076},
doi = {https://doi.org/10.1016/j.jeconom.2023.105558},
url = {https://www.sciencedirect.com/science/article/pii/S0304407623002749},
author = {Ivan A. Canay and Gastón Illanes and Amilcar Velez}
}

@article{rosen2008confidence,
title = {Confidence sets for partially identified parameters that satisfy a finite number of moment inequalities},
journal = {Journal of Econometrics},
volume = {146},
number = {1},
pages = {107--117},
year = {2008},
issn = {0304-4076},
doi = {https://doi.org/10.1016/j.jeconom.2008.08.001},
url = {https://www.sciencedirect.com/science/article/pii/S0304407608000791},
author = {Adam M. Rosen}
}

@article{romano2008inference,
title = {Inference for identifiable parameters in partially identified econometric models},
journal = {Journal of Statistical Planning and Inference},
volume = {138},
number = {9},
pages = {2786--2807},
year = {2008},
note = {Special Issue in Honor of Theodore Wilbur Anderson, Jr. on the Occasion of his 90th Birthday},
issn = {0378-3758},
doi = {https://doi.org/10.1016/j.jspi.2008.03.015},
url = {https://www.sciencedirect.com/science/article/pii/S0378375808000694},
author = {Joseph P. Romano and Azeem M. Shaikh},
}

@article{andrews2010inference,
author = {Andrews, Donald W. K. and Soares, Gustavo},
title = {Inference for Parameters Defined by Moment Inequalities Using Generalized Moment Selection},
journal = {Econometrica},
volume = {78},
number = {1},
pages = {119--157},
doi = {https://doi.org/10.3982/ECTA7502},
url = {https://onlinelibrary.wiley.com/doi/abs/10.3982/ECTA7502},
eprint = {https://onlinelibrary.wiley.com/doi/pdf/10.3982/ECTA7502},
year = {2010}
}

@article{romano2010inference,
  author = {Romano, Joseph P. and Shaikh, Azeem M.},
  title = {Inference for the Identified Set in Partially Identified Econometric Models},
  journal = {Econometrica},
  volume = {78},
  number = {1},
  pages = {169--211},
  doi = {https://doi.org/10.3982/ECTA6706},
  url = {https://onlinelibrary.wiley.com/doi/abs/10.3982/ECTA6706},
  eprint = {https://onlinelibrary.wiley.com/doi/pdf/10.3982/ECTA6706},
  year = {2010}
}

@article{romano2014practical,
 ISSN = {00129682, 14680262},
 URL = {http://www.jstor.org/stable/24029299},
 author = {Joseph P. Romano and Azeem M. Shaikh and Michael Wolf},
 journal = {Econometrica},
 number = {5},
 pages = {1979--2002},
 publisher = {The Econometric Society},
 title = {A PRACTICAL TWO-STEP METHOD FOR TESTING MOMENT INEQUALITIES},
 urldate = {2026-04-11},
 volume = {82},
 year = {2014}
}

\end{document}